\documentclass[prd,aps,nofootinbib,twocolumn,floatfix,10 pt]{revtex4}
\usepackage{amssymb}
\usepackage{amsmath,graphicx,color,epsfig}
\begin{document}

\title{$K_{1}(1270)-K_{1}(1400)$ mixing and the fourth generation SM effects in $B\rightarrow K_{1}\ell^{+}\ell^{-}$ decays}
\author{Aqeel Ahmed$\footnote{aqeel@ncp.edu.pk}^{1}$, Ishtiaq Ahmed$\footnote{ishtiaq@ncp.edu.pk}^{1,2}$, M. Ali Paracha$\footnote{ali@ncp.edu.pk}^{1,2}$ and Abdur Rehman$\footnote{rehman@ncp.edu.pk}^{1}$}
\affiliation{
$^{1}$National Centre for Physics,
Quaid-i-Azam University Campus,
Islamabad, 45320 Pakistan\\
$^{2}$Physics Department, Quaid-i-Azam University, Islamabad, 45320 Pakistan}

\date{\today}
\begin{abstract}
The implications of the fourth generation quarks in the $B\rightarrow
K_{1}(1270,1400)\ell^{+}\ell^{-}$ with $\ell=\mu , \tau$ decays are studied,
where the mass eigenstates $K_{1}(1270)$ and $K_{1}(1400)$ are the mixture of $^{1}{P}_{1}$ and $^{3}{P}_{1}$ states with the mixing
angle $\theta_{K}$. In this context, we have studied various observables like branching ratio $(\mathcal{BR})$, forward backward
asymmetry $(\mathcal{A}_{FB})$ and longitudinal and transverse helicity fractions $(f_{L,T})$ of $K_{1}$ meson in $B\to K_{1}\ell^{+}\ell^{-}$ decays. To study these observables, we have used the Light Cone QCD sum rules form factors and set the mixing angle $\theta_{K}=-34^{\circ}$. It is noticed that the $\mathcal{BR}$ is suppressed for $K_{1}(1400)$
as a final state meson compared to that of $K_{1}(1270)$. Same is the case when the final state leptons are tauons rather than muons. In both the situations all the above mentioned observables are quite sensitive to the fourth generation effects. Hence the measurements of these observables at LHC, for the above mentioned processes can serve as a good tool to investigate the indirect searches for the existence of fourth generation quarks.
\end{abstract}
\maketitle



\section{Introduction}\label{intro}

The Standard Model (SM) with one Higgs boson is the simplest and has been tested with great precision. But with all of its successes, it has some
theoretical shortcomings which impede its status as a fundamental
theory. One such shortcoming is the so-called hierarchy problem, the various extensions and the SM differ in the solutions of this problem. These extensions are: the two Higgs doublet models
(2HDM), Minimal Supersymmetric SM (MSSM), Universal Extra Dimension
(UED) model and SM4. SM4, implying a fourth family of quarks and leptons seems to be the most economical in number of additional particles and simpler in the sense that it does not introduce any new operators. It thus provides a natural extension of the SM which has been searched for previously by the
LEP and Tevatron and now will be investigated at the LHC \cite{4sm}.
If a fourth family is discovered, it is likely to have consequences
at least as profound as those that have emerged from the discovery
of the third family. The fourth generation SM is not only provide a
simple explanation of the experimental results which are difficult
to reconcile with SM including CP violation anomaly \cite{15, 16} but also give enough CP asymmetries to facilitate baryogenesis \cite{18}. By the addition of fourth generation Cabibbo-Kobayashi-Maskawa (CKM) matrix become $4\times
4$ unitary matrix which requires six real parameters and three
phases. These two extra phases imply the possibility of extra
sources of CP violation. In addition, the fact that the heavier
quarks ($t^{\prime },$ $b^{\prime }$) and leptons ($\nu ^{\prime
},$ $\ell^{\prime }$) of the fourth generation can play a crucial role
in dynamical electroweak symmetry breaking (DEWSB) \cite{dewsb} as an economical way to address the hierarchy puzzle
which renders this extension of SM.
Furthermore the LHC will provide a suitable amount of data which
enlighten these puzzles more clearly as well as decide the faith
of the extra generation to undimmed the smog from theoretical
picture and help us to enhance our theoretical understanding.

In the past few years a number of analysis showed: (a) SM with fourth generation is
consistent with the electroweak precision test (EWPT) \cite{21, 22, 24} and it is pointed out in \cite{22, 24, 27} that in
the presence of fourth generation a heavy Higgs boson does not
contradict with EWPT, (b) SU(5) gauge coupling unification could be
achieved without
supersymmetry \cite{28}, (c) Electroweak baryogenesis can be accomodated \cite%
{29} and (d) As mentioned earlier that the DEWSB might be
actuated by the presence of extra generation \cite{dewsb}. Moreover, the fourth generation SM, in principle, could resolve the
certain anomalies present in flavor changing processes \cite{39}. Also the mismatch in the $CP$ asymmetry in $B\to K\pi$ data
\cite{hfag} with the SM \cite{bashi2} as well
 as CP violation in $B\rightarrow
\phi K_{s}$ decay may also provide some hint of new physics (NP) \cite{abds1}.
Henceforth the measurement of different observables in the rare $B$
decays can also be very helpful to put or check the constraints on
the $4$th generation parameters.

In general there are two ways to search the NP: one is the direct search where we can produce
the new particles by raising the energy of colliders and the other
one is indirect search, i.e. to increase the experimental precision
on the data of different SM processes where the NP effects can manifest themselves. The processes that are
suitable for indirect searches of NP  are those which are forbidden
or very rare in the SM and can be measured precisely. In this
context the rare $B$ decays mediated through the flavor changing
neutral current (FCNC) processes provide a potentially effective
testing ground to look for the physics in and beyond the SM. In the
SM, these FCNC transitions are not allowed at tree level but are
allowed at loop level through Glashow-Iliopoulos-Maiani (GIM)
mechanism \cite{GIM}. In the context of SM, the rare $B$ decays are
quite interesting because they provide a quantitative determination
of the quark flavor rotation matrix, in particular the matrix
elements $V_{tb}$, $V_{ts}$ and $V_{td}$ \cite{AAli}.

The exploration of Physics beyond the SM through various inclusive
$B$ meson decays like $B\to X_{s,d}\ell^{+}\ell^{-}$ and their
corresponding exclusive processes, $B\to M\ell^{+}\ell^{-}$ with $M
= K,K^{\ast},K_{1},\rho$ etc., have already been done  \cite{bst, 63}.
These studies showed that the above mentioned inclusive and
exclusive decays of $B$ meson are very sensitive to the flavor
structure of the SM and provide a windowpane for any NP including
the fourth generation SM. Therefore, the direct searches like the study of production, decay channels and the signals of the
existence of fourth generation quarks and leptons in the present
colliders are being performed. Since it is expected that $m_{t^{\prime}}>m_{t}$ thus the fourth generation quark can be manifest
their indirect existence in the loop diagrams. Due to this reason
FCNC transitions are at the forefront and one of the main research
direction of all operating $B$ factories including CLEO, Belle,
Tevatron and LHCb \cite{4sm}. However, the studies that involve the direct
searches of the fourth generation quarks or their indirect searches
via FCNC processes require the values of the quark masses and mixing
elements which are not free parameters but rather they are
constrained by experiments \cite{abds}.

There are two different ways to incorporate the NP effects
in the rare decays, one through the Wilson coefficients and the
other through new operators which are absent in the SM. In the
fourth generation SM the NP arises due to the modified Wilson
coefficients $C_{7}^{eff}$ ,$C_{9}^{eff}$ and $C_{10}^{eff}$ as the fourth generation quark $(t^{\prime})$ contributes in $b\to s(d)$ transition at the loop
level along with other quarks $u$, $c$ and $t$ of SM. It is
necessary to mention here the FCNC decay modes like $B\rightarrow
X_{s}\ell^{+}\ell^{-}$,$B\rightarrow K^{\ast}\ell^{+}\ell^{-}$ \cite{Kast} and
$B\rightarrow K\ell^{+}\ell^{-}$ are also very useful in the
determination of precise values of $C_{7}^{eff}$ ,$C_{9}^{eff}$ and
$C_{10}^{eff}$ Wilson coefficients \cite{inali} as well as sign
information on $C_{7}^{eff}$. Moreover, the measured branching ratio
$b\rightarrow s\gamma$ by CLEO \cite{inali1} has been used to
constraint the Wilson coefficient $C_{7}^{eff}$ \cite{bashi10}.

With our motivation stated above, the complementary information from the rare B decays is necessary for the indirect searches of NP including fourth generation. This complementary investigation improve the precession of SM
parameters which are helpful in discovery of the NP. In this
connection, like the rare semileptonic decays involving $B\rightarrow (X_{s},K^{\ast},K)\ell^{+}\ell^{-}$, the $B\rightarrow K_{1}(1270,1400)\ell^{+}\ell^{-}$ decays are also rich in phenomenology for the NP \cite{IPJ}. In some sense they are more interesting and more sophisticated to NP since they are mixture of $K_{1A}$ and $K_{1B}$, where $K_{1A}$ and $K_{1B}$ are $^3P_1$ and
$^1P_1$ states, respectively. The physical states $K_{1}(1270)$ and $K_{1}(1400)$ can be obtained by the mixing of $K_{1A}$ and $K_{1B}$ as
\begin{subequations}
\begin{eqnarray}
\vert K_{1}(1270)\rangle &=& \vert K_{1A}\rangle\sin{\theta_{K}}+\vert K_{1B}\rangle\cos{\theta_{K}},\label{mix1}\\
\vert K_{1}(1400)\rangle &=& \vert K_{1A}\rangle\cos{\theta_{K}}-\vert K_{1B}\rangle\sin{\theta_{K}},\label{mix2}
\end{eqnarray}
\end{subequations}
where the magnitude of mixing angel $\theta_{K}$ has been estimated to be $34^{\circ}\leq |\theta_{K}|\leq 58^{\circ}$ in Ref. \cite{Suzuki}. Recently, from the study of $B\to K_{1}(1270)\gamma$ and $\tau \to K_{1}(1270)\nu_{\tau}$, the value of $\theta_{K}$ has been estimated to be $\theta_{K}=-(34\pm13)^{\circ}$, where the minus sign of $\theta_{K}$ is related to the chosen phase of $\vert K_{1A}\rangle$ and $\vert K_{1B}\rangle$ \cite{HY}. Getting an independent conformation of this value of mixing angle $\theta_{K}$ is by itself interesting. As we shall see that this particular choice suppresses the ${\cal BR}$ for $K_{1}(1400)$ in the final state compared to $K_{1}(1270)$, which can be tested.

Many studies have already shown \cite{IPJ} that the observables like branching ratio
$(\mathcal{BR})$, forward-backward asymmetry $(\mathcal{A}_{FB})$ and helicity fractions $f_{L,T}$ for semileptonic B decays are greatly influenced under different scenarios beyond the SM. Therefore, the precise measurement of these observables will play an important role in the indirect searches of NP. In this respect, it is natural to ask how these observables are influenced by the fourth generation parameters. The purpose of present study addresses this question i.e. investigate the possibility of searching NP due to the fourth generation SM in $B\rightarrow K_{1}(1270, 1400)\ell^{+}\ell^{-}$ decays with $\ell=\mu,\tau$ using the above mentioned observables.

The plan of the manuscript is as follows. In sec. \ref{tf}, we fill our toolbox with the theoretical framework needed to study the said process in the fourth generation SM. In Sec. \ref{mix}, we present the mixing of $K_{1}(1270)$ and $K_{1}(1400)$ and the form factors used in this study. In Sec. \ref{obs}, we discuss the observables of $B\rightarrow K_{1}\ell^{+}\ell^{-}$ in detail. In Sec. \ref{num}, we give the numerical analysis of our observables and discuss the sensitivity of these observables with the fourth generation SM scenario. We conclude our findings in Sec. \ref{conc}.

\section{Theoretical Framework}\label{tf}

At the quark level $B\rightarrow K_{1}(1270,1400)\ell^{+}\ell^{-}$ decays are induced by the transition $b\rightarrow s\ell^{+}\ell^{-}$, which in the SM, is described by the following effective Hamiltonian
\begin{eqnarray}
\mathcal{H}_{eff} &=&-\frac{4 G_{F}}{\sqrt{2}}V_{tb}V_{ts}^{\ast }\bigg[{%
\sum\limits_{i=1}^{10}}C_{i}({\mu })O_{i}({\mu })\bigg],
\label{effective haniltonian 1}
\end{eqnarray}
where $O_{i}({\mu })$ $(i=1,\ldots ,10)$ are the four-quark operators and $%
C_{i}({\mu })$ are the corresponding Wilson coefficients at the energy scale
${\mu }$ \cite{Goto}. Using renormalization group equations to resum the QCD
corrections, Wilson coefficients are evaluated at the energy scale ${\mu =m}%
_{b}$. The theoretical uncertainties associated with the renormalization
scale can be considerably reduced when the next-to-leading-logarithm
corrections are included.

The explicit expressions of the operators responsible for exclusive $B \to
K_{1} (1270,1400)\ell^{+}\ell^{-}$ decays are given by
\begin{eqnarray}
O_{7} &=&\frac{e^{2}}{16\pi ^{2}}m_{b}\left( \bar{s}\sigma _{\mu \nu
}R b\right) F^{\mu \nu },\  \\
O_{9} &=&\frac{e^{2}}{16\pi ^{2}}(\bar{s}\gamma _{\mu }L b)(\bar{\ell}\gamma
^{\mu }\ell),\  \\
O_{10} &=&\frac{e^{2}}{16\pi ^{2}}(\bar{s}\gamma _{\mu }L b)(\bar{\ell}%
\gamma ^{\mu }\gamma _{5}\ell),  \label{wc}
\end{eqnarray}
with $R,L=\left( 1\pm \gamma _{5}\right)/2$. In terms of the above
Hamiltonian, the free quark decay amplitude for $b\rightarrow s\ell^{+}\ell^{-}
$ can be written as:

\begin{align}  \label{qa}
&\mathcal{M}_{SM}(b \rightarrow s \ell^{+}\ell^{-})
= -\frac{G_{F}\alpha}{\sqrt{2}%
\pi } V_{tb}V_{ts}^{\ast } \bigg\{ C_{9}^{effSM}(\bar{s}\gamma _{\mu
}L b)(\bar{\ell} \gamma ^{\mu
}\ell)\notag\\
&+C_{10}(\bar{s}\gamma _{\mu }L b)(\bar{\ell}\gamma ^{\mu }\gamma
_{5}\ell)
-2m_{b}C_{7}^{effSM}(\bar{s}i\sigma _{\mu \nu }\frac{q^{\nu }}{q^{2}}R b)(\bar{ \ell%
}\gamma ^{\mu }\ell) %
\bigg\}
\end{align}
where $q$ is the momentum transfer. The operator $O_{10}$ can not be
induced by the insertion of four-quark operators because of the
absence of the neutral $Z$ boson in the effective theory. Hence, the
Wilson coefficient $C_{10}$ is not renormalized under QCD
corrections and therefore it is independent of the energy scale. In
addition to this, the above quark level decay amplitude can take
contributions from the matrix elements of four-quark operators,
$\sum_{i=1}^{6}\langle \ell^{+}\ell^{-}s|O_{i}|b\rangle $,
which are usually absorbed into the effective Wilson coefficient $%
C_{9}^{effSM}(\mu )$, which can be decomposed into the following three parts
\cite{bst, 63}
\begin{equation*}
C_{9}^{effSM}(\mu )=C_{9}(\mu )+Y_{SD}(z,s^{\prime })+Y_{LD}(z,s^{\prime }),
\end{equation*}
where the parameters $z$ and $s^{\prime }$ are defined as $%
z=m_{c}/m_{b},\,\,\,s^{\prime }=q^{2}/m_{b}^{2}$. $Y_{SD}(z,s^{\prime })$
describe the short-distance contributions from four-quark operators far
away from the $c\bar{c}$ resonance regions, which can be calculated reliably
in the perturbative theory. The long-distance contributions $%
Y_{LD}(z,s^{\prime })$ from four-quark operators near the $c\bar{c}$
resonance cannot be calculated from first principles of QCD and are usually
parameterized in the form of a phenomenological Breit-Wigner formula, by making
use of the vacuum saturation approximation and quark-hadron duality.
The explicit expressions for $Y_{SD}(z,s^{\prime })$ and $Y_{LD}(z,s^{\prime })$ are
\begin{align}
Y_{SD}(z,s^{\prime })& =h(z,s^{\prime })\{3C_{1}(\mu )+C_{2}(\mu
)+3C_{3}(\mu ) \notag \\&+C_{4}(\mu ) +3C_{5}(\mu )+C_{6}(\mu )\}
 \notag \\&-\frac{1}{2}h(1,s^{\prime })\{4C_{3}(\mu ) +4C_{4}(\mu )+3C_{5}(\mu)\}
 \notag \\&-\frac{1}{2}h(0,s^{\prime })\{C_{3}(\mu )+3C_{4}(\mu )\}\notag \\&+{\frac{2}{9}}%
\{3C_{3}(\mu )+C_{4}(\mu )+3C_{5}(\mu )+C_{6}(\mu )\},
\end{align}
with
\begin{align}
h(z,s^{\prime }) &=-{\frac{8}{9}}\mathrm{ln}z+{\frac{8}{27}}+{\frac{4}{9}}x-%
{\frac{2}{9}}(2+x)|1-x|^{1/2}\notag \\&\times\left\{
\begin{array}{l}
\ln \left| \frac{\sqrt{1-x}+1}{\sqrt{1-x}-1}\right| -i\pi \quad \mathrm{for}{%
{\ }x\equiv 4z^{2}/s^{\prime }<1} \\
2\arctan \frac{1}{\sqrt{x-1}}\qquad \mathrm{for}{{\ }x\equiv
4z^{2}/s^{\prime }>1}%
\end{array}
\right.,
\\
h(0,s^{\prime })
&={\frac{8}{27}}-{\frac{8}{9}}\mathrm{ln}{\frac{m_{b}}{\mu
}}-{\frac{4}{9}}\mathrm{ln}s^{\prime }+{\frac{4}{9}}i\pi,
\end{align}
and
\begin{equation}
Y_{LD}\left( z,s^{\prime }\right) =\frac{3\pi }{\alpha ^{2}}%
C^{(0)}\sum\limits_{V_{i}=\psi _{i}}\kappa _{i}\frac{m_{V_{i}}\Gamma \left(
V_{i}\rightarrow l^{+}l^{-}\right) }{m_{V_{i}}^{2}-s^{\prime
}m_{b}^{2}-im_{V_{i}}\Gamma _{V_{i}}}
\end{equation}
where $C^{(0)}=3C_{1}+C_{2}+3C_{3}+C_{4}+3C_{5}+C_{6}$.

Irrespective to this, the non-factorizable effects \cite{bs1} from the charm loop can bring about further corrections
to the radiative $b\rightarrow s\gamma $ transition, which can be absorbed
into the effective Wilson coefficient $C_{7}^{effSM}$. Specifically, the
Wilson coefficient $C^{effSM}_{7}$ takes the form \cite{chen}
\begin{equation*}
C_{7}^{eff}(\mu )=C_{7}(\mu )+C_{b\rightarrow s\gamma }(\mu ),
\end{equation*}
with
\begin{align}
C_{b\rightarrow s\gamma }(\mu ) &=i\alpha
_{s}\bigg[{\frac{2}{9}}\eta
^{14/23}(G_{1}(x_{t})-0.1687)-0.03C_{2}(\mu )\bigg],
\\
G_{1}(x) &={\frac{x(x^{2}-5x-2)}{8(x-1)^{3}}}+{\frac{3x^{2}\mathrm{ln}^{2}x%
}{4(x-1)^{4}}},
\end{align}
where $\eta =\alpha _{s}(m_{W})/\alpha _{s}(\mu )$, $x=m_{t}^{2}/m_{W}^{2}$. $C_{b\rightarrow s\gamma }$ is the absorptive part for the $b\rightarrow sc%
\bar{c}\rightarrow s\gamma $ re-scattering and we have dropped out
the small contributions proportional to CKM sector
$V_{ub}V_{us}^{\ast }$. Furthermore, in the SM, the zero position of the forward-backward asymmetry depends only on the Wilson coefficients \cite{new2} which correspond to the short distance physics. In the present study, our focus is to determine the effects of the fourth family of quarks on different observables. As we will see the NP effects modify only the Wilson coefficients. Therefore, we will ignore the long distance charmonium $c\bar{c}$ contributions in our numerical calculation.

As noted in section \ref{intro}, the NP scenario provided by the fourth
generation quarks is introduced on the same pattern as the three generations in
the SM. Therefore, the operator basis are exactly the same as that of the SM,
while the values of Wilson coefficients in Eq. \eqref{qa} alter
according to
\begin{eqnarray}
C_{7}^{eff}&=&C_{7}^{effSM}+\frac{\lambda_{t^{\prime}}}{\lambda_{t}}C_{7}^{new},\notag\\
C_{9}^{eff}&=&C_{9}^{effSM}+\frac{\lambda_{t^{\prime}}}{\lambda_{t}}C_{9}^{new},\label{effwc}\\
C_{10}^{eff}&=&C_{10}+\frac{\lambda_{t^{\prime}}}{\lambda_{t}}C_{10}^{new},\notag
\end{eqnarray}
where $\lambda_{t}=V^{\ast}_{tb}V_{ts}$ and $\lambda_{t^{\prime}}$
can be parameterized as:
\begin{equation}
\lambda_{t^{\prime}}=\vert
V^{\ast}_{t^{\prime}b}V_{t^{\prime}s}\vert
e^{i\phi},\label{lambdatp}
\end{equation}
where $\phi$ is the phase factor corresponding to the $b\to s$
transition in the fourth generation SM, which we set $90^{\circ}$ \cite{phi} in the forthcoming
numerical analysis of different physical observables. Here
$V_{t^{\prime}b}$ and $V_{t^{\prime}s}$ are the elements of
$4\times4$ CKM extended matrix.  The new
contributions of the fourth generation up quark $t^{\prime}$ at
loop level in $C_{7}^{eff}$, $C_{9}^{eff}$ and $C_{10}^{eff}$ in Eq.
\eqref{effwc} can be obtained from the corresponding SM
counterparts by trading, $m_{t}\to m_{t^{\prime}}$. The
unitarity condition of $4\times 4$ CKM matrix now takes the form
\begin{equation}
V_{t^{\prime }b}V_{t^{\prime }s}^{\ast }=-(V_{ub}V_{us}^{\ast }+
V_{cb}V_{cs}^{\ast }+ V_{tb}V_{ts}^{\ast }).
\end{equation}
If we define $\lambda_{f}=V_{fb}V_{fs}^{\ast }$ then
the unitarity relation can be written in more elegant form
\begin{equation}
\lambda _{t^{\prime }}=-\left(\lambda _{u}+\lambda _{c}+\lambda _{t}\right)
\end{equation}
Notice that this unitarity relation relates the unknown
parameters in terms of the known parameters. Current theoretical
bound on $\lambda_{t^{\prime}}$ value is $\lambda_{t^{\prime}}\leq
1.5\times10^{-2}$ \cite{bound}.

There are different limits on the lower bound of the fourth
generation quark masses. The direct searches at the Tevatron
constrained the $t^{\prime}$ mass, $m_{t^{\prime}}>256$GeV at $90\%$
C.L. \cite{CDF2} and by the decay of $b^{\prime}$ quark to t and
$W^{-}$, they set a limit on $b^{\prime}$ mass,
$m_{b^{\prime}}>338GeV$ at $95\%$ C.L. \cite{CDF1}. Present searches
by CDF(D0) of fourth generation $t^{\prime}$ in their decays to
$Wq$, have excluded $t^{\prime}$ quark with a mass below 335(296)GeV
at $95\%$ CL \cite{CDF3}. In near future, we will see that these
bounds could be considerably improved at LHC. Moreover, the
fourth-generation quark masses are constrained by the perturbative
unitarity of heavy-fermion scattering amplitudes \cite{chano} to be
$m_{t^{\prime}} \leq 500\sim 600$ GeV. However, in our numerical
calculations, we set the bounds $300\leq m_{t^{\prime}}\leq600$ GeV.

\subsection{Form Factors and Mixing of $K_{1}(1270)-K_{1}(1400)$}\label{mix}

The exclusive $B\rightarrow K_{1}(1270,1400)\ell^{+}\ell^{-}$ decays involve the
hadronic matrix elements of quark operators given in Eq. (\ref{qa})
which can be parameterized in terms of the form factors as:
\begin{align}
\left\langle K_{1}(k,\varepsilon )\left\vert V_{\mu }\right\vert
B(p)\right\rangle  &=\varepsilon _{\mu }^{\ast }\left(
M_{B}+M_{K_{1}}\right) V_{1}(q^{2})  \notag \\
&-(p+k)_{\mu }\left( \varepsilon ^{\ast }\cdot q\right) \frac{V_{2}(q^{2})}{%
M_{B}+M_{K_{1}}}  \notag \\
&-q_{\mu }\left( \varepsilon \cdot q\right)
\frac{2M_{K_{1}}}{q^{2}}\left[ V_{3}(q^{2})-V_{0}(q^{2})\right]\notag\\
 \label{tf6} \\
\left\langle K_{1}(k,\varepsilon )\left\vert A_{\mu }\right\vert
B(p)\right\rangle  &=\frac{2i\epsilon _{\mu \nu \alpha \beta }}{%
M_{B}+M_{K_{1}}}\varepsilon ^{\ast \nu }p^{\alpha }k^{\beta }A(q^{2})
\label{tf7}\end{align}
where $V_{\mu }=\bar{s}\gamma _{\mu }b$ and $A_{\mu }=\bar{s}\gamma
_{\mu
}\gamma _{5}b$ are the vectors and axial vector currents, involved in the transition matrix, respectively. Also $%
p(k)$ are the momenta of the $B(K_{1})$ mesons and $\varepsilon
_{\mu }$ correspond to the polarization of the final state axial vector
$K_{1}$ meson. In Eq.(\ref{tf6}) we have
\begin{equation}
V_{3}(q^{2})=\frac{M_{B}+M_{K_{1}}}{2M_{K_{1}}}V_{1}(q^{2})-\frac{M_{B}-M_{K_{1}}}{%
2M_{K_{1}}}V_{2}(q^{2})  \label{tf8}
\end{equation}%
with
\begin{equation*}
V_{3}(0)=V_{0}(0)
\end{equation*}%
In addition, there is also a contribution from the
Penguin form factors which can be written as
\begin{align}
&\left\langle K_{1}(k,\varepsilon )\left\vert \bar{s}i\sigma _{\mu
\nu }q^{\nu }b\right\vert B(p)\right\rangle \notag\\ &=\left[ \left(
M_{B}^{2}-M_{K_{1}}^{2}\right) \varepsilon _{\mu }-(\varepsilon
\cdot
q)(p+k)_{\mu }\right] F_{2}(q^{2})  \notag \\
&+(\varepsilon ^{\ast }\cdot q)\left[ q_{\mu }-\frac{q^{2}}{%
M_{B}^{2}-M_{K_{1}}^{2}}(p+k)_{\mu }\right] F_{3}(q^{2})  \label{tf9}
\end{align}
\begin{equation} \left\langle K_{1}(k,\varepsilon )\left\vert \bar{s}i\sigma
_{\mu \nu }q^{\nu }\gamma _{5}b\right\vert
B(p)\right\rangle=-i\epsilon _{\mu \nu \alpha \beta }\varepsilon
^{\ast \nu }p^{\alpha }k^{\beta }F_{1}(q^{2}) \label{tf10}
\end{equation}%
with $F_{1}(0)=2F_{2}(0).$

As the physical states $K_{1}(1270)$ and $K_{1}(1400)$ are mixed states
of the $K_{1A}$ and $K_{1B}$ with mixing angle $\theta_{K}$ as
defined in Eqs. (\ref{mix1}-\ref{mix2}). The $B\to K_{1}$ form factors can be
parameterized as
\begin{widetext}
\begin{eqnarray}
\left(\begin{array}{c}\langle K_{1}(1270)\vert
\bar{s}\gamma_{\mu}(1-\gamma_{5})b\vert B\rangle\\
\langle K_{1}(1400)\vert \bar{s}\gamma_{\mu}(1-\gamma_{5})b\vert
B\rangle\end{array}\right)&=& M\left(\begin{array}{c}\langle
K_{1A}\vert
\bar{s}\gamma_{\mu}(1-\gamma_{5})b\vert B\rangle\\
\langle K_{1B}\vert \bar{s}\gamma_{\mu}(1-\gamma_{5})b\vert
B\rangle\end{array}\right),\label{m1}\\
\left(\begin{array}{c}\langle K_{1}(1270)\vert
\bar{s}\sigma_{\mu\nu}q^{\mu}(1+\gamma_{5})b\vert B\rangle\\
\langle K_{1}(1400)\vert
\bar{s}\sigma_{\mu\nu}q^{\mu}(1+\gamma_{5})b\vert
B\rangle\end{array}\right)&=& M\left(\begin{array}{c}\langle
K_{1A}\vert
\bar{s}\sigma_{\mu\nu}q^{\mu}(1+\gamma_{5})b\vert B\rangle\\
\langle K_{1B}\vert
\bar{s}\sigma_{\mu\nu}q^{\mu}(1+\gamma_{5})b\vert
B\rangle\end{array}\right),\label{m2}
\end{eqnarray}
\end{widetext}
where the mixing matrix $M$ is
\begin{equation}
M=\left(\begin{array}{cc}\sin\theta_{K}&\cos\theta_{K}\\
\cos\theta_{K}&-\sin\theta_{K}\end{array}\right).\label{m3}
\end{equation}

So the form factors $A^{K_{1}}$, $V_{0,1,2}^{K_{1}}$ and
$F_{0,1,2}^{K_{1}}$ satisfy the following relation
\begin{widetext}
\begin{eqnarray}\left(\begin{array}{c}\frac{A^{K_{1}(1270)}}{m_{B}+m_{K_{1}(1270)}}\\
\frac{A^{K_{1}(1400)}}{m_{B}+m_{K_{1}(1400)}}\end{array}\right) &=&
M\left(\begin{array}{c}\frac{A^{K_{1A}}}{m_{B}+m_{K_{1A}}}\\
\frac{A^{K_{1B}}}{m_{B}+m_{K_{1B}}}\end{array}\right),\label{m4}\\
\left(\begin{array}{c}(m_{B}+m_{K_{1}(1270)})V_{1}^{K_{1}(1270)}\\
(m_{B}+m_{K_{1}(1400)})V_{1}^{K_{1}(1400)}\end{array}\right)&=&
M\left(\begin{array}{c}(m_{B}+m_{K_{1A}})V_{1}^{K_{1A}}\\
(m_{B}+m_{K_{1B}})V_{1}^{K_{1B}}\end{array}\right),\label{m5}\\
\left(\begin{array}{c}\frac{V_{2}^{K_{1}(1270)}}{m_{B}+m_{K_{1}(1270)}}\\
\frac{V_{2}^{K_{1}(1400)}}{m_{B}+m_{K_{1}(1400)}}\end{array}\right)
&=&
M\left(\begin{array}{c}\frac{V_{2}^{K_{1A}}}{m_{B}+m_{K_{1A}}}\\
\frac{V_{2}^{K_{1B}}}{m_{B}+m_{K_{1B}}}\end{array}\right),\label{m6}\\
\left(\begin{array}{c}m_{K_{1}(1270)}V_{0}^{K_{1}(1270)}\\
m_{K_{1}(1400)}V_{0}^{K_{1}(1400)}\end{array}\right) &=&
M\left(\begin{array}{c}m_{K_{1A}}V_{0}^{K_{1A}}\\
m_{K_{1B}}V_{0}^{K_{1B}}\end{array}\right),\label{m7}\\
\left(\begin{array}{c}F_{1}^{K_{1}(1270)}\\
F_{1}^{K_{1}(1400)}\end{array}\right) &=&
M\left(\begin{array}{c}F_{1}^{K_{1A}}\\
F_{1}^{K_{1B}}\end{array}\right),\label{m8}\\
\left(\begin{array}{c}(m^{2}_{B}-m^{2}_{K_{1}(1270)})F_{2}^{K_{1}(1270)}\\
(m^{2}_{B}+m^{2}_{K_{1}(1400)})F_{2}^{K_{1}(1400)}\end{array}\right)
&=&
M\left(\begin{array}{c}(m^{2}_{B}+m^{2}_{K_{1A}})F_{2}^{K_{1A}}\\
(m^{2}_{B}+m^{2}_{K_{1B}})F_{2}^{K_{1B}}\end{array}\right),\label{m9}\\
\left(\begin{array}{c}F_{3}^{K_{1}(1270)}\\
F_{3}^{K_{1}(1400)}\end{array}\right) &=&
M\left(\begin{array}{c}F_{3}^{K_{1A}}\\
F_{3}^{K_{1B}}\end{array}\right),\label{m10}
\end{eqnarray}
\end{widetext}
where we have supposed that $k^{\mu}_{K_{1}(1270),K_{1}(1400)}\simeq
k^{\mu}_{K_{1A},K_{1B}}$.

For the numerical analysis we have used the light-cone QCD sum rules form factors \cite{fmf}, summarized in Table
\ref{tabel1}, where the momentum dependence dipole parametrization is:
\begin{equation}
\mathcal{T}^{X}_{i}(q^{2})=\frac{\mathcal{T}^{X}_{i}(0)}{1-a_{i}^{X}\left(q^{2}/m^{2}_{B}\right)+b_{i}^{X}\left(q^{2}/m^{2}_{B}\right)^{2}}\label{m11}.
\end{equation}
where $\mathcal{T}$ is $A$, $V$ or $F$ form factors and the subscript
$i$ can take a value 0, 1, 2 or 3 the superscript $X$  belongs to
$K_{1A}$ or $K_{1B}$ state.
\begin{table*}[tbp]
\begin{tabular}{|p{.7in}p{.7in}p{.7in}p{.4in}||p{.7in}p{.7in}p{.7in}p{.4in}|}
\hline \hline
$\mathcal{T}^{X}_{i}(q^{2})$&$\mathcal{T}(0)$&$a$&$b$&$\mathcal{T}^{X}_{i}(q^{2})$&$\mathcal{T}(0)$&$a$&$b$\\
\hline
$V_{1}^{K_{1A}}$&$0.34$&$0.635$&$0.211$&$V_{1}^{K_{1B}}$&$-0.29$&$0.729$&$0.074$\\
$V_{2}^{K_{1A}}$&$0.41$&$1.51$&$1.18$&$V_{1}^{K_{1B}}$&$-0.17$&$0.919$&$0.855$\\
$V_{0}^{K_{1A}}$&$0.22$&$2.40$&$1.78$&$V_{0}^{K_{1B}}$&$-0.45$&$1.34$&$0.690$\\
$A^{K_{1A}}$&$0.45$&$1.60$&$0.974$&$A^{K_{1B}}$&$-0.37$&$1.72$&$0.912$\\
$F_{1}^{K_{1A}}$&$0.31$&$2.01$&$1.50$&$F_{1}^{K_{1B}}$&$-0.25$&$1.59$&$0.790$\\
$F_{2}^{K_{1A}}$&$0.31$&$0.629$&$0.387$&$F_{2}^{K_{1B}}$&$-0.25$&$0.378$&$-0.755$\\
$F_{3}^{K_{1A}}$&$0.28$&$1.36$&$0.720$&$F_{3}^{K_{1B}}$&$-0.11$&$1.61$&$10.2$\\
\hline\hline
\end{tabular}
\caption{$B\to K_{1A,1B}$ form factors \cite{fmf}, where $a$ and $b$
are the parameters of the form factors in dipole parametrization.}
\label{tabel1}
\end{table*}

\section{Physical Observables}\label{obs}

In this section, we calculate some interesting observables like
the branching ratio $(\mathcal{BR})$, forward-backward asymmetry
$(\mathcal{A}_{FB})$ as well as the helicity fractions of the final
state $K_{1}$ meson and their sensitivity for the NP due to
fourth generation SM,. From Eq. (\ref{qa}), one can get the decay
amplitudes for $B\rightarrow K_{1}(1270)\ell^{+}\ell^{-} $ and
$B\rightarrow K_{1}(1400)\ell^{+}\ell^{-}$ as
\begin{equation}
\mathcal{M}(B\rightarrow K_{1}\ell^{+}\ell^{-})=-\frac{%
G_{F}\alpha }{2\sqrt{2}\pi }V_{tb}V_{ts}^{\ast }
\left[ T_{V}^{\mu }\overline{%
\ell}\gamma _{\mu }\ell+T_{A}^{\mu }\overline{\ell}\gamma _{\mu }\gamma _{5}\ell\right]\label{35}
\end{equation}%
where the functions $T_{A}^{\mu }$ and $T_{V}^{\mu }$ can be written in terms of matrix elements and then in auxiliary
functions, as
\begin{align}
T_{A}^{\mu } &=C_{10}^{tot}\left\langle K_{1}(k,\epsilon )\left\vert
\bar{s}\gamma ^{\mu }\left( 1-\gamma ^{5}\right)
b\right\vert B(p)\right\rangle   \\
T_{V}^{\mu } &=C_{9}^{tot}\left\langle K_{1}(k,\epsilon )\left\vert
\bar{s}\gamma ^{\mu }\left( 1-\gamma ^{5}\right) b\right\vert %
B(p)\right\rangle \notag \\
&-C_{7}^{tot}\frac{2im_{b}}{q^{2}}\langle K_{1}(k,\epsilon
)\left\vert \bar{s}\sigma ^{\mu \nu }\left( 1+\gamma ^{5}\right)
q_{\nu }b\right\vert B(p)\rangle
\\
T_{V}^{\mu } &=f_{1}\epsilon
^{\mu \nu \rho \sigma }\varepsilon _{\nu}^{\ast }p_{\rho }k_{\sigma }-if_{2}%
\varepsilon ^{\ast\mu }-f_{3}(q\cdot\varepsilon)(p^{\mu }+k^{\mu })\label{TV}\\
T_{A}^{\mu } &= \bigg\{f_{4}\epsilon ^{\mu \nu \rho
\sigma }\varepsilon _{\mu}^{\ast}p_{\rho }k_{\sigma
}+if_{5}\varepsilon ^{\ast
\mu}\notag\\
&-if_{6}(q\cdot\varepsilon)(p^{\mu }+k^{\mu })
+if_{0}(q\cdot\varepsilon)q^{\mu }\bigg\}
\label{VA}
\end{align}%
One can notice that by using the following Dirac equations of motion, the last term in the expression of $T_{V}^{\mu}$ will vanish,
\begin{eqnarray}
q^{\mu }(\bar{\psi}_{1}\gamma _{\mu }\psi _{2}) =(m_{2}-m_{1})\bar{\psi}%
_{1}\psi _{2}  \label{eq-motion1} \\
q^{\mu }(\bar{\psi}_{1}\gamma _{\mu }\gamma _{5}\psi _{2}) =-(m_{1}+m_{2})%
\bar{\psi}_{1}\gamma _{5 }\psi _{2}  \label{eq-motion}
\end{eqnarray}

The auxiliary functions appearing in Eqs. (\ref{TV}) and (\ref{VA}) are defined as:
\begin{widetext}
\begin{eqnarray}
f_{1}&=&4(m_{b}+m_{s})\frac{C_{7}^{eff}}{q^{2}}\bigg\{F_{1}^{K_{1A}}\sin\theta_{K}+F_{1}^{K_{1B}}\cos\theta_{K}\bigg\}+2C_{9}^{eff}\left\{\frac{A_{1}^{K_{1A}}\sin\theta_{K}}{m_{B}+m_{K_{1A}}}+\frac{A_{1}^{K_{1B}}\cos\theta_{K}}{m_{B}+m_{K_{1B}}}\right\}\label{ax-a}\\
f_{2}&=&2(m_{b}+m_{s})\frac{C_{7}^{eff}}{q^{2}}\bigg\{(m_{B}^{2}-m^{2}_{K_{1A}})F_{2}^{K_{1A}}\sin{\theta_{K}}+(m_{B}^{2}-m^{2}_{K_{1B}})F_{2}^{K_{1B}}\cos{\theta_{K}}\bigg\}\notag\\
&&+C_{9}^{eff}\bigg\{(m_{B}+m_{K_{1A}})V_{1}^{K_{1A}}\sin{\theta_{K}}+(m_{B}+m_{K_{1B}})V_{1}^{K_{1B}}\cos{\theta_{K}}\bigg\}\\
f_{3}&=&2(m_{b}+m_{s})\frac{C_{7}^{eff}}{q^{2}}\left\{\left(F_{2}^{K_{1A}}+\frac{q^{2}F_{3}^{K_{1A}}}{m_{B}^{2}-m^{2}_{K_{1A}}}\right)\sin{\theta_{K}}+\left(F_{2}^{K_{1B}}+\frac{q^{2}F_{3}^{K_{1B}}}{m_{B}^{2}-m^{2}_{K_{1B}}}\right)\cos{\theta_{K}}\right\}\notag\\
&&+C_{9}^{eff}\left(\frac{V_{2}^{K_{1A}}\sin{\theta_{K}}}{m_{B}+m_{K_{1A}}}+\frac{V_{2}^{K_{1B}}\cos{\theta_{K}}}{m_{B}+m_{K_{1B}}}\right)\\
f_{4}&=&2C_{10}^{eff}\left(\frac{A^{K_{1A}}\sin{\theta_{K}}}{m_{B}+m_{K_{1A}}}+\frac{A^{K_{1B}}\cos{\theta_{K}}}{m_{B}+m_{K_{1B}}}\right)\\
f_{5}&=&C_{10}^{eff}\bigg\{(m_{B}+m_{K_{1A}})V_{1}^{K_{1A}}\sin{\theta_{K}}
+(m_{B}+m_{K_{1B}})V_{1}^{K_{1B}}\cos{\theta_{K}}\bigg\}\\
f_{6}&=&C_{10}^{eff}\left(\frac{V_{2}^{K_{1A}}\sin{\theta_{K}}}{m_{B}+m_{K_{1A}}}+\frac{V_{2}^{K_{1B}}\cos{\theta_{K}}}{m_{B}+m_{K_{1B}}}\right)\\
f_{0}&=&2\frac{C_{10}^{eff}}{q^{2}}\bigg\{m_{K_{1A}}\left(V_{3}^{K_{1A}}-V_{0}^{K_{1A}}\right)\sin{\theta_{K}}+m_{K_{1B}}\left(V_{3}^{K_{1B}}-V_{0}^{K_{1B}}\right)\cos{\theta_{K}}\bigg\}\label{ax-d0}
\end{eqnarray}
\end{widetext}

\subsection{Branching Ratio}

The double differential decay rate for $B\rightarrow K_{1}\ell^{+}\ell^{-}$
can be written as \cite{HY,Colangelo}%
\begin{eqnarray}
\frac{d\Gamma }{dq^{2}d\cos\theta} &=&\frac{G_{F}^{2}\alpha ^{2}}{2^{11}\pi ^{5}m_{B}^{3}}\left\vert V_{tb}V_{ts}^{\ast }\right\vert^{2} u(q^{2})\times \left\vert\mathcal{M}\right\vert^2 \label{decayrate}
\end{eqnarray}%
with
\begin{equation}
\left\vert\mathcal{M}\right\vert^2 =\mathcal{A}(q^2)\cos^{2}\theta+\mathcal{B}(q^2)\cos\theta+\mathcal{C}(q^2) \label{hq}
\end{equation}
and
\begin{equation}
u(q^{2})\equiv
\sqrt{\lambda\left(1-\frac{4m_{\ell}^{2}}{q^{2}}\right)},\label{uq}\\
\end{equation}
where
\begin{eqnarray}
\lambda &\equiv
&\lambda\left(m_{B}^{2},m_{K_{1}}^{2},q^{2}\right)\notag\\
&=&m_{B}^{4}+m_{K_{1}}^{4}+q^{4}-2m_{K_{1}}^{2}m_{B}^{2}-2q^{2}m_{B}^{2}-2q^{2}m_{K_{1}}^{2}.\notag\\
&&\label{lambda}
\end{eqnarray}%

One can get the differential decay rate by performing the integration on $\cos\theta$ in Eq. (\ref{decayrate}), so
\begin{equation}
\frac{d\Gamma}{dq^{2}}=\frac{G_{F}^{2}\alpha ^{2}}{2^{11}\pi ^{5}m_{B}^{3}}\left\vert V_{tb}V_{ts}^{\ast }\right\vert^{2}\frac{1}{3}[2\mathcal{A}(q^{2})+6\mathcal{B}(q^{2})]
\end{equation}
where
\begin{widetext}
\begin{align}
\mathcal{A}(q^{2})&=\frac{1}{2}\lambda(q^{2}-4m^{2})\left[|f_{1}|^{2}+|f_{4}|^2\right]-\frac{1}{m_{K_{1}}^{2}q^{2}}\left[|f_{2}|^{2}+|f_{5}|^{2}\right] -\frac{\lambda}{m_{K_{1}}^{2}q^{2}}\left[|f_{3}|^{2}+|f_{6}|^{2}\right]\notag\\
&+\frac{2\left(m_{B}^{2}-
m_{K_{1}}^{2}-q^{2}\right)}{m_{K_{1}}^{2}q^{2}}\left\{\lambda\Re\left[f_{2}f_{3}^{\ast}\right]+\Re\left[f_{5}f_{6}^{\ast}\right]\right\}\label{aq}\\
\mathcal{B}(q^{2})&=4\Re\left[f_{1}f_{5}^{\ast}+f_{2}f_{4}^{\ast}\right]\sqrt{q^{2}(q^{2}-4m^{2})\lambda}\label{bq}\\
\mathcal{C}(q^{2})&=\frac{1}{2}(q^{2}-4m^{2})\lambda\left[|f_{1}|^{2}+|f_{4}|^2+8|f_{5}|^2\right]+4|f_{2}|^2(2m^{2}+q^{2})
+\frac{\lambda}{m_{K_{1}}^{2}q^{2}}\left[|f_{2}+|f_{5}|^2+\lambda(|f_{3}|^{2}+|f_{6}|^2)\right]\notag\\
&-2\Re(f_{2}f_{3}^{\ast})+|f_{0}|^{2}4m^{2}q^{2} +2\Re(f_{5}f_{6}^{\ast})\left[m_{B}^{2}-M_{K_{1}}^{2}-(4m^{2}-q^{2})\right]\notag\\
&-8m^{2}\Re(f_{5}f_{6}^{\ast})-\Re(f_{0}f_{6}^{*}(m_{B}^{2}+m_{K_{1}}^{2}))+\frac{1}{m_{K_{1}}^{2}}\left[|f_{6}|^{2}2m^{2}(2(m_{B}^{2}+m_{K_{1}}^{2}-q^{2}))\right]\label{cq}
\end{align}%
\end{widetext}
The kinematical variables used in above equations are defined as
$u=\left( p-p_{l^{-}}\right) ^{2}-\left( p-p_{l^{+}}\right) ^{2}$,
$u=-u(q^{2})\cos \theta $.
Here $\lambda$ is defined in Eq. \eqref{lambda} and $\theta $ is the angle between the moving direction of
$\ell^{+}$ and $B$ meson in the centre of mass frame of
$\ell^{+}\ell^{-}$ pair.

It is also very useful to define the branching fractions $\mathcal{R}_{\ell}$ as:
\begin{equation}
\mathcal{R}_{\ell}=\frac{\mathcal{BR}(B\rightarrow K_{1}(1400)\ell ^{+}\ell ^{-})}{\mathcal{BR}(B\rightarrow K_{1}(1270)\ell ^{+}\ell ^{-})}
\end{equation}
where $\ell=\mu,\ \tau$.

\subsection{Forward-Backward Asymmetries}

In this section we investigate the forward-backward asymmetry ($\mathcal{A}_{FB}$)
of leptons. The measurement of the $\mathcal{A}_{FB}$ at LHC is significant due to
the minimal form factors \cite{new2} hence this observable has great
importance to check the more clear signals of any NP than the other
observables such as branching ratio etc. In the context of fourth
generation, the $\mathcal{A}_{FB}$ can also play a crucial role because it is
driven by the loop top quark so it is sensitive to the fourth
generation up type quark $t^{\prime}$ \cite{63}.

The differential $\mathcal{A}_{FB}$ of final state lepton for the said decays can be
written as
\begin{equation}
{\frac{d\mathcal{A}_{FB}(q^{2})}{dq^{2}}}=\int_{0}^{1}\frac{d^{2}\Gamma }{dq^{2}d\cos \theta }%
d\cos \theta -\int_{-1}^{0}\frac{d^{2}\Gamma }{dq^{2}d\cos \theta
}d\cos \theta  \label{FBformula}
\end{equation}%

From experimental point of view the normalized forward-backward
asymmetry is more useful, which is defined as
\begin{equation*}
\mathcal{A}_{FB}=\frac{\int_{0}^{1}\frac{d^{2}\Gamma }{dq^{2}d\cos \theta }%
d\cos \theta -\int_{-1}^{0}\frac{d^{2}\Gamma }{dq^{2}d\cos \theta
}d\cos \theta  }{\int_{-1}^{1}\frac{d^{2}\Gamma }{dq^{2}d\cos \theta
} d\cos \theta }
\end{equation*}%
The differential $\mathcal{A}_{FB}$ for $B\rightarrow K_{1}\ell^{+}\ell^{-}$ decays can be obtained from Eq. (\ref{decayrate}), as%
\begin{eqnarray}
\frac{d\mathcal{A}_{FB}(q^{2})}{dq^{2}}&=&-\frac{G_{F}^{2}\alpha ^{2}}{2^{11}\pi ^{5}m_{B}^{3}}
\left\vert V_{tb}V_{ts}^{\ast }\right\vert
^{2}u\left(
q^{2}\right) \notag\\
&&\times\frac{3\mathcal{B}(q^{2})}{2\mathcal{A}(q^{2})+6\mathcal{C}_{1}(q^{2})}\label{FBA}
\end{eqnarray}%
where $\mathcal{A}(q^{2})$, $\mathcal{B}(q^{2})$ and $\mathcal{C}(q^{2})$ are defined in Eqs.
$(\ref{aq}, \ref{bq},\ref{cq})$.

\subsection{Helicity Fractions of $K_{1}$ meson}

We now discuss helicity fractions of $K_{1}(1270,1400)$ meson in $B\rightarrow
K_{1}\ell^{+}\ell ^{-}$ which are interesting observables and are insensitive to the uncertainties arising due to form factors
and other input parameters. Thus the helicity fractions can be a good
tool to test the NP beyond the SM. The final state meson
helicity fractions were already discussed in the literature for
$B\rightarrow K^{\ast }\left( K_{1}\right) \ell^{+}\ell^{-}$ decays
\cite{Colangelo, paracha}.

The explicit expression of the longitudinal $(f_{L})$ and the transverse$(f_{T})$ helicity fractions for $B\rightarrow
K_{1}\ell^{+}\ell ^{-}$ decay can be obtained by trading $\left\vert\mathcal{M}\right\vert$ to $\left\vert\mathcal{M}_{L}\right\vert$ and $\left\vert\mathcal{M}_{\pm}\right\vert$, respectively, in Eq. \eqref{decayrate}. Here
\begin{eqnarray}
\left\vert\mathcal{M}_{L}\right\vert^{2}&=&\mathcal{D}_{L}\cos^{2}\theta+\mathcal{E}_{L}\\
\left\vert\mathcal{M}_{\pm}\right\vert^{2}&=&\mathcal{D}_{\pm}\cos^{2}\theta+\mathcal{E}_{\pm}
\end{eqnarray}
By performing the integration on $\cos\theta$ in Eq. (\ref{decayrate}), we get
\begin{align}
\frac{d\Gamma_{L} }{dq^{2}} &=\frac{G_{F}^{2}\alpha ^{2}}{2^{11}\pi ^{5}}%
\frac{\left\vert V_{tb}V_{ts}^{\ast }\right\vert ^{2}}{m_{B}^{3}}
u(q^{2}) \frac{2}{3}\left[\mathcal{D}_{L}(q^{2})+3\mathcal{E}_{L}(q^{2})\right]\\
\frac{d\Gamma_{\pm} }{dq^{2}} &=\frac{G_{F}^{2}\alpha ^{2}}{2^{11}\pi ^{5}}%
\frac{\left\vert V_{tb}V_{ts}^{\ast }\right\vert ^{2}}{m_{B}^{3}}
u(q^{2}) \frac{2}{3}[\mathcal{D}_{\pm}(q^{2})+3\mathcal{E}_{\pm}(q^{2})]
\end{align}%
where $\mathcal{D}_{L}(q^{2})$, $\mathcal{D}_{\pm}(q^{2})$, $\mathcal{E}_{L}(q^{2})$ and $\mathcal{E}_{\pm}(q^{2})$ can be parameterized in terms of the auxiliary functions [c.f. Eqs. $(\ref{ax-a}-\ref{ax-d0})$] as
\begin{widetext}
\begin{align}
\mathcal{D}_{L}(q^{2})&=\frac{1}{2m^{2}_{K_{1}}}\left\{|f_{5}|^{2}\left[\left(m^{2}_{B}-m^{2}_{K_{1}}-q^{2}\right)^{2}
-16m^{2}m^{2}_{K_{1}}\right]+|f_{2}|^{2}\left(2m^{2}_{K_{1}}q^{2}+\lambda\right)+\lambda^{2}|f_{3}|^{2}-8m^{2}\lambda|f_{5}|^{2}+4m^{2}q^{2}\lambda|f_{0}|^{2}\right.\notag\\
&\left.-2\Re(f_{5}f_{6}^{\ast})\left(m_{B}^{2}-m^{2}_{K_{1}}-q^{2}\right)\lambda-4\lambda\Re(f{2}f_{3}^{\ast})\left(m_{B}^{2}-m^{2}_{K_{1}}-q^{2}\right)+\lambda|f_{6}|^{2}\left[8\left(m_{B}^{2}+m^{2}_{K_{1}}-4q^{2}\right)m^{2}+\lambda\right]\right\}\\
\mathcal{D}_{+}(q^{2})&=\frac{1}{4}\left\{\left(q^{2}+4m^{2}\right)\left[\lambda|f_{4}|^2+4|f_{5}|^2+4\sqrt\lambda\left(\Im(f_{1}^{\ast}f_{2})+\Im(f_{4}^{\ast}f_{5})\right)\right]+(q^{2}-4m^{2})\left(\lambda|f_{1}|^2+4|f_{2}|^{2}\right)\right\}\\
\mathcal{D}_{-}(q^{2})&=\frac{1}{4}\left\{\left(q^{2}+4m^{2}\right)\left[\lambda|f_{1}|^2+|f_{2}|^2\right]+(q^{2}-4m^{2})\left[\lambda|f_{4}|^{2}+4\sqrt\lambda\left(\Im(f_{1}f_{2}^{\ast})+\Im(f_{4}f_{5}^{\ast})\right)\right]\right\}\\
\mathcal{E}_{L}(q^{2})&=\frac{1}{2m^{2}_{K_{1}}}\left\{\left(4m^2-q^2\right)\left[|f_{5}|^{2}\left(m^{4}_{B}-(m^{2}_{K_{1}}+q^{2})m_{B}^{2}\right)
+|f_{2}|^{2}\left(m_{B}^{2}-m^{2}_{K_{1}}-q^{2}\right)^{2}+2\lambda^{2}\left(|f_{3}|^{2}+|f_{6}|^{2}\right)\right.\right.\notag\\
&\left.-4\lambda\left(m_{B}^{2}-m^{2}_{K_{1}}-q^{2}\right)\Re(f_{2}f_{3}^{\ast})\right]+q^{2}\lambda\Re(f_{5}f_{6}^{\ast})\left[\left(m_{B}^{2}-m^{2}_{K_{1}}\right)+4m^{2}-q^{2}\right]-4m^{2}\lambda\Im(f{5}f_{6}^{\ast})\left(m_{B}^{2}-m^{2}_{K_{1}}\right)\notag\\
&\left.+\left(m^{2}_{K_{1}}+q^{2}\right)|f_{5}|^{2}\left[q^{2}\left(4m^{2}-m^{2}_{K_{1}}-q^{2}\right)+4m^{2}\left(m^{2}_{K_{1}}+q^{2}\right)\right]\right\}\\
\mathcal{E}_{+}(q^{2})&=\frac{1}{4}\left(q^{2}-4m^{2}\right)\left\{\lambda\left(|f_{1}|^2+|f_{4}|^2\right)+\left(|f_{2}|^{2}+|f_{5}|^{2}\right)+4\sqrt\lambda\left(\Im(f_{1}^{\ast}f_{2})+\Im(f_{4}^{\ast}f_{5})\right)\right\}\\
\mathcal{E}_{-}(q^{2})&=\frac{1}{4}\left(q^{2}-4m^{2}\right)\left\{\lambda\left(|f_{1}|^2+|f_{4}|^2\right)+\left(|f_{2}|^{2}+|f_{5}|^{2}\right)+4\sqrt\lambda\left(\Im(f_{1}f_{2}^{\ast})+\Im(f_{4}f_{5}^{\ast})\right)\right\}
\end{align}
\end{widetext}
Finally the longitudinal and transverse helicity fractions become
\begin{eqnarray}
f_{L}(q^{2}) &=&\frac{d\Gamma _{L}(q^{2})/dq^{2}}{d\Gamma
(q^{2})/dq^{2}} \\
f_{\pm }(q^{2}) &=&\frac{d\Gamma _{\pm }(q^{2})/dq^{2}}{d\Gamma
(q^{2})/dq^{2}} \\
f_{T}(q^{2}) &=&f_{+}(q^{2})+f_{-}(q^{2})
\end{eqnarray}%
so that the sum of the longitudinal and transverse helicity
amplitudes is equal to one i.e. $f_{L}(q^{2})+f_{T}(q^{2})=1$ for
each value of $q^{2}$ \cite{Colangelo}.

\section{Numerical Results and Discussion}\label{num}

We present here our numerical results of the branching ratio
($\mathcal{BR}$), the forward backward asymmetry $(\mathcal{A}_{FB})$ and the helicity
fractions ($f_{L,T})$ of $K_{1}(1270,1400)$ meson for the $B\rightarrow
K_{1}(1270,1400)\ell^{+}\ell^{-}$ decays with $\ell=\mu,\tau$. Here we have taken the central values of all the input parameters. We first give the
numerical values of input parameters which are used in our numerical
calculations \cite{pdg}:
\begin{equation*}
\begin{tabular}{l}
\hline
$m_{B}=5.28$ GeV, $m_{b}=4.28$ GeV, $m_{\mu}=0.105$ GeV,\\
$m_{\tau}=1.77$ GeV, $f_{B}=0.25$ GeV, $|V_{tb}V_{ts}^{\ast}|=45\times
10^{-3}$,\\ $\alpha^{-1}=137$, $G_{F}=1.17\times 10^{-5}$ GeV$^{-2}$,\\
$\tau_{B}=1.54\times 10^{-12}$ sec, $m_{K_{1}(1270)}=1.270$ GeV,\\
$m_{K_{1}(1400)}=1.403$ GeV, $m_{K_{1A}}=1.31$ GeV,\\ $m_{K_{1B}}=1.34$ GeV.\\
\hline
\end{tabular}
\end{equation*}

Besides these input parameters, the form factors (the scalar functions of the square of the momentum transfer), the non-perturbative
quantities, which are also very important. To study the above mentioned physical
observables we use the light cone QCD sum rules (LCQCD) form factors
which are given in Table \ref{tabel1}. In our numerical calculations, we set the mixing angle $\theta_{K}=-(34\pm13)^{0}$ \cite{HY} where
we have taken the central value $\theta_{K}=-34^{0}$ and the values of the SM Wilson
Coefficients at $\mu\sim m_{b}$ are given in Table \ref{wc table}.
\begin{table*}[ht]
\begin{tabular}{cccccccccc}
\hline\hline
$C_{1}$&$C_{2}$&$C_{3}$&$C_{4}$&$C_{5}$&$C_{6}$&$C_{7}$&$C_{9}$&$C_{10}$
\\ \hline
 \ \ \ 1.107 \ \ \ & \ \ \ -0.248 \ \ \ & \ \ \ -0.011 \ \ \ & \ \ \ -0.026 \ \ \ & \ \ \ -0.007 \ \ \ & \ \ \ -0.031 \ \ \ & \ \ \ -0.313 \ \ \ & \ \ \ 4.344 \ \ \ & \ \ \ -4.669 \ \ \ \\
\hline\hline
\end{tabular}\caption{The Wilson coefficients $C_{i}^{\mu}$ at the scale $\mu\sim m_{b}$ in the SM.}
\label{wc table}
\end{table*}

\begin{figure*}[ht]
\begin{tabular}{cc}
\hspace{0.6cm}($\mathbf{a}$)&\hspace{1.2cm}($\mathbf{b}$)\\
\includegraphics[scale=0.4]{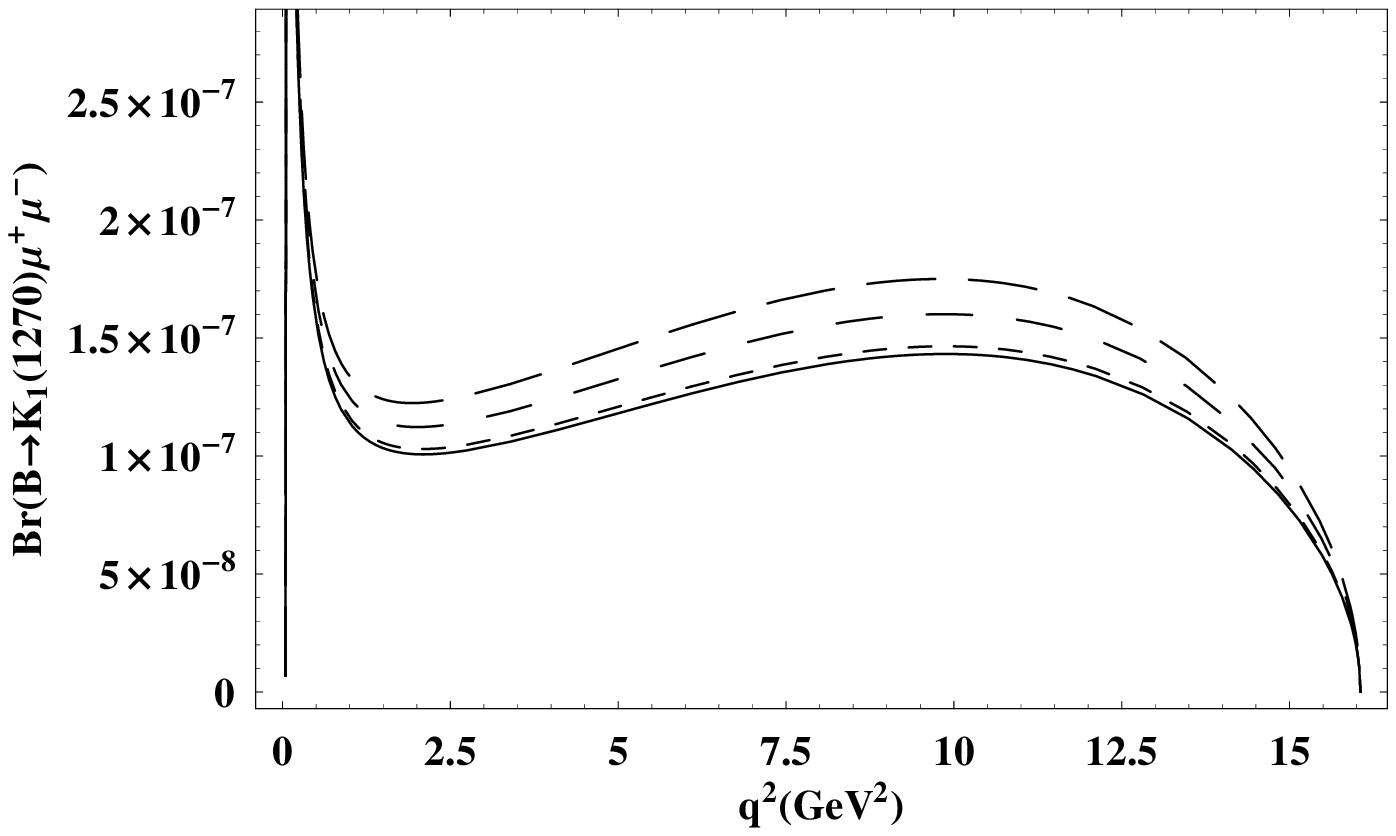}  \ \ \
&  \ \ \  \includegraphics[scale=0.4]{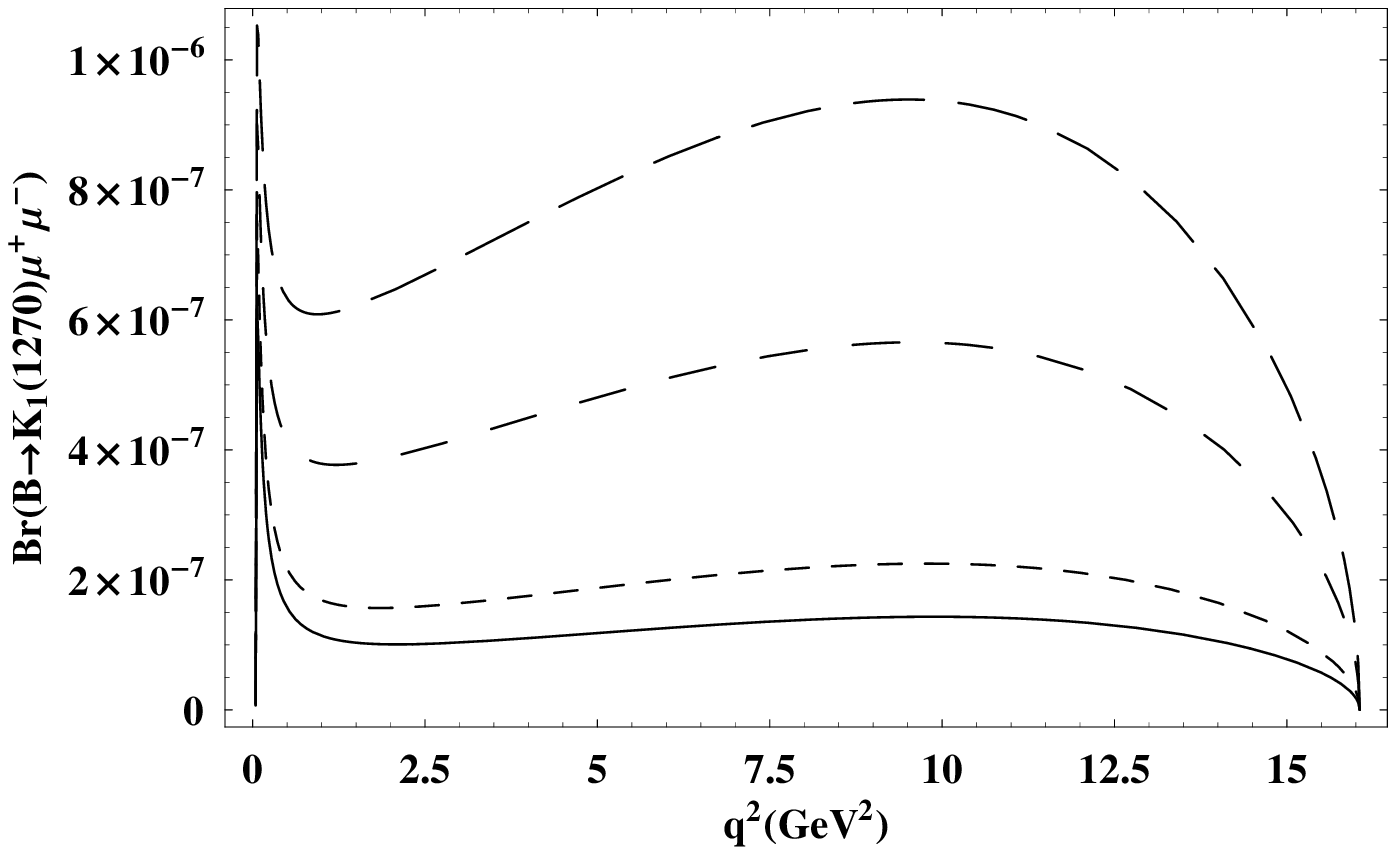}\end{tabular}
\caption{The dependence of branching ratio of $B\to K_{1}(1270)\mu^{+}\mu^{-}$ on $q^{2}$ for different values of $m_{t^{\prime}}$ and $\left\vert V^{\ast}_{t^{\prime}b}V_{t^{\prime}s}\right\vert$. In all the graphs, the solid line corresponds to the SM, small dashed , medium dashed, long dashed correspond, $m_{t^{\prime}}=$ 300 GeV, 500 GeV and 600 GeV respectively. $\left\vert V^{\ast}_{t^{\prime}b}V_{t^{\prime}s}\right\vert$ has the value 0.003 and 0.015 in $(a)$ and $(b)$ respectively.} \label{Branching ratios for muons 12}
\end{figure*}

First, we discuss the $\mathcal{BR}$s of $B\rightarrow
K_{1}(1270,1400)\mu^{+}\mu^{-}(\tau^{+}\tau^{-})$ decays which we
have plotted as a function of $q^2$ (GeV$^{2})$, shown in Figs.
\ref{Branching ratios for muons 12}-\ref{Branching ratios for tauons
14}, both in the SM and in the fourth generation scenario. Figures \ref{Branching ratios for muons 12} and \ref{Branching ratios for
tauons 12} show the $\mathcal{BR}$ of $B \to K_{1}(1270)$ with
$\mu^{+}\mu^{-}$ and $\tau^{+}\tau^{-}$ respectively and Figs.
\ref{Branching ratios for muons 14} and \ref{Branching ratios for
tauons 14} show the same for $B\to K_{1}(1400)$. These figures
depict that the values of $\mathcal{BR}$ strongly depend on the fourth
generation effects which come through the new parameters (i.e the
Wilson coefficients with $m_{t^{\prime}}$ instead of $m_{t}$ as well
as from the $V_{t^{\prime} b}V_{t^{\prime} s}$ are encapsulated in
Eq. (\ref{effwc})). One can see clearly from these graphs that the
increment in the values of the fourth generation parameters, increase
the value of the branching ratio accordingly, i.e. the
$\mathcal{BR}$ is an increasing function of both $m_{t^{\prime}}$
and $V_{t^{\prime} b}V_{t^{\prime} s}$. Moreover, this constructive
characteristic of the fourth generation effects to the $\mathcal{BR}$
manifest throughout the $q^{2}$ region irrespective to the mass of the final particles.
In addition, one can also extract the constructive behavior of the
fourth generation to the $\mathcal{BR}$ from Table \ref{br table}.
However, the quantitative analysis of the $\mathcal{BR}$  shows that
the NP effects due to the fourth generation are comparatively more
sensitive to the case of $B\rightarrow K_{1}(1270)\ell^{+}\ell^{-}$
than the case of $B\rightarrow K_{1}(1400)\ell^{+}\ell^{-}$.

Moreover, Table \ref{br table} shows that the maximum deviation (when we
set $m_{t^{\prime}}=600$ GeV, $V_{t^{\prime}
b}V_{t^{\prime}s}=1.5\times10^{-3}$) from the SM value due to the
fourth generation: for the case of $B\rightarrow
K_{1}(1270)\mu^{+}\mu^{-}$ is approximately $6$ times, for the case
of $B\rightarrow K_{1}(1270)\tau^{+}\tau^{-}$ is about $3.3$ times,
for  $B\rightarrow K_{1}(1400)\mu^{+}\mu^{-}$ is approximately $5.9$
time and for $B\rightarrow K_{1}(1400)\tau^{+}\tau^{-}$ is about
$2.9$ times than that of SM values.
\begin{table*}[ht]
\begin{tabular}{|c|c|c|}
\hline\hline
& $\mathcal{BR}(B\rightarrow K_{1}(1270)\mu ^{+}\mu ^{-})$, SM value: $1.97 \times10^{-6}$& $\mathcal{BR}%
(B\rightarrow K_{1}(1400)\mu ^{+}\mu ^{-})$, SM value: $5.76\times10^{-8}$ \\ \hline
\begin{tabular}{c}
$\  \ \ \ |V_{t^{\prime }b}V_{t^{\prime }s}|$ \ \ \  \ \\ \hline
$3\times 10^{-3}$ \\
$1.5\times 10^{-2}$%
\end{tabular}
&
\begin{tabular}{c|c|c}
\ \ \ \ $ m_{t^{\prime }}=300$\ \ \ \  &\ \  \ \ $m_{t^{\prime }}=500$ \ \ \ \ & \ \ \ \ $m_{t^{\prime }}=600$\ \  \ \ \\ \hline
$2.01\times 10^{-6}$ & $2.18\times 10^{-6}$ & $2.38\times 10^{-6}$ \\
$3.04\times 10^{-6}$ & $7.43\times 10^{-6}$ & $1.22\times 10^{-5}$%
\end{tabular}
&
\begin{tabular}{c|c|c}
\ \ \ \ $m_{t^{\prime }}=300$ \ \ \ \ &\ \  \ \ $m_{t^{\prime }}=500$\ \ \ \  &\ \  \ \ $m_{t^{\prime }}=600$ \ \ \ \ \\ \hline
$5.88\times 10^{-8}$ & $6.36\times 10^{-8}$ & $6.90\times 10^{-8}$ \\
$8.78\times 10^{-8}$ & $2.09\times 10^{-7}$ & $3.44\times 10^{-7}$%
\end{tabular} \\ \hline\hline
& $\mathcal{BR}(B\rightarrow K_{1}(1270)\tau ^{+}\tau ^{-})$, SM value: $6.06 \times10^{-8}$ & $\mathcal{BR}%
(B\rightarrow K_{1}(1400)\tau ^{+}\tau ^{-})$, SM value: $9.39 \times10^{-10}$\\ \hline
\begin{tabular}{c}
$\ \ \ \ |V_{t^{\prime }b}V_{t^{\prime }s}|$\ \ \ \  \\ \hline
$3\times 10^{-3}$ \\
$1.5\times 10^{-2}$%
\end{tabular}
&
\begin{tabular}{c|c|c}
$\ \ \ \ m_{t^{\prime }}=300$\ \  \ \ &\ \ \ \  $m_{t^{\prime }}=500$\ \  \ \ &\ \  \ \ $m_{t^{\prime }}=600$\ \ \ \  \\ \hline
$6.14\times 10^{-8}$ & $6.38\times 10^{-8}$ & $6.62\times 10^{-8}$ \\
$8.12\times 10^{-8}$ & $1.39\times 10^{-7}$ & $2.01\times 10^{-7}$%
\end{tabular}
&
\begin{tabular}{c|c|c}
\ \ \ \ $m_{t^{\prime }}=300$\ \  \ \ &\ \  \ \ $m_{t^{\prime }}=500$\ \ \ \  &\ \  \ \ $m_{t^{\prime }}=600$ \ \ \ \ \\ \hline
$9.51\times 10^{-10}$ & $9.80\times 10^{-10}$ & $1.01\times 10^{-9}$ \\
$1.24\times 10^{-9}$ & $1.98\times 10^{-9}$ & $2.74\times 10^{-9}$%
\end{tabular}
\\ \hline\hline
\end{tabular}\caption{The values of branching ratio of $B\to K_{1}(1270,1400)\ell^{+}\ell^{-}$ with $\ell=\mu,\ \tau$ for different values of $m_{t^{\prime}}$ and $\left\vert V^{\ast}_{t^{\prime}b}V_{t^{\prime}s}\right\vert$.}
\label{br table}
\end{table*}
\begin{table*}[ht]
\begin{tabular}{|c|c|c|}
\hline\hline
& $\mathcal{R}_{\mu}=\frac{\mathcal{BR}(B\rightarrow K_{1}(1400)\mu ^{+}\mu ^{-})}{\mathcal{BR}(B\rightarrow K_{1}(1270)\mu ^{+}\mu ^{-})}$, SM value: $2.92 \times10^{-2}$&  $\mathcal{R}_{\tau}=\frac{\mathcal{BR}(B\rightarrow K_{1}(1400)\tau ^{+}\tau ^{-})}{\mathcal{BR}(B\rightarrow K_{1}(1270)\tau ^{+}\tau ^{-})}$, SM value: $1.54 \times10^{-2}$\\ \hline
\begin{tabular}{c}
$\  \ \ \ |V_{t^{\prime }b}V_{t^{\prime }s}|$ \ \ \  \ \\ \hline
$3\times 10^{-3}$ \\
$1.5\times 10^{-2}$%
\end{tabular}
&
\begin{tabular}{c|c|c}
\ \ \ \ $ m_{t^{\prime }}=300$\ \ \ \  &\ \  \ \ $m_{t^{\prime }}=500$ \ \ \ \ & \ \ \ \ $m_{t^{\prime }}=600$\ \  \ \ \\ \hline
$2.92\times 10^{-2}$ & $2.91\times 10^{-2}$ & $2.90\times 10^{-2}$ \\
$2.88\times 10^{-2}$ & $2.81\times 10^{-2}$ & $2.81\times 10^{-2}$%
\end{tabular}
&
\begin{tabular}{c|c|c}
\ \ \ \ $m_{t^{\prime }}=300$ \ \ \ \ &\ \  \ \ $m_{t^{\prime }}=500$\ \ \ \  &\ \  \ \ $m_{t^{\prime }}=600$ \ \ \ \ \\ \hline
$1.54\times 10^{-2}$ & $1.53\times 10^{-2}$ & $1.52\times 10^{-2}$ \\
$1.52\times 10^{-2}$ & $1.42\times 10^{-2}$ & $1.36\times 10^{-2}$%
\end{tabular} \\ \hline\hline
\end{tabular}\caption{The values of branching fractions $\mathcal{R}_{\ell}$, with $\ell=\mu,\tau$, for different values of $m_{t^{\prime}}$ and $\left\vert V^{\ast}_{t^{\prime}b}V_{t^{\prime}s}\right\vert$.}
\label{brf table}
\end{table*}

This is important to emphasis here that the increment in the
branching ratio due to the fourth generation effect is optimally
well separated than that of SM value. Furthermore the change in
branching ratios due to the hadronic uncertainties  as well as the
uncertainty of the mixing angle $\theta_{K}$ are negligible in
comparison of the NP effects. Therefore, any dramatically increment
in the measurement of the branching ratio at present experiments
will be a clear indication of NP. So the precise measurement of
branching ratio is very handy tool to extract the information about
the fourth generation parameters.

To observe the variation which comes through the fourth generation parameters
in the branching fractions
$\mathcal{R}_{\ell}=\mathcal{BR}(B\rightarrow K_{1}(1400)\ell
^{+}\ell ^{-})/\mathcal{BR}(B\rightarrow K_{1}(1270)\ell ^{+}\ell
^{-})$, with $\ell=\mu,\ \tau$, we draw the graph of
$\mathcal{R}_{\mu}$ and $\mathcal{R}_{\tau}$ as a function of
$q^{2}$ in Figs. \ref{brf for muons} and \ref{brf for tauons}.
 We have also summarized the numerical values of the branching fractions corresponding to the values of
 $m_{t^{\prime}}$ and $|V_{t^{\prime}b}V_{t^{\prime}s}|$ in Table \ref{brf
 table}. These numerical analysis shows that the
 branching fraction are insensitive to the NP. So this analysis support the argument that this observable is suitable to fix the value of $\theta_{K}$ \cite{HY}.
\begin{figure*}[ht]
\begin{tabular}{cc}
\hspace{.6cm}($\mathbf{a}$)&\hspace{1.2cm}($\mathbf{b}$)\\
\includegraphics[scale=0.4]{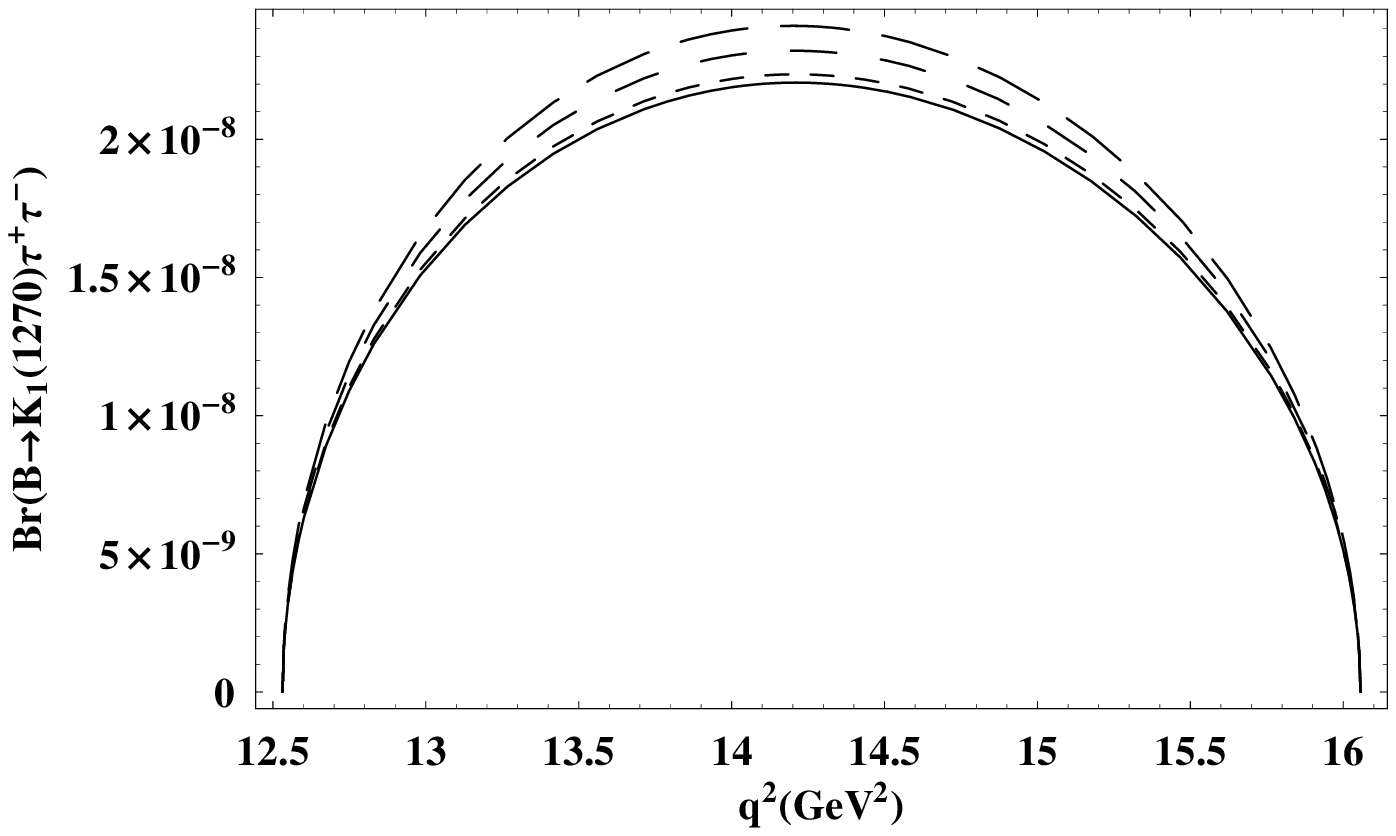} \ \ \
& \ \ \  \includegraphics[scale=0.4]{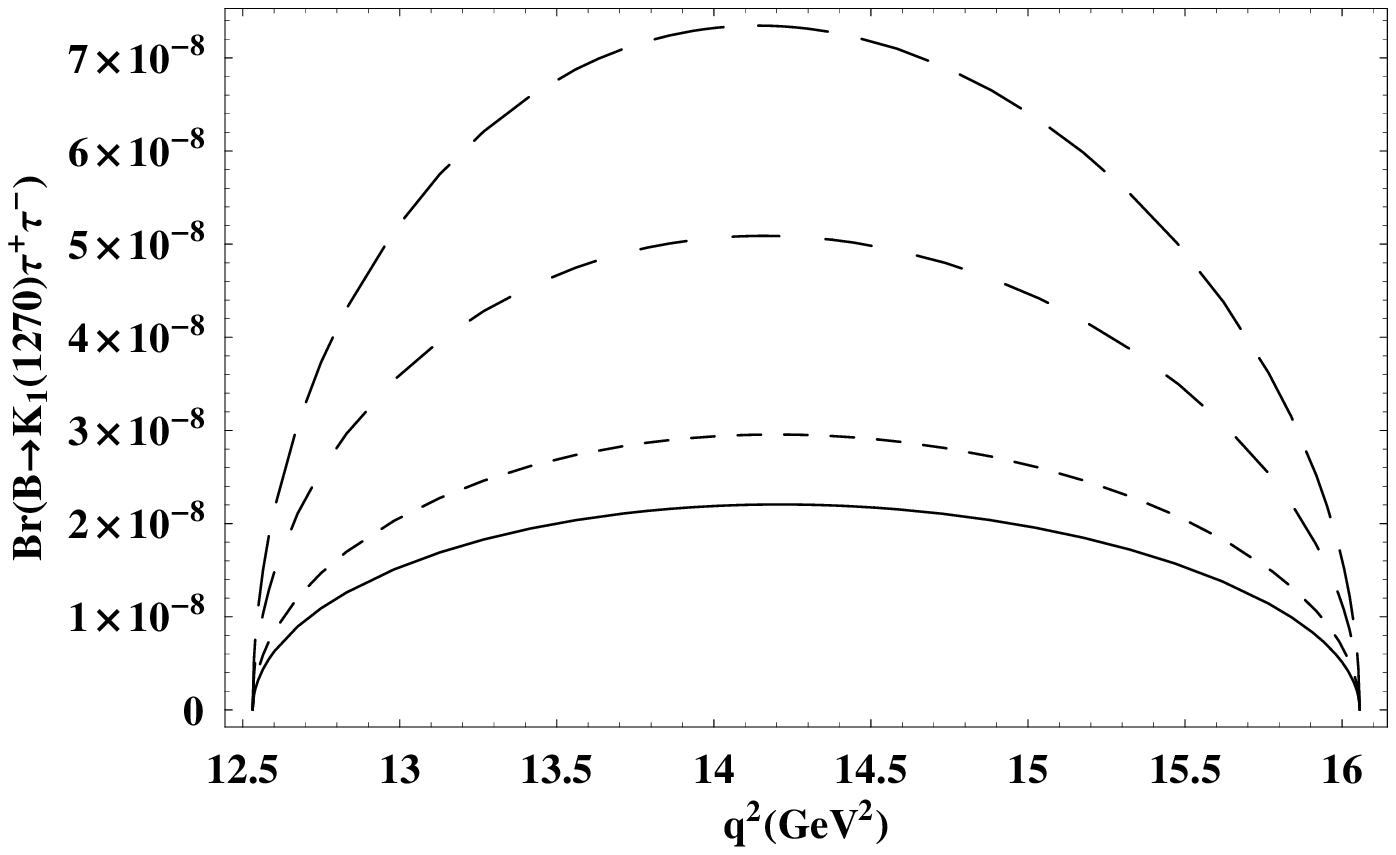}\end{tabular}
\caption{The dependence of branching ratio of $B\to K_{1}(1270)\tau^{+}\tau^{-}$ on $q^{2}$ for different values of $m_{t^{\prime}}$ and $\left\vert V^{\ast}_{t^{\prime}b}V_{t^{\prime}s}\right\vert$. The legends and the values
of fourth generation parameters are same as in Fig. \ref{Branching ratios for muons 12}.} \label{Branching ratios for tauons 12}
\end{figure*}
\begin{figure*}[ht]
\begin{tabular}{cc}
\hspace{.6cm}($\mathbf{a}$)&\hspace{1.2cm}($\mathbf{b}$)\\
\includegraphics[scale=0.4]{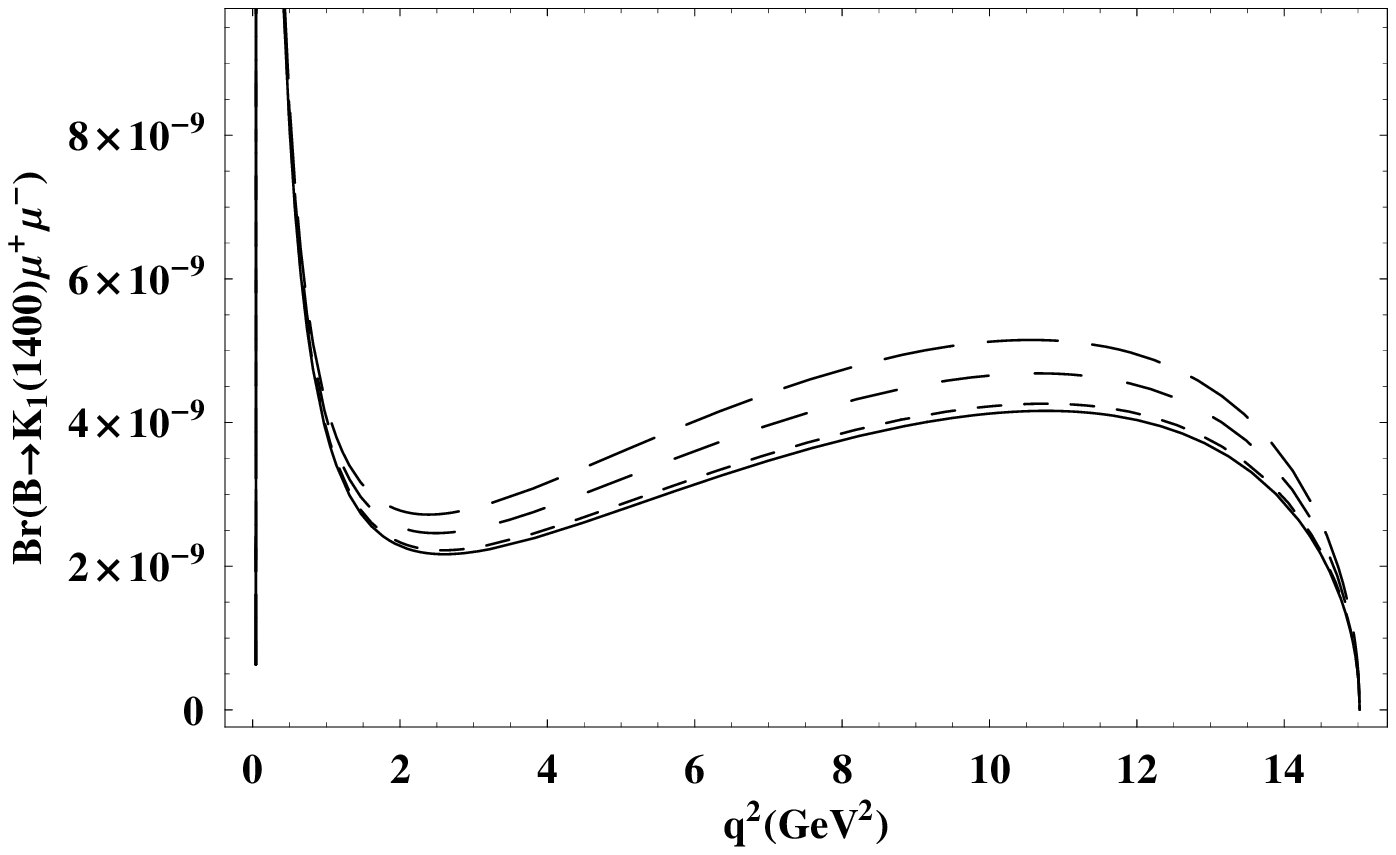} \ \ \
& \ \ \  \includegraphics[scale=0.4]{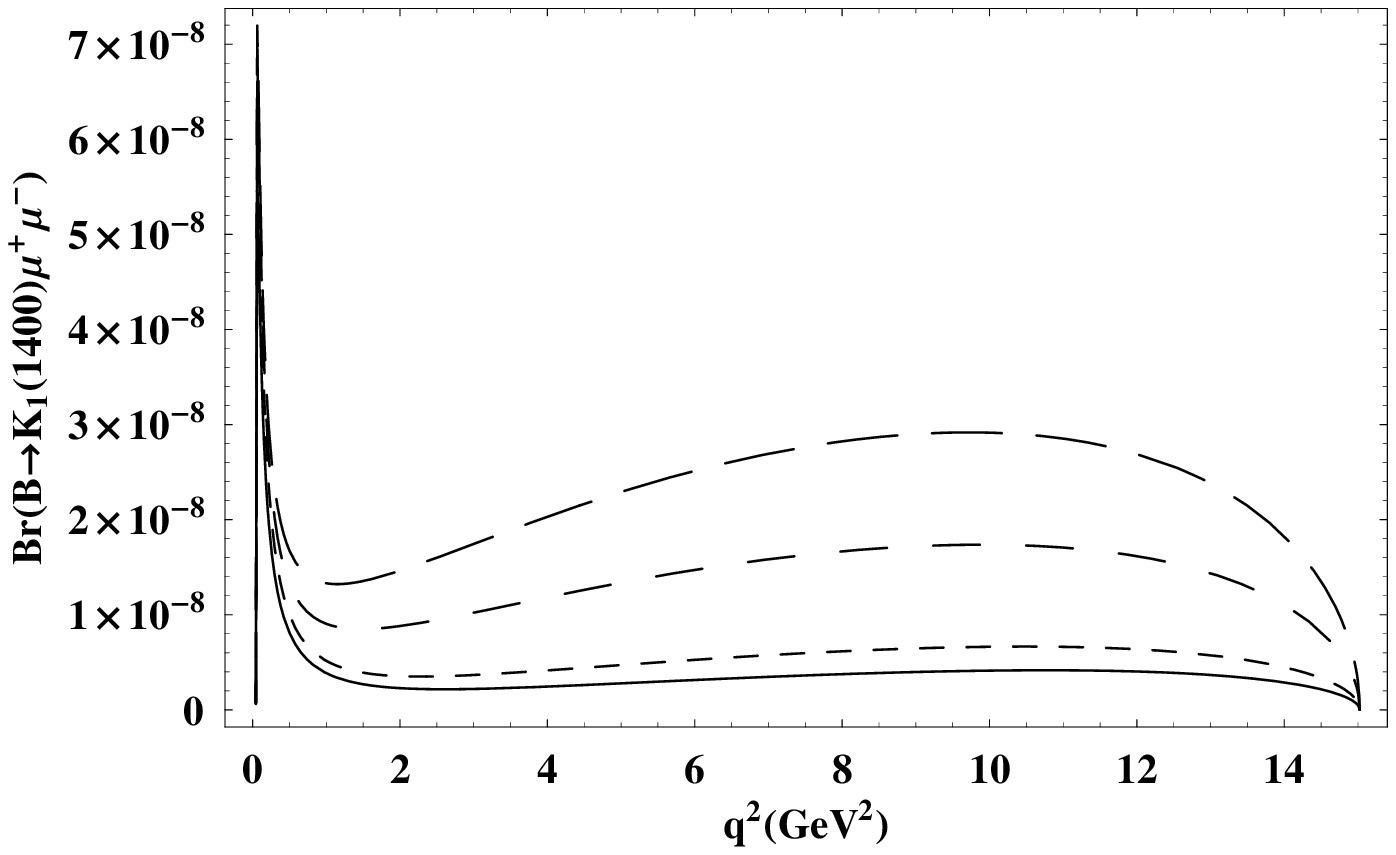}\end{tabular}
\caption{The dependence of branching ratio of $B\to K_{1}(1400)\mu^{+}\mu^{-}$ on $q^{2}$ for different values of $m_{t^{\prime}}$ and $\left\vert V^{\ast}_{t^{\prime}b}V_{t^{\prime}s}\right\vert$. The legends and the values
of fourth generation parameters are same as in Fig. \ref{Branching ratios for muons 12}.}  \label{Branching ratios for muons 14}
\end{figure*}
\begin{figure*}[ht]
\begin{tabular}{cc}
\hspace{.6cm}($\mathbf{a}$)&\hspace{1.2cm}($\mathbf{b}$)\\
\includegraphics[scale=0.4]{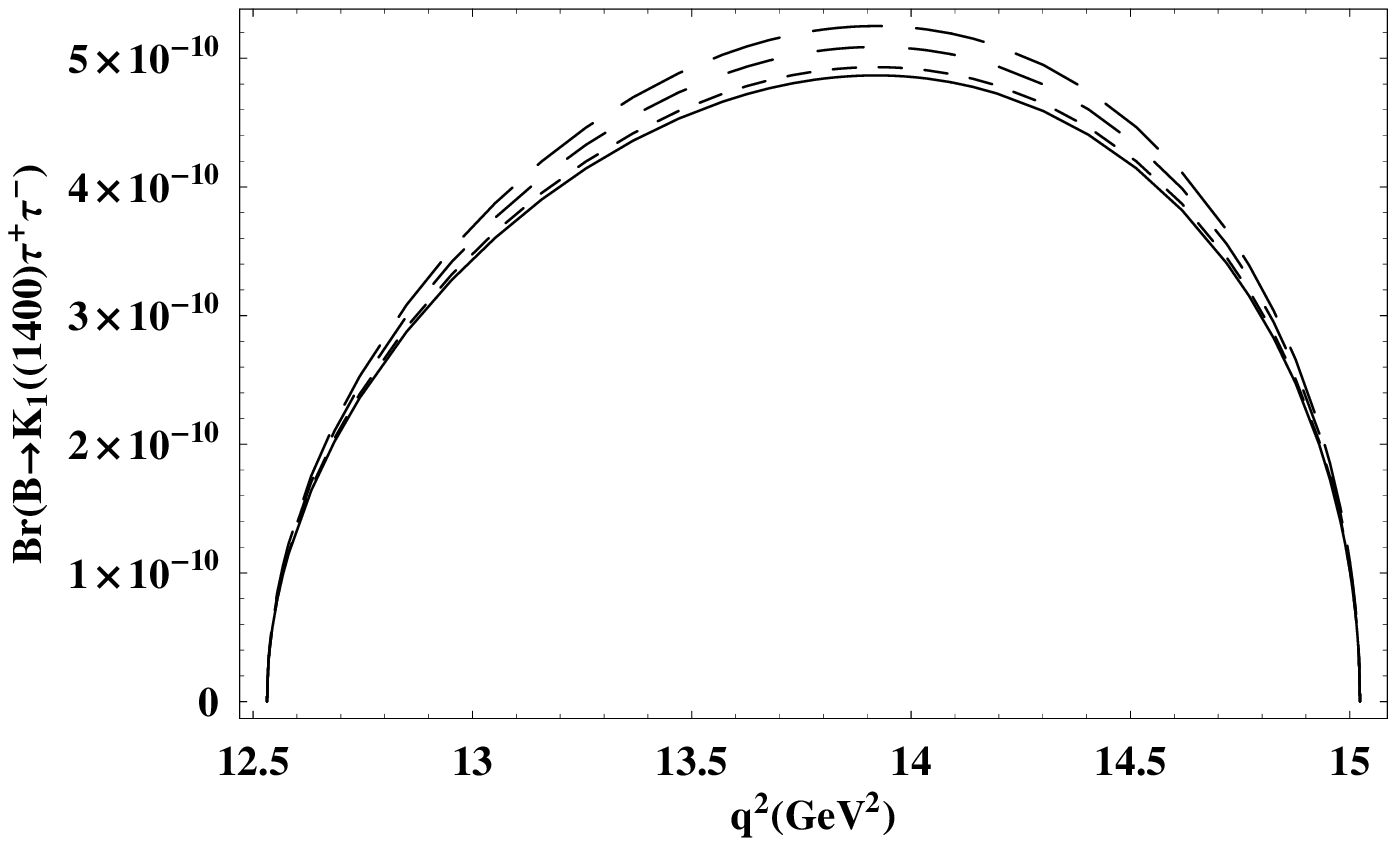}\ \ \
& \ \ \  \includegraphics[scale=0.4]{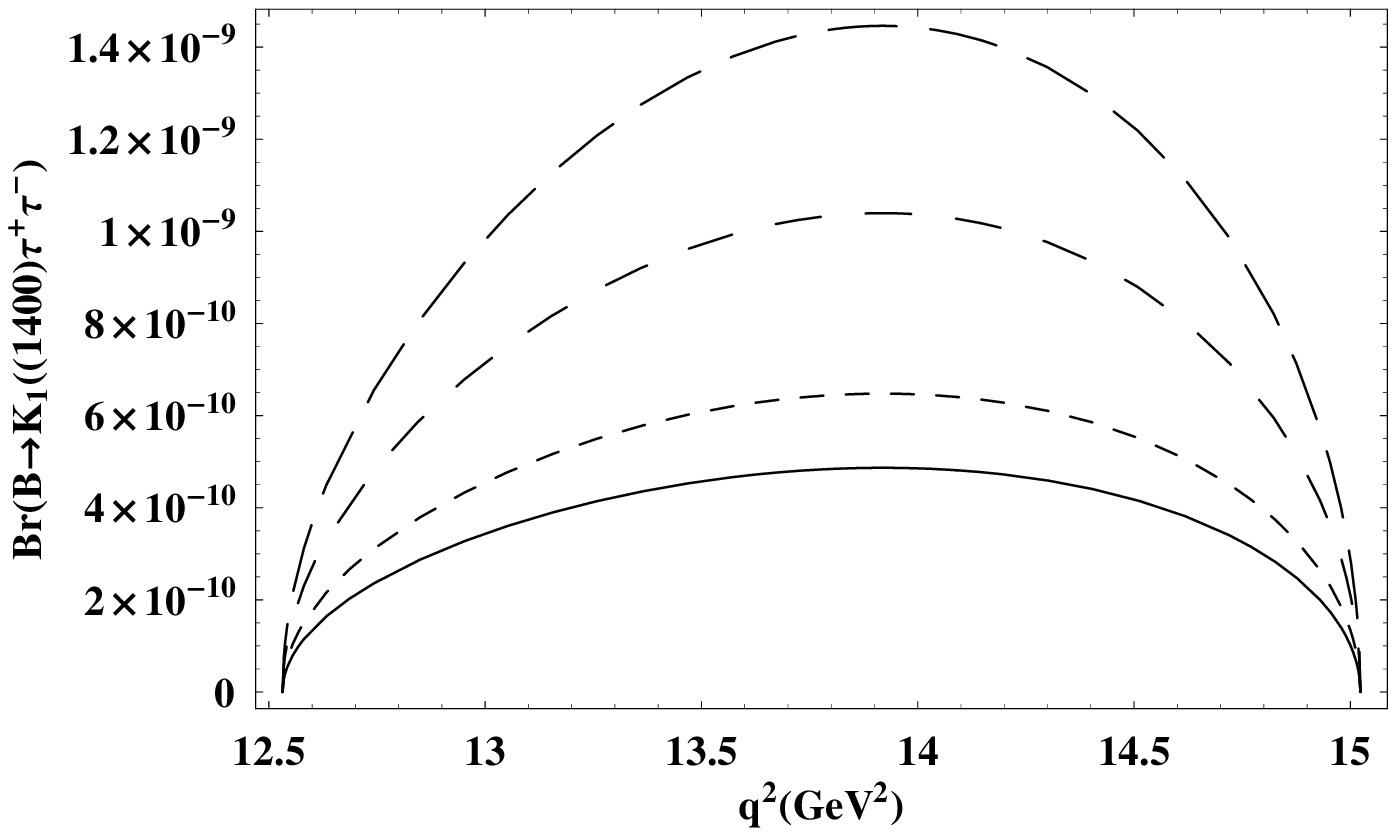}\end{tabular}
\caption{The dependence of branching ratio of $B\to K_{1}(1400)\tau^{+}\tau^{-}$ on $q^{2}$ for different values of $m_{t^{\prime}}$ and $\left\vert V^{\ast}_{t^{\prime}b}V_{t^{\prime}s}\right\vert$. The legends and the values
of fourth generation parameters are same as in Fig. \ref{Branching ratios for muons 12}.}  \label{Branching ratios for tauons 14}
\end{figure*}
\begin{figure*}[ht]
\begin{tabular}{cc}
\hspace{.6cm}($\mathbf{a}$)&\hspace{1.2cm}($\mathbf{b}$)\\
\includegraphics[scale=0.4]{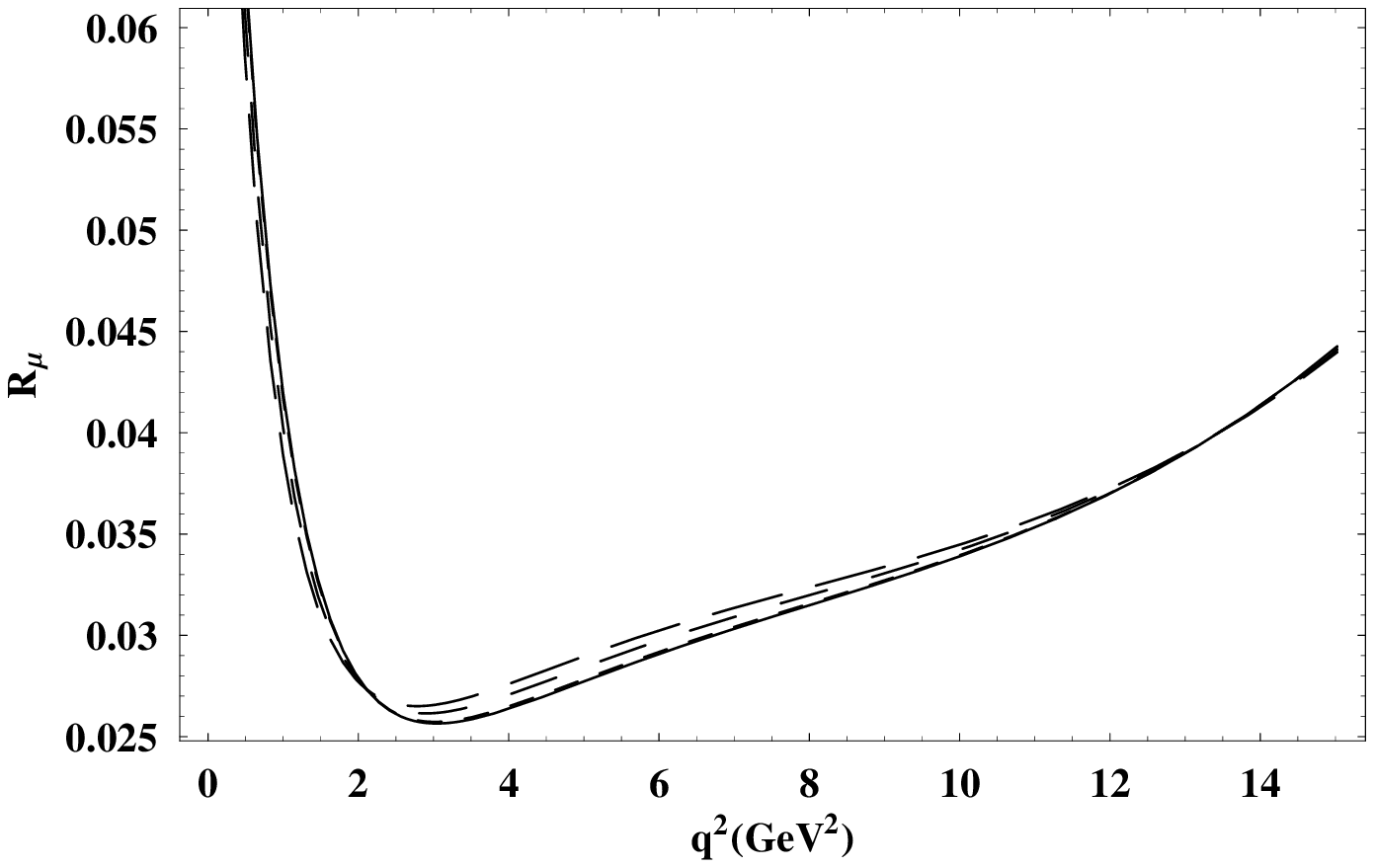} \ \ \
& \ \ \  \includegraphics[scale=0.4]{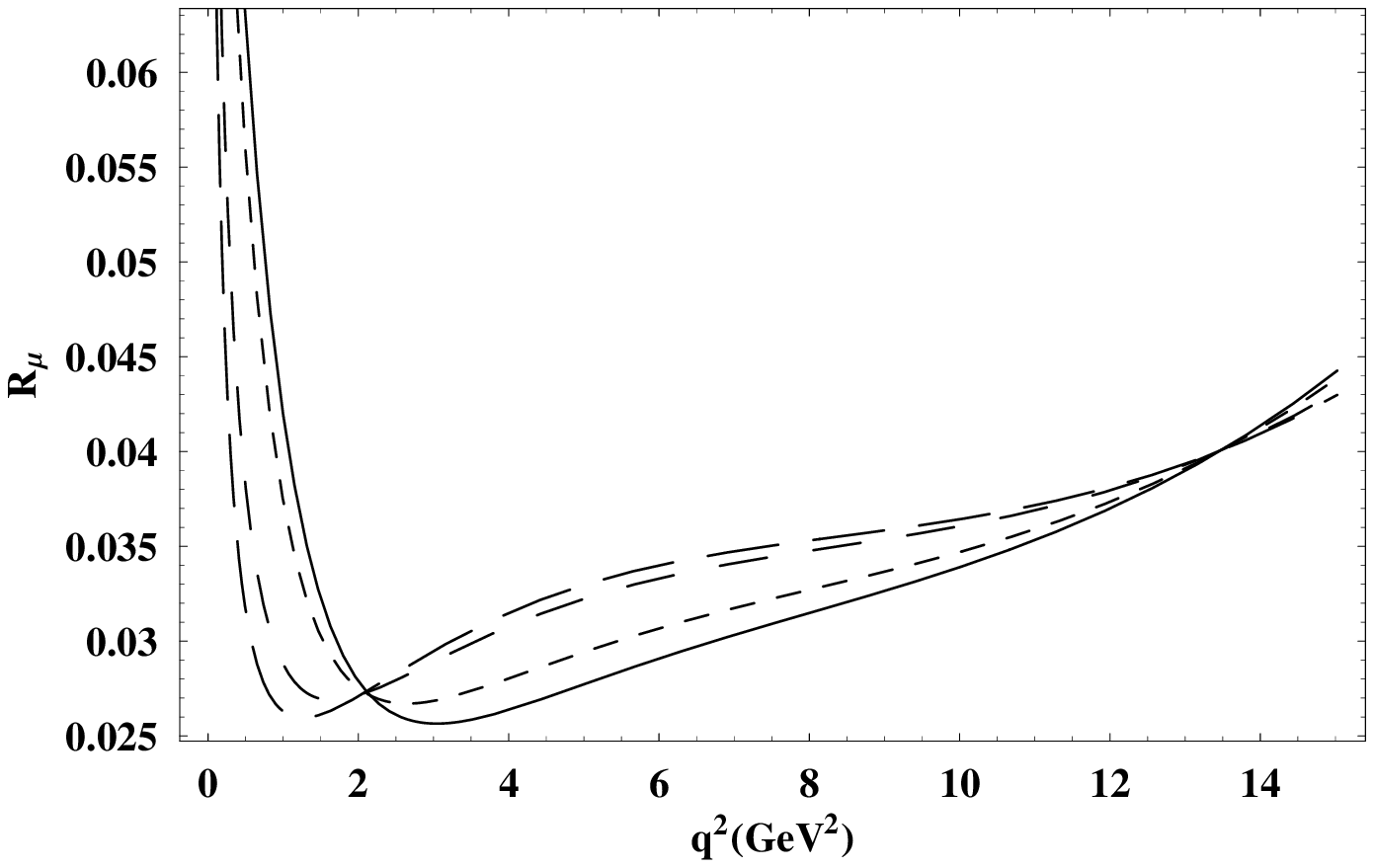}\end{tabular}
\caption{The dependence of branching fraction $\mathcal{R}_{\mu}=\mathcal{BR}(B\rightarrow K_{1}(1400)\mu ^{+}\mu ^{-})/\mathcal{BR}(B\rightarrow K_{1}(1270)\mu ^{+}\mu ^{-})$ on $q^{2}$ for different values of $m_{t^{\prime}}$ and $\left\vert V^{\ast}_{t^{\prime}b}V_{t^{\prime}s}\right\vert$. The legends and the values
of fourth generation parameters are same as in Fig. \ref{Branching ratios for muons 12}.}  \label{brf for muons}
\end{figure*}
\begin{figure*}[ht]
\begin{tabular}{cc}
\hspace{.6cm}($\mathbf{a}$)&\hspace{1.2cm}($\mathbf{b}$)\\
\includegraphics[scale=0.4]{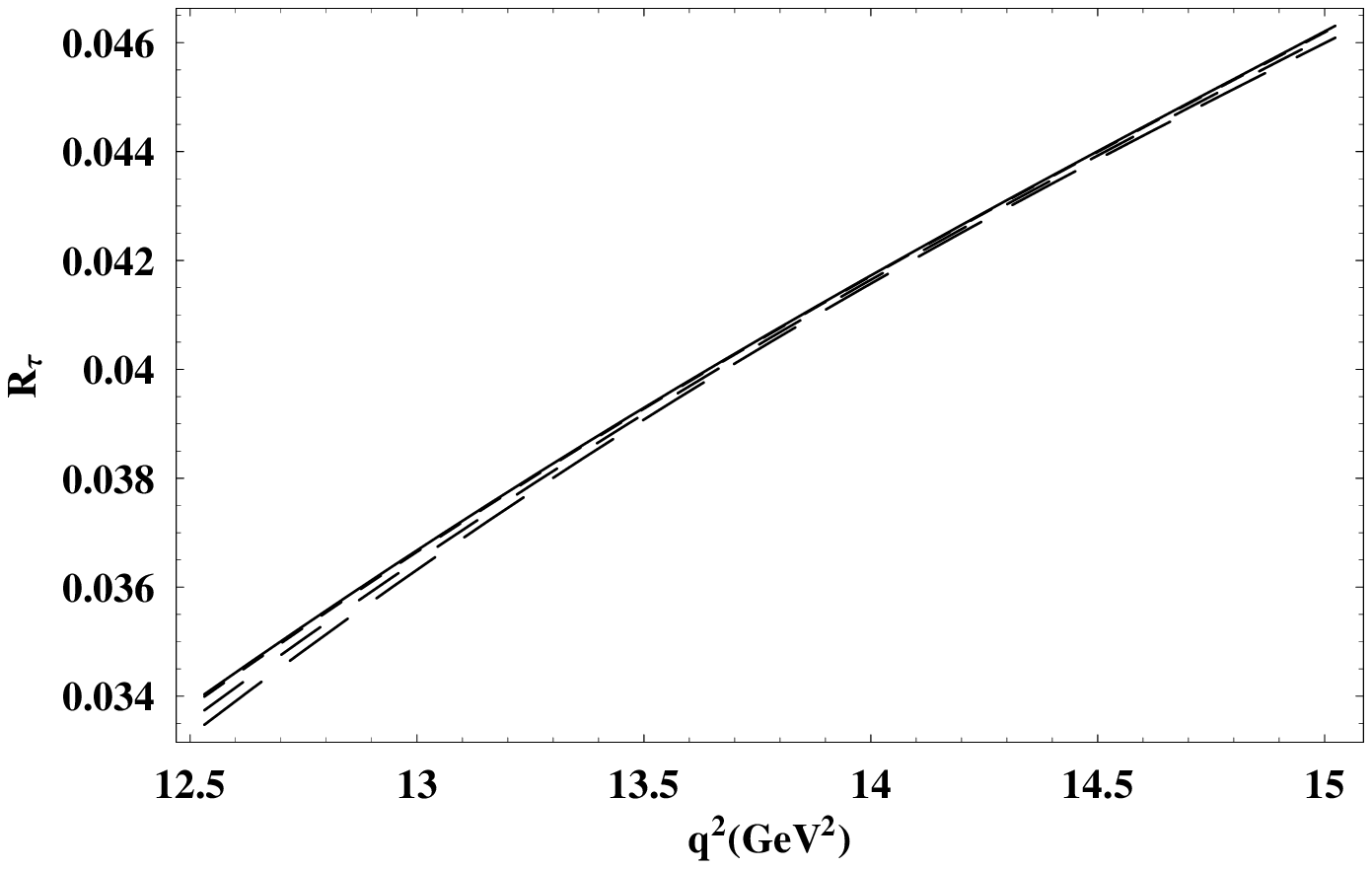} \ \ \
& \ \ \  \includegraphics[scale=0.4]{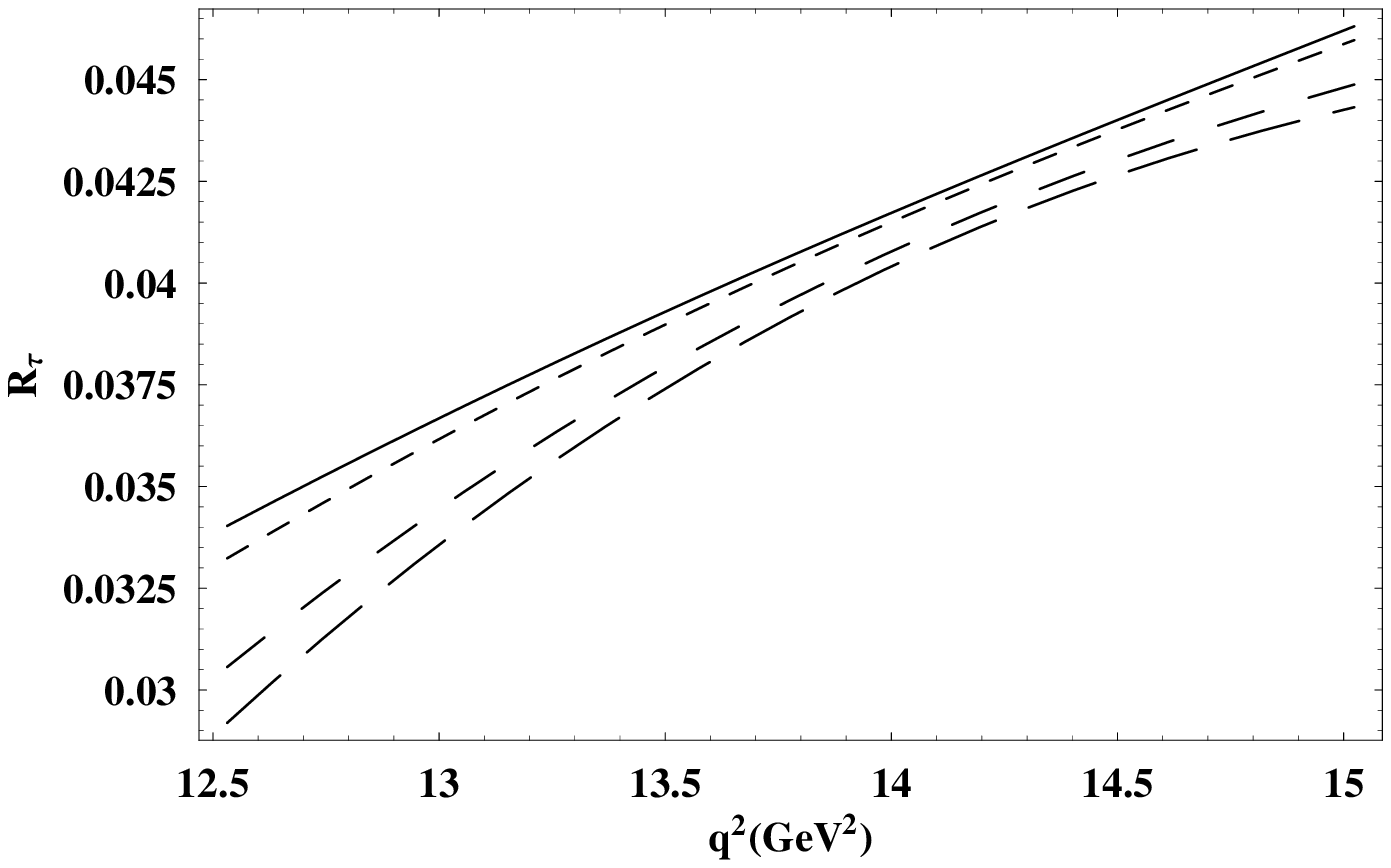}\end{tabular}
\caption{The dependence of branching fraction $\mathcal{R}_{\tau}=\mathcal{BR}(B\rightarrow K_{1}(1400)\tau ^{+}\tau ^{-})/\mathcal{BR}(B\rightarrow K_{1}(1270)\tau ^{+}\tau ^{-})$ on $q^{2}$ for different values of $m_{t^{\prime}}$ and $\left\vert V^{\ast}_{t^{\prime}b}V_{t^{\prime}s}\right\vert$. The legends and the values
of fourth generation parameters are same as in Fig. \ref{Branching ratios for muons 12}.}  \label{brf for tauons}
\end{figure*}

\begin{figure*}[ht]
\begin{tabular}{cc}
\hspace{.6cm}($\mathbf{a}$)&\hspace{1.2cm}($\mathbf{b}$)\\
\includegraphics[scale=0.4]{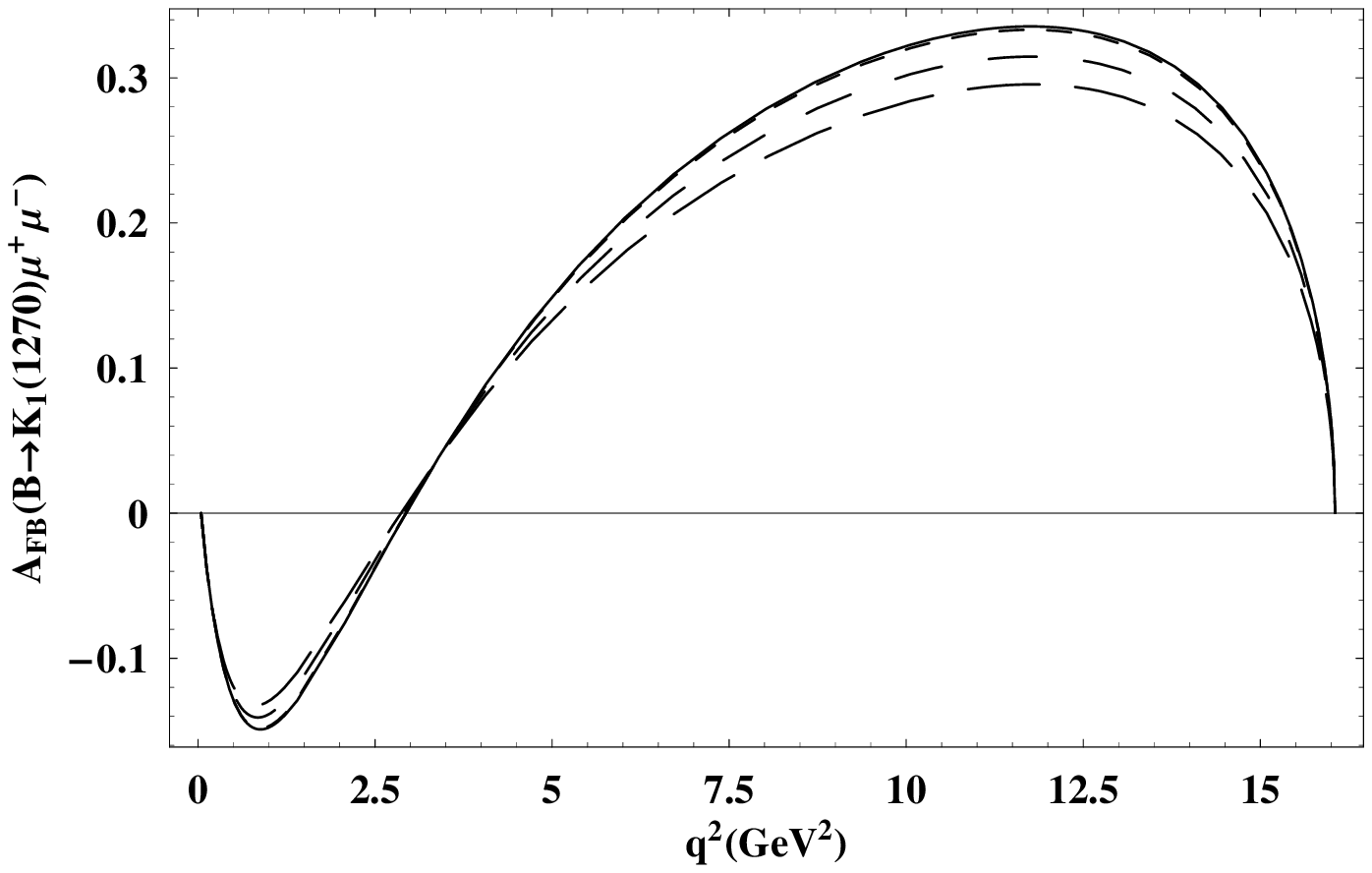} \ \ \
& \ \ \  \includegraphics[scale=0.4]{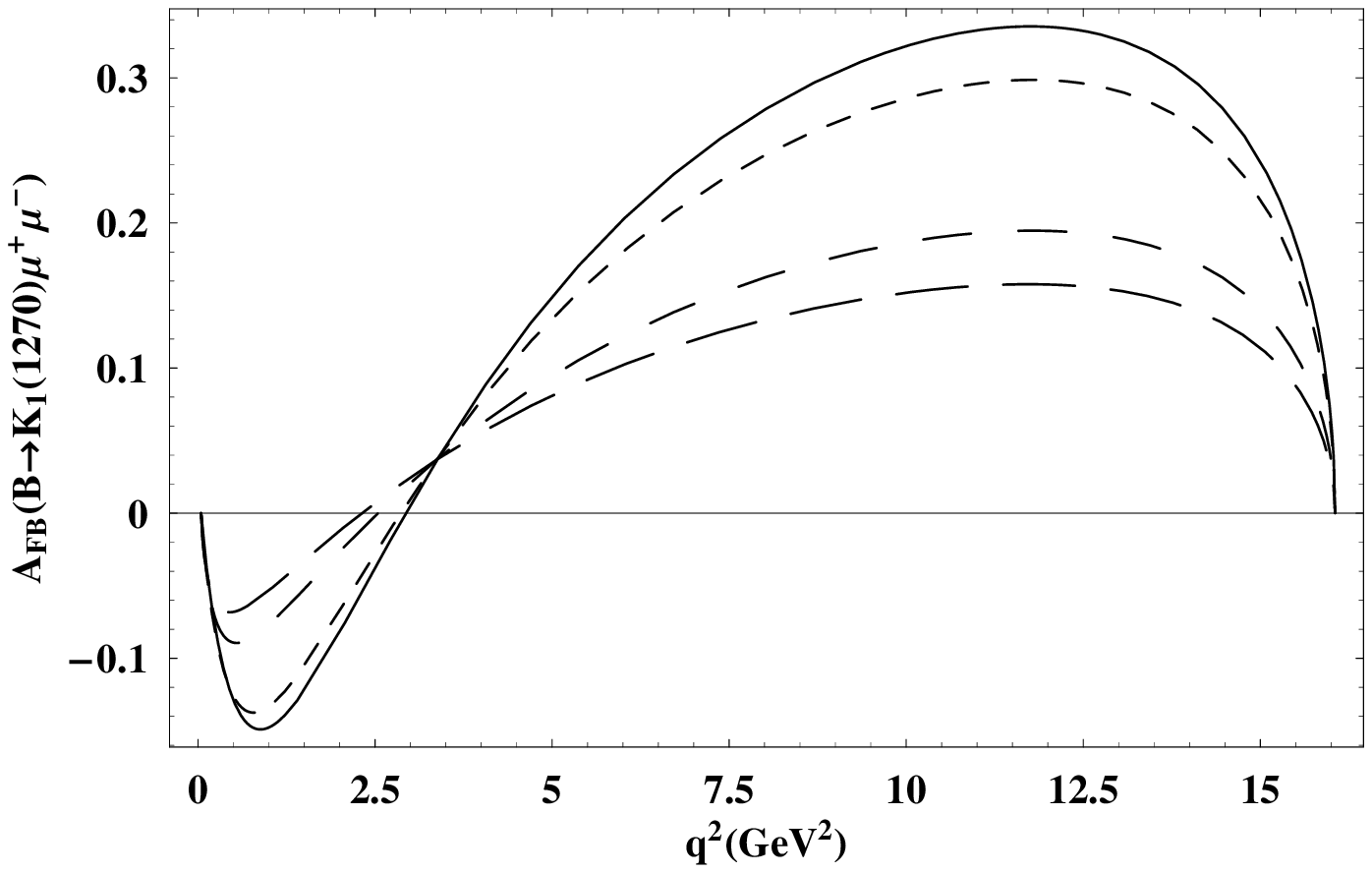}\end{tabular}
\caption{The dependence of forward backward asymmetry of $B\to K_{1}(1270)\mu^{+}\mu^{-}$ on $q^{2}$ for different values of $m_{t^{\prime}}$ and $\left\vert V^{\ast}_{t^{\prime}b}V_{t^{\prime}s}\right\vert$. The legends and the values
of fourth generation parameters are same as in Fig. \ref{Branching ratios for muons 12}.}  \label{FBA for muons 12}
\end{figure*}

\begin{figure*}[ht]
\begin{tabular}{cc}
\hspace{.6cm}($\mathbf{a}$)&\hspace{1.2cm}($\mathbf{b}$)\\
\includegraphics[scale=0.4]{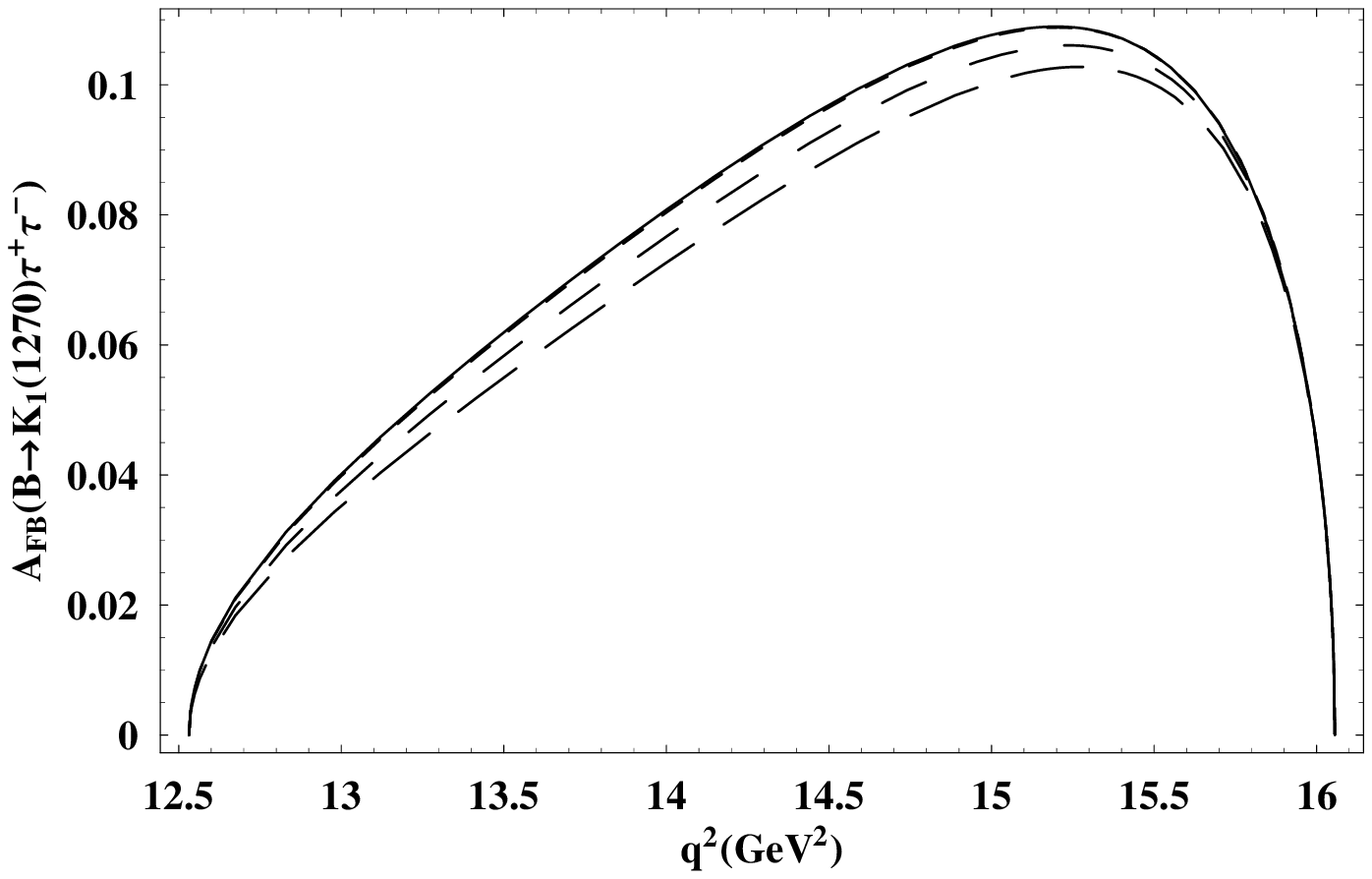} \ \ \
& \ \ \  \includegraphics[scale=0.4]{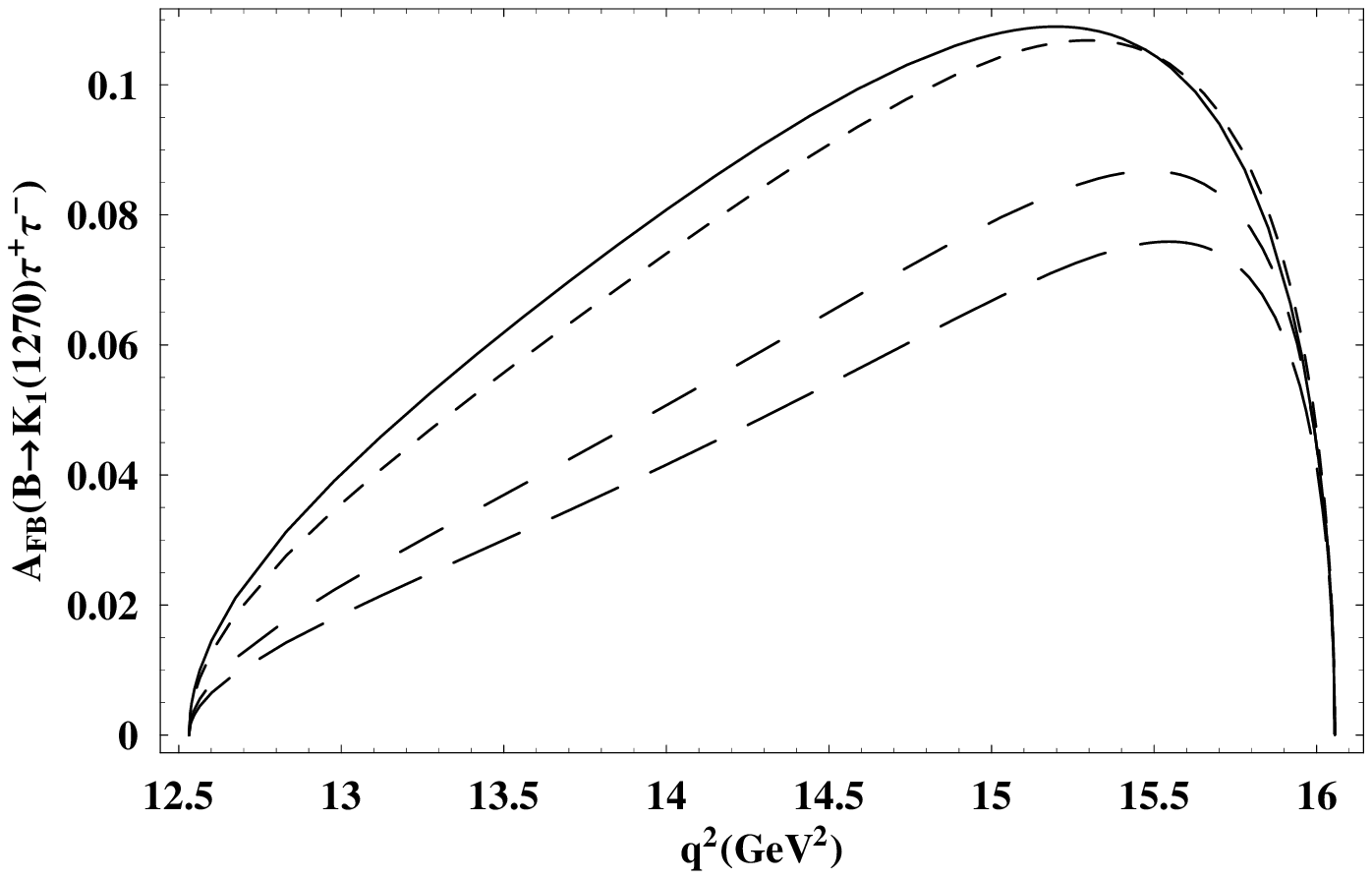}\end{tabular}
\caption{The dependence of forward backward asymmetry of $B\to K_{1}(1270)\tau^{+}\tau^{-}$ on $q^{2}$ for different values of $m_{t^{\prime}}$ and $\left\vert V^{\ast}_{t^{\prime}b}V_{t^{\prime}s}\right\vert$. The legends and the values
of fourth generation parameters are same as in Fig. \ref{Branching ratios for muons 12}.} \label{FBA for tauons 12}
\end{figure*}

\begin{figure*}[ht]
\begin{tabular}{cc}
\hspace{.6cm}($\mathbf{a}$)&\hspace{1.2cm}($\mathbf{b}$)\\
\includegraphics[scale=0.4]{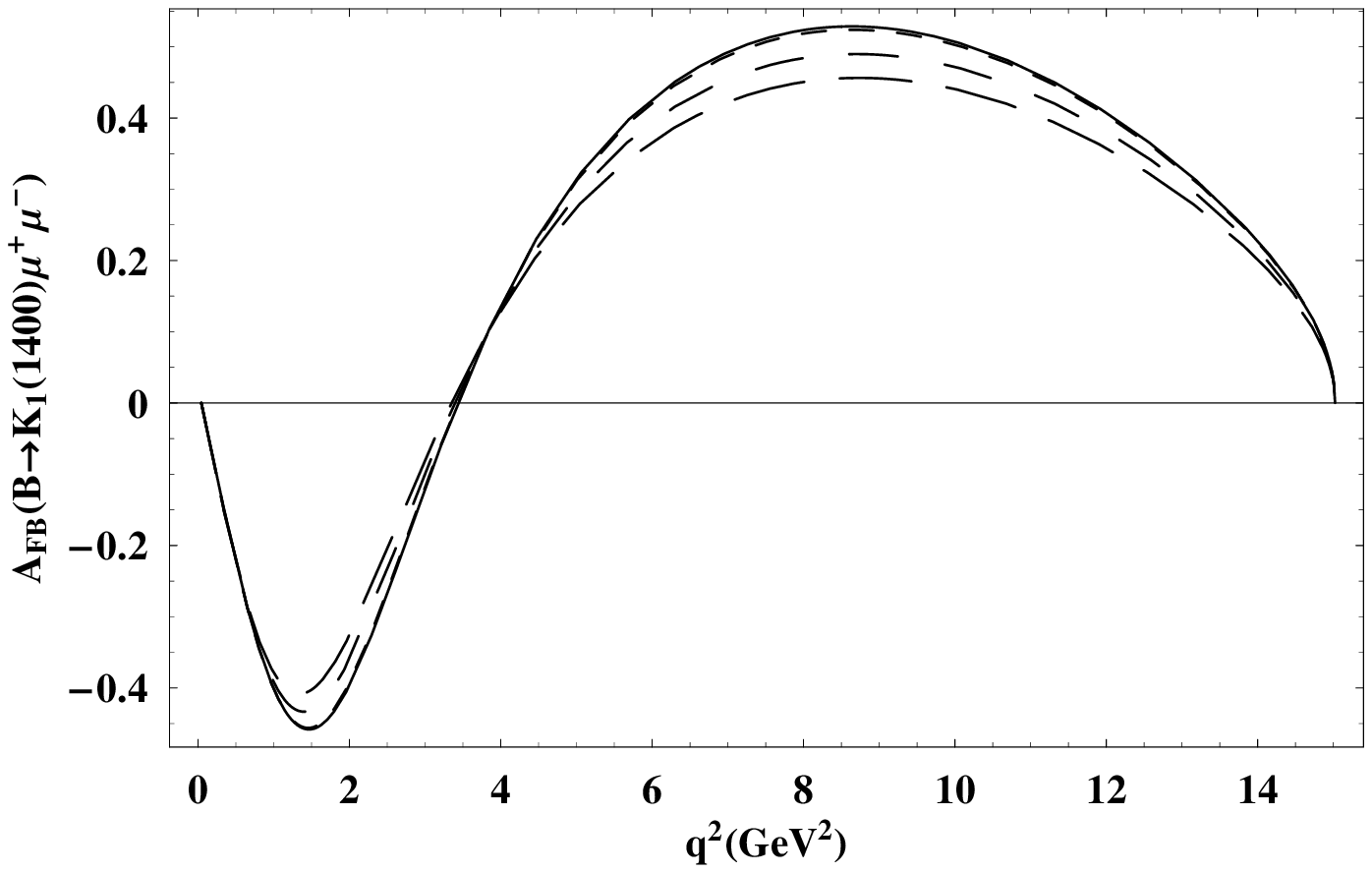} \ \ \
& \ \ \  \includegraphics[scale=0.4]{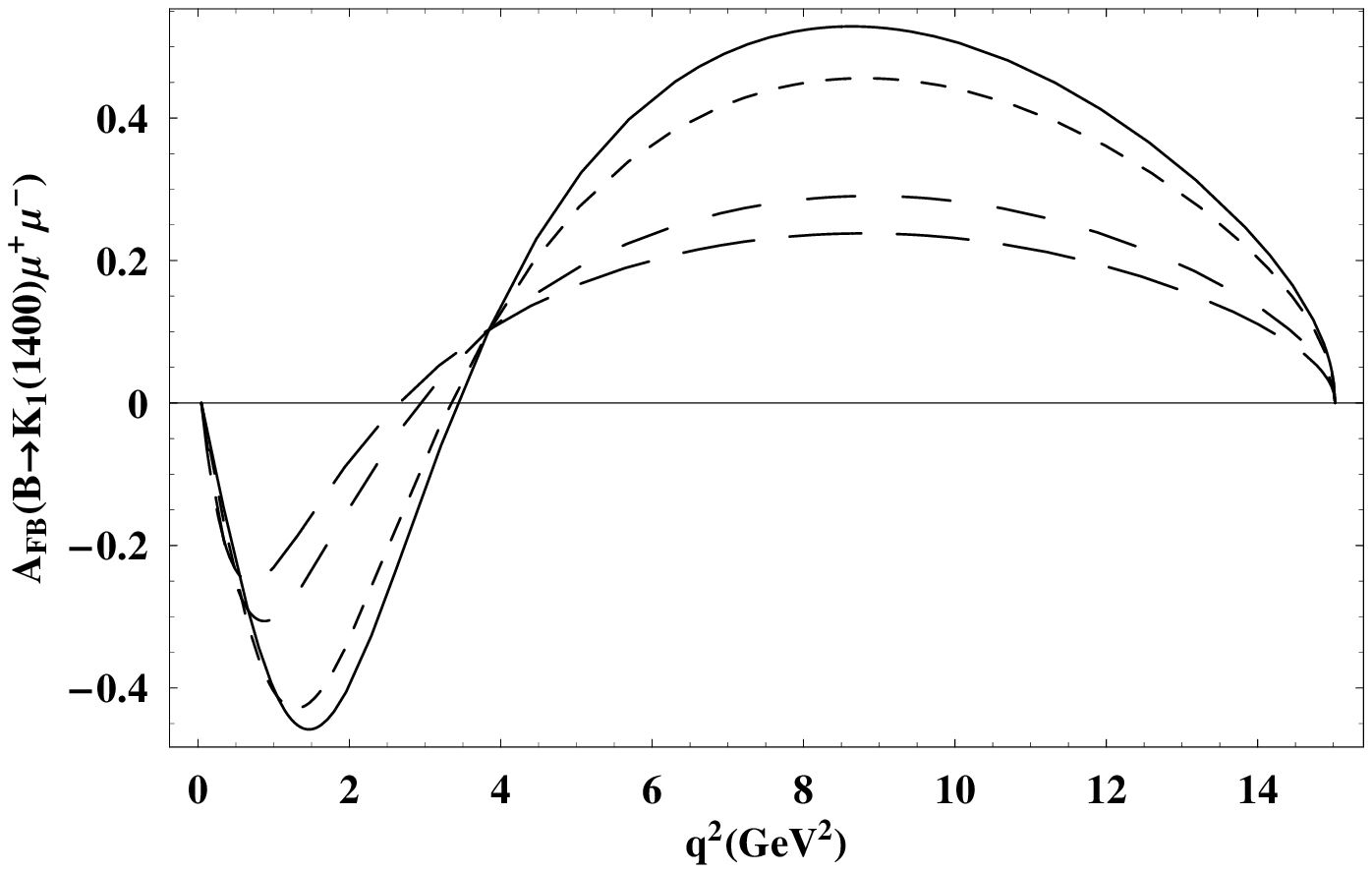}\end{tabular}
\caption{The dependence of forward backward asymmetry of $B\to K_{1}(1400)\mu^{+}\mu^{-}$ on $q^{2}$ for different values of $m_{t^{\prime}}$ and $\left\vert V^{\ast}_{t^{\prime}b}V_{t^{\prime}s}\right\vert$. The legends and the values
of fourth generation parameters are same as in Fig. \ref{Branching ratios for muons 12}.} \label{FBA for muons 14}
\end{figure*}

\begin{figure*}[ht]
\begin{tabular}{cc}
\hspace{.6cm}($\mathbf{a}$)&\hspace{1.2cm}($\mathbf{b}$)\\
\includegraphics[scale=0.4]{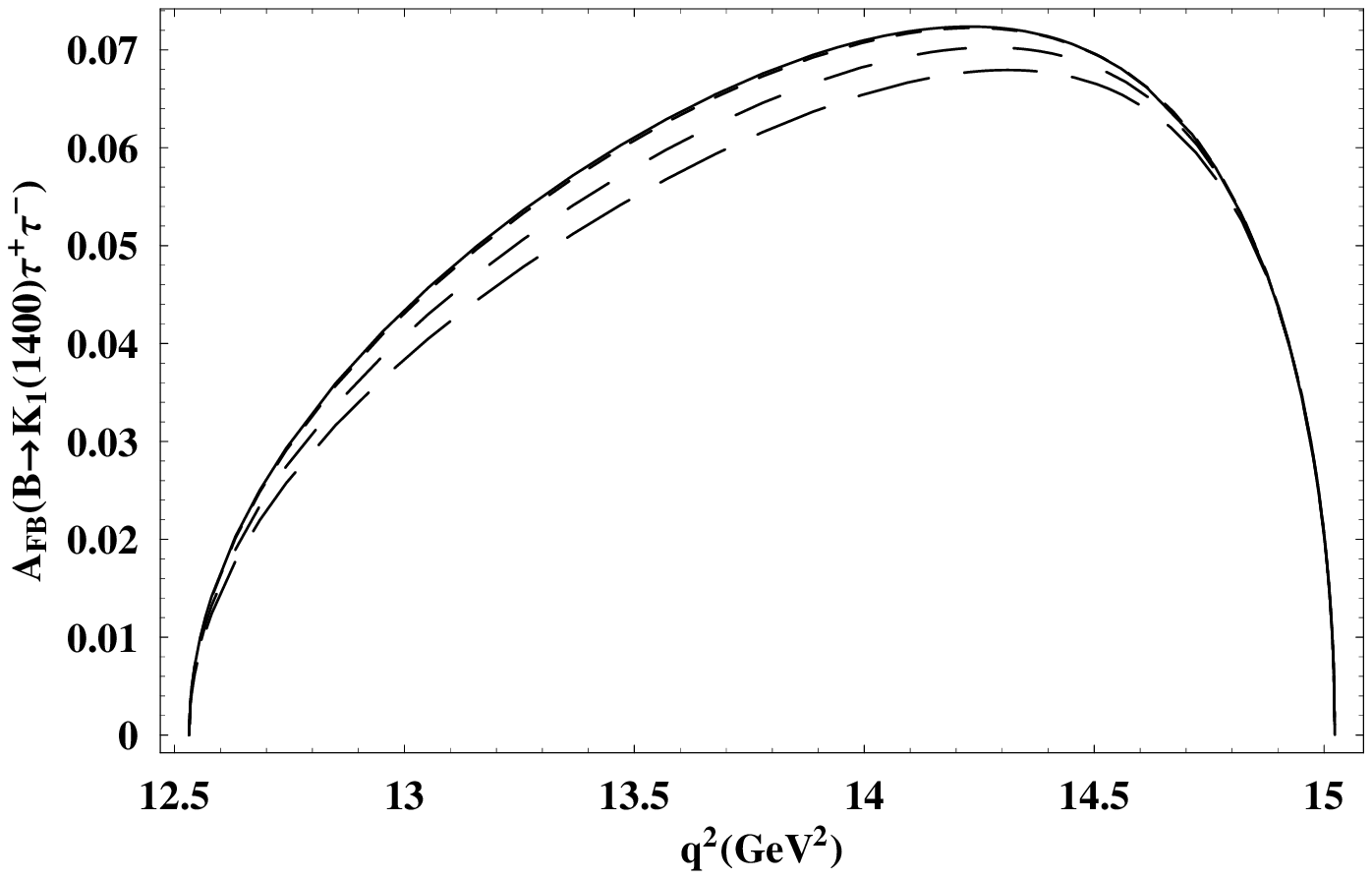} \ \ \
& \ \ \  \includegraphics[scale=0.4]{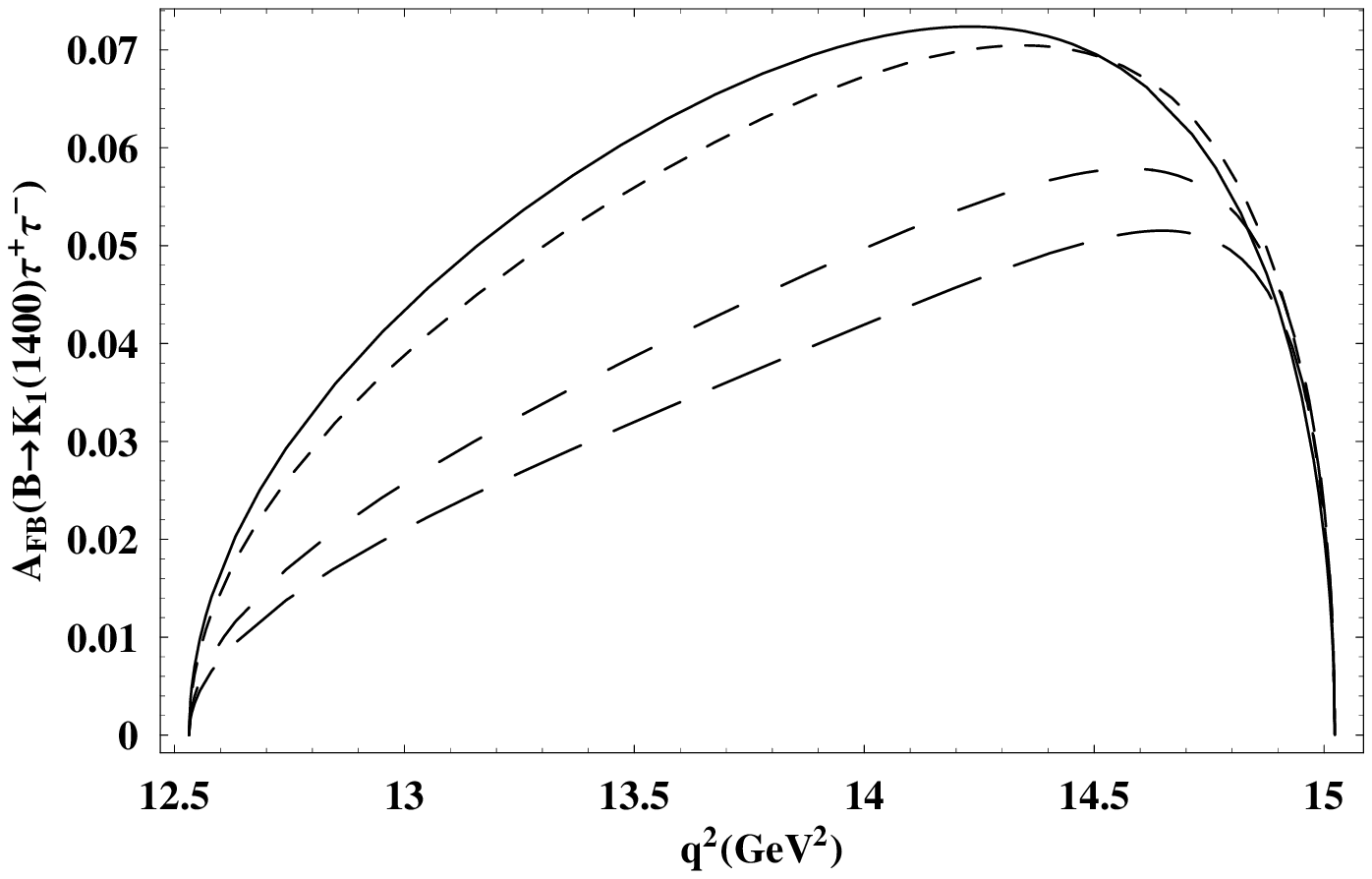}\end{tabular}
\caption{The dependence of forward backward asymmetry of $B\to K_{1}(1400)\tau^{+}\tau^{-}$ on $q^{2}$ for different values of $m_{t^{\prime}}$ and $\left\vert V^{\ast}_{t^{\prime}b}V_{t^{\prime}s}\right\vert$. The legends and the values
of fourth generation parameters are same as in Fig. \ref{Branching ratios for muons 12}.} \label{FBA for tauons 14}
\end{figure*}

To illustrate the generic effects due to the fourth generation quarks
on the forward-backward asymmetry $\mathcal{A}_{FB}$, we plot
$\frac{d(\mathcal{A}_{FB})}{dq^{2}}$ as a function of $q^{2}$ in
Figs. \ref{FBA for muons 12}-\ref{FBA for tauons 14}. As it is shown
in Ref. \cite{HY} that the zero position of the $\mathcal{A}_{FB}$
depends weakly on the value of $\theta_{K}$ but can be changed due
to the variation of the NP scenarios. For the zero position of
$\mathcal{A}_{FB}$ it is also argued that the uncertainty in the
zero position of the $\mathcal{A}_{FB}$ is due to the hadronic
uncertainties (form factors) is negligible \cite{new2}. Therefore,
the zero position of the $\mathcal{A}_{FB}$ could also provide a
stringent test for the NP effects.

In the present study Figs. \ref{FBA for muons 12} and \ref{FBA for
muons 14} show the case of muons as final state leptons, the
increment in the $|V_{t^{\prime}b}V_{t^{\prime}s}|$ and
$m_{t^{\prime}}$ values shift the zero position of the
$\mathcal{A}_{FB}$ towards the low $q^{2}$ region, this behavior is
compatible with $B\to K^{\ast}\mu^{+}\mu^{-}$ decay \cite{vb}. Moreover, the maximum values of
$|V_{t^{\prime}b}V_{t^{\prime}s}|$ and $m_{t^{\prime}}$, shift the SM
value ($2.8$ GeV$^{2}$) of zero position of the $\mathcal{A}_{FB}$
for the case of $B\rightarrow K_{1}(1270)\mu^{+}\mu^{-}$ (see Fig.
\ref{FBA for muons 12}-b) to the value $2.1$ GeV$^{2}$. For the case
of $B\rightarrow K_{1}(1400)\mu^{+}\mu^{-}$ (see Fig. \ref{FBA for
muons 14}-b) the zero position of the $\mathcal{A}_{FB}$ is shifted from its SM value (3.4 GeV$^{2}$) to the value 2.4 GeV$^{2}$.

Besides the zero position of $\mathcal{A}_{FB}$, the magnitude of
$\mathcal{A}_{FB}$ is also important tool (particularly, when the
tauons are the final state leptons where the zero of the
$\mathcal{A}_{FB}$ is absent) to investigate the NP. A closer look
on the pattern of Figs. \ref{FBA for muons 12}-\ref{FBA for tauons
14} tells us that the fourth generation parameters decrease the
magnitude of $\mathcal{A}_{FB}$ from its SM value. The analysis of
$\mathcal{A}_{FB}$ also demonstrate that in contrast to the
$\mathcal{BR}$, the magnitude of the $\mathcal{A}_{FB}$ is
decreasing function of the fourth generation parameters. It is clear
from these graphs that decreasing behavior of the magnitude of
$\mathcal{A}_{FB}$ is irrespective of the final state particles. It is suitable to comment here that just like the zero
position of the $\mathcal{A}_{FB}$, the magnitude of
$\mathcal{A}_{FB}$ depends on the values of the Wilson coefficient
$C_{7}, C_{9}$ and $C_{10}$. Thus the effects on the magnitude of
$\mathcal{A}_{FB}$ are almost insensitive due to the uncertainties
in the form factors. We noticed that the uncertainty due to the
mixing angle $\theta_{K}$, magnitude of $\mathcal{A}_{FB}$ is mildly
effected. On the other hand the change in the magnitude of
$\mathcal{A}_{FB}$ due to the fourth generation are very prominent and easy to measure at the experiment. In the last, precise
measurement of the zero position and the magnitude of
$\mathcal{A}_{FB}$ are very good observables to yield any
indirect imprints of NP including fourth generation.

We now discuss another interesting observable to get the complementary information about NP in $B\rightarrow
 K_{1}(1270,1400)\ell^{+}\ell^{-}$ transitions i.e. the helicity
 fractions of $K_{1}(1270,1400)$ produced in the final state. The measurement of longitudinally $K^{*}$ helicity fractions
 ($f_{L}$) in the decay modes $B\rightarrow
 K^{*}\ell^{+}\ell^{-}$ by BABAR collaboration with experimental error \cite{bab} put
 enormous interest in this observable. Additionally, this is also shown that the helicity fractions of final state meson, just
like $\mathcal{BR}$ and $\mathcal{A}_{FB}$, are also very good
observables to dig out the NP \cite{Colangelo, paracha}.
Current and future $B$ factories will accumulate more data on this
observable which will be helpful not only to reduce the experimental errors but also to
get any possible hint of NP from this observable. In this regard, it
is natural to study the helicity fractions for the complementary
FCNC processes like $B\rightarrow K_{1}(1270,1400)\ell^{+}\ell^{-}$
in and beyond the SM. For this purpose, we have plotted the
longitudinal ($f_{L}$) and transverse ($f_{T}$) helicity fractions of
$K_{1}(1270,1400)$  for SM and with different values of fourth generation parameters in
Figs.(11-14). In these graphs the values of the longitudinal
($f_{L}$) and transverse ($f_{T}$) helicity fractions of
$K_{1}(1270,1400)$ are plotted against $q^{2}$ and one can see
clearly that at each value of $q^{2}$ the sum of $f_{L}$ and
$f_{T}$ is equal to one.

\begin{figure*}[ht]
\begin{tabular}{cc}
\hspace{.6cm}($\mathbf{a}$)&\hspace{1.2cm}($\mathbf{b}$)\\
\includegraphics[scale=0.4]{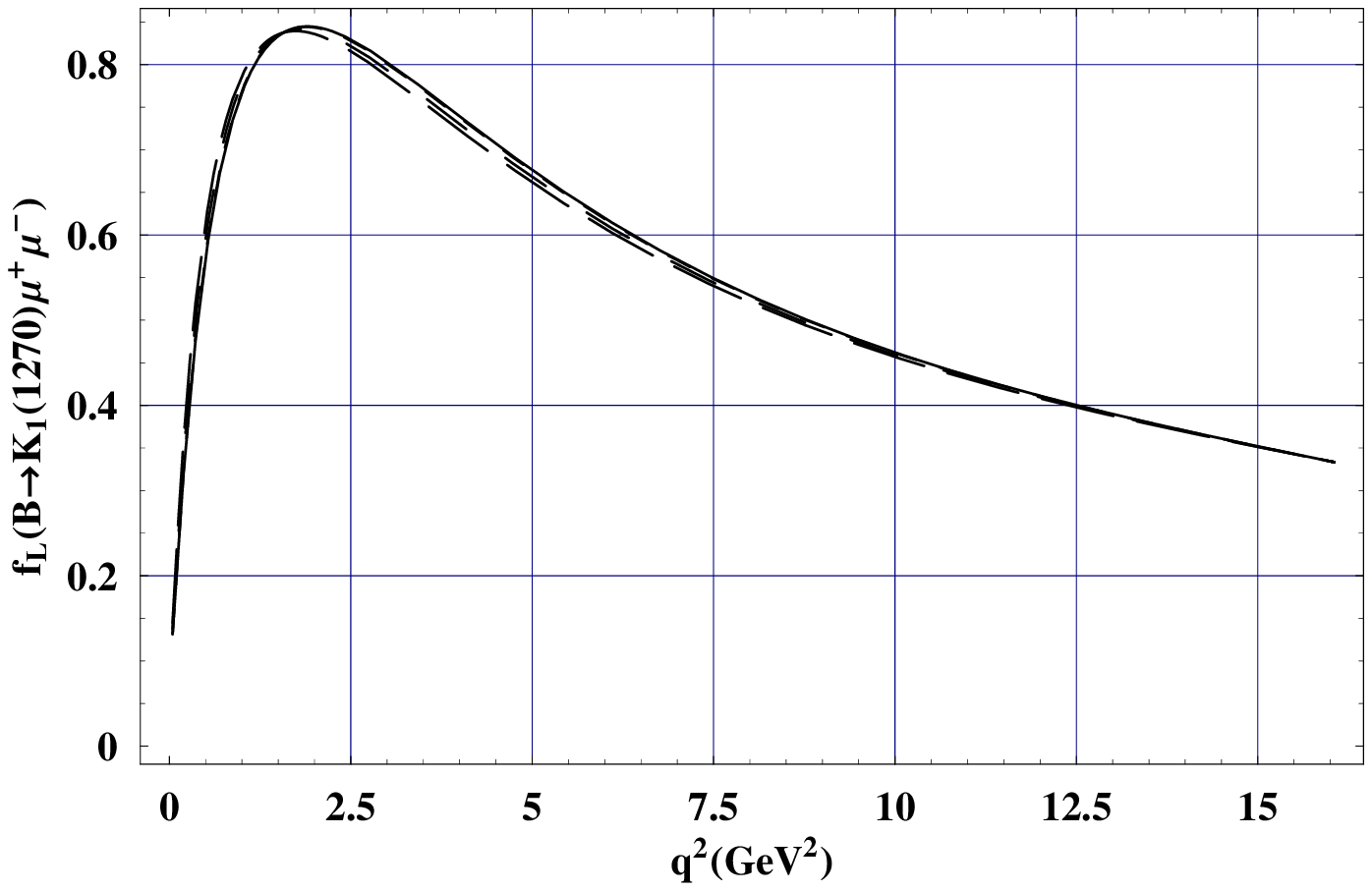} \ \ \
& \ \ \  \includegraphics[scale=0.4]{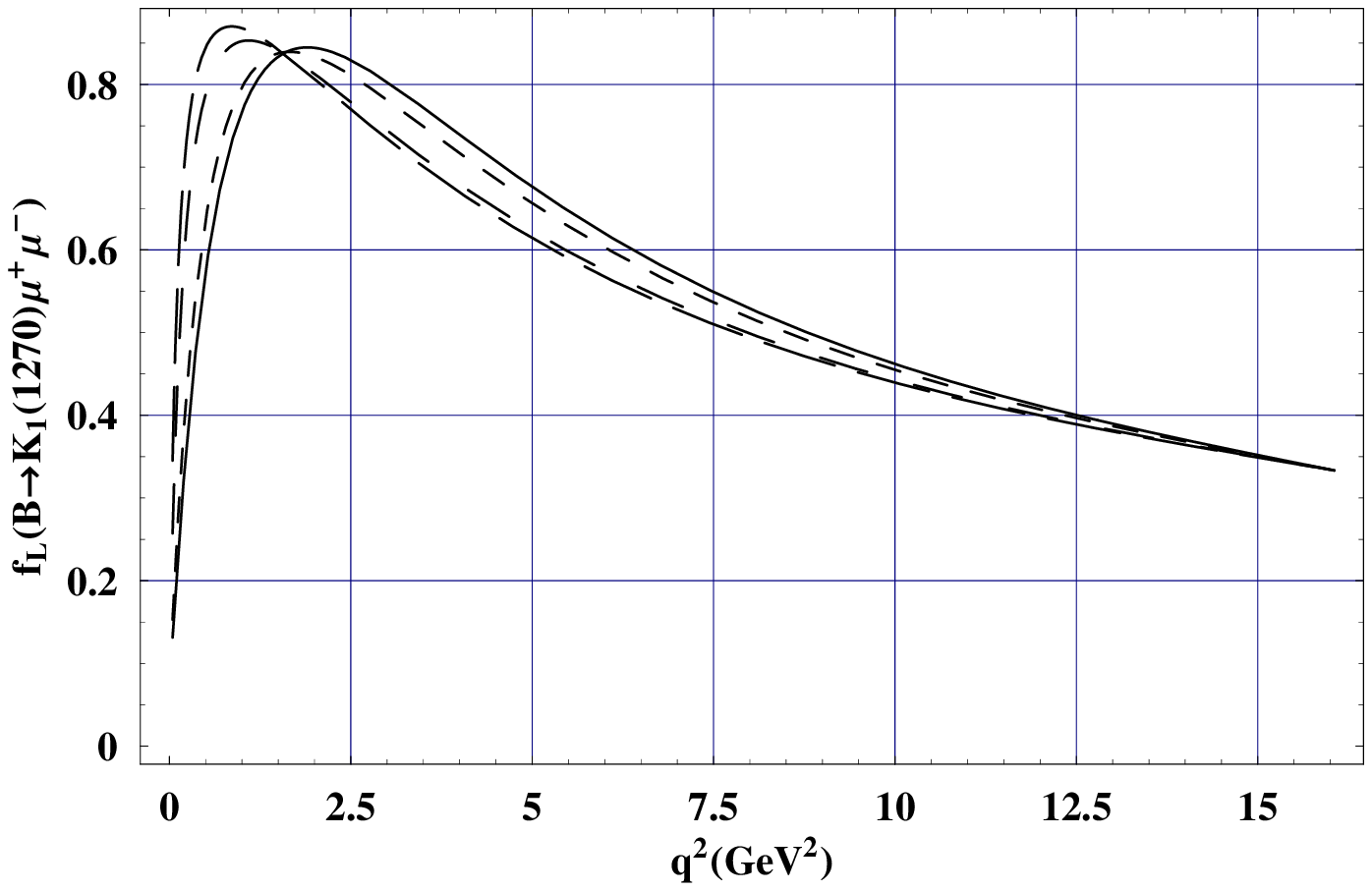}\\
\hspace{.6cm}($\mathbf{c}$)&\hspace{1.2cm}($\mathbf{d}$)\\
\includegraphics[scale=0.4]{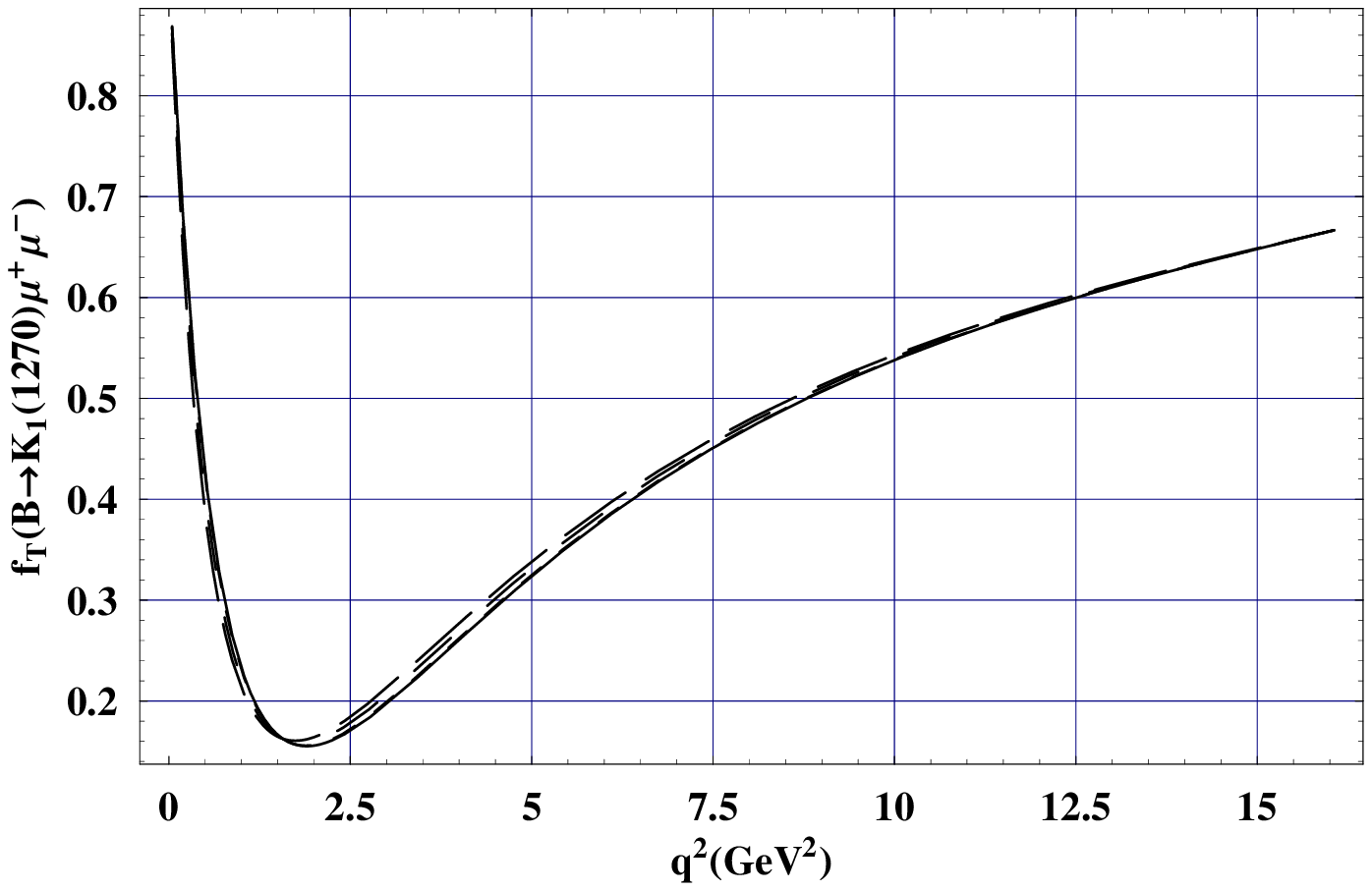} \ \ \
& \ \ \  \includegraphics[scale=0.4]{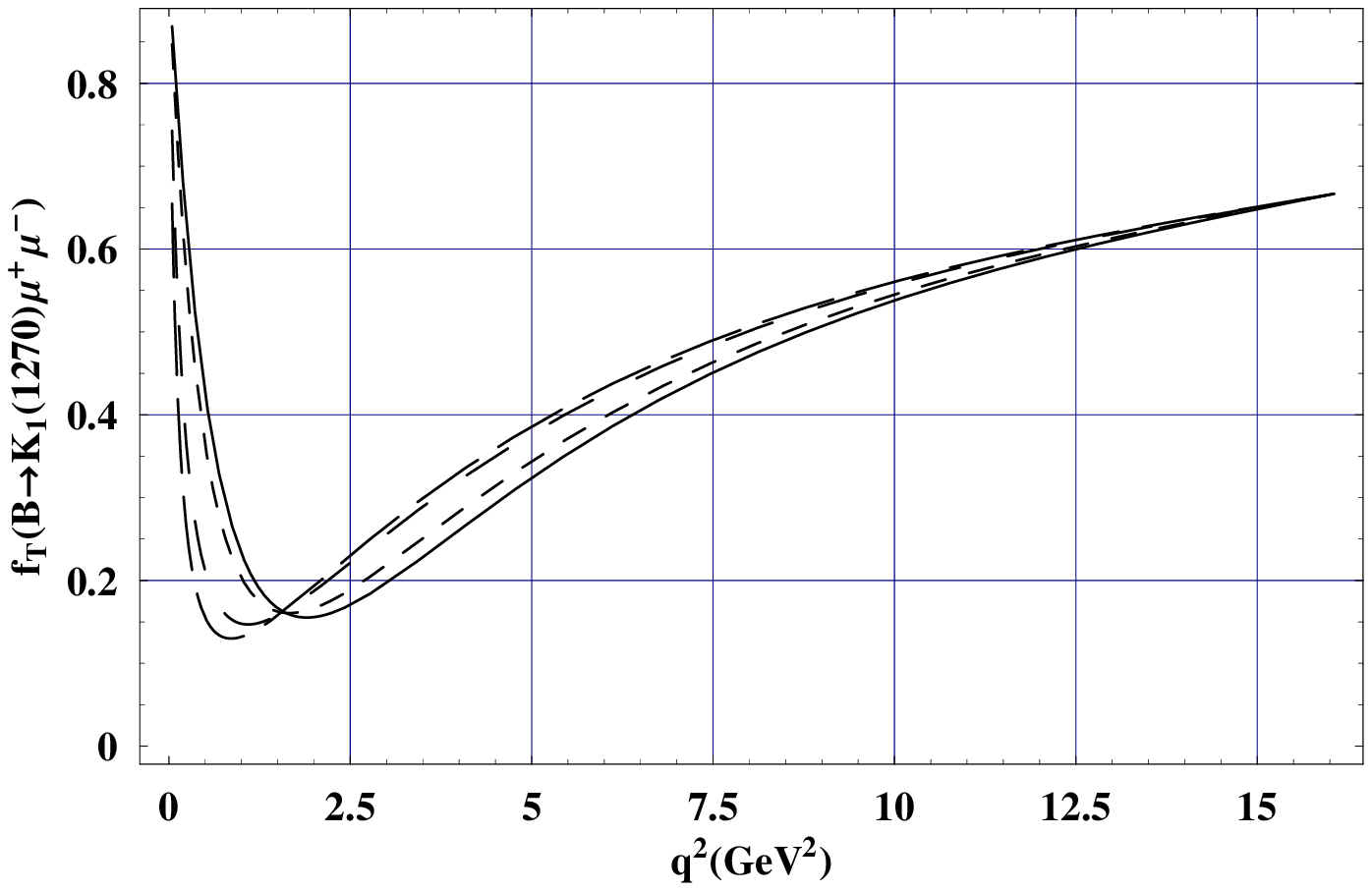}
\end{tabular}
\caption{The dependence the probabilities of the longitudinal $(a,\ b)$ and transverse $(c,\ d)$ helicity fractions, $f_{L,T}$, of $K_{1}$ in $B\to K_{1}(1270)\mu^{+}\mu^{-}$ decays on $q^{2}$ for different values of $m_{t^{\prime}}$ and $\left\vert V^{\ast}_{t^{\prime}b}V_{t^{\prime}s}\right\vert$. In all the graphs, the solid line corresponds to the SM, small dashed , medium dashed, long dashed correspond, $m_{t^{\prime}}=$ 300 GeV, 500 GeV and 600 GeV respectively. $\left\vert V^{\ast}_{t^{\prime}b}V_{t^{\prime}s}\right\vert$ has the value 0.003 and 0.015 in $(a,c)$ and $(b,d)$ respectively.} \label{HFL for muons 12}
\end{figure*}

Fig. \ref{HFL for muons 12} and \ref{HFL for muons 14} show the case
of muons as final state leptons, the effects of the fourth
generation on the longitudinal (transverse) helicity fractions of
$K_{1}(1270)$ are marked up in the $0<q^{2}\leq12$ GeV$^{2}$ region.
On the other hand, for $K_{1}(1400)$ the effected region is
$0<q^{2}\leq6$ GeV$^{2}$. Here one can notice that the $q^2$ region
of $K_{1}(1400)$ is smaller than that of $K_{1}(1270)$ but the
fourth generation effects are more prominent.  It is clear from
these figures that although the influence of the fourth generation
parameters on the maximum (minimum) values of the $K_{1}(1270,
1400)$ helicity fractions are not very much effected (One can see from Figs. \ref{HFL for muons 12} and \ref{HFL for muons 14} that for the case of $B\to K_{1}(1270)\mu^{+}\mu^{-}$, the difference in the extremum values of helicity fractions , even at the maximum values of fourth generation parameters, is negligible to the SM value and for $B\to K_{1}(1400)\mu^{+}\mu^{-}$ the difference to the SM value is 0.09) but there is a reasonable shift in the position of these values which lies roughly
at $q^{2}\simeq1.8$ GeV$^{2}$ for SM. Figs. \ref{HFL for muons 14}
and \ref{HFL for tauons 14} also show that how the position of
the maximum (minimum) values of $f_{L}$ ($f_{T}$) varies with the
change in $m_{t^{\prime}}$ and $|V_{t^{\prime}b}V_{t^{\prime}s}|$
values. Furthermore, the position of these extremum values are
shifted towards the low $q^{2}$ region and on setting the maximum
values of the fourth generation parameters this shift in the position
is approximately $0.9$ GeV$^{2}$. One more comment is necessary to
mention here that like the zero position of the $\mathcal{A}_{FB}$,
the position of the extremum values of the helicity fractions are not
effected due to the uncertainty of the mixing angle $\theta_{K}$.

\begin{figure*}[ht]
\begin{tabular}{cc}
\hspace{.6cm}($\mathbf{a}$)&\hspace{1.2cm}($\mathbf{b}$)\\
\includegraphics[scale=0.4]{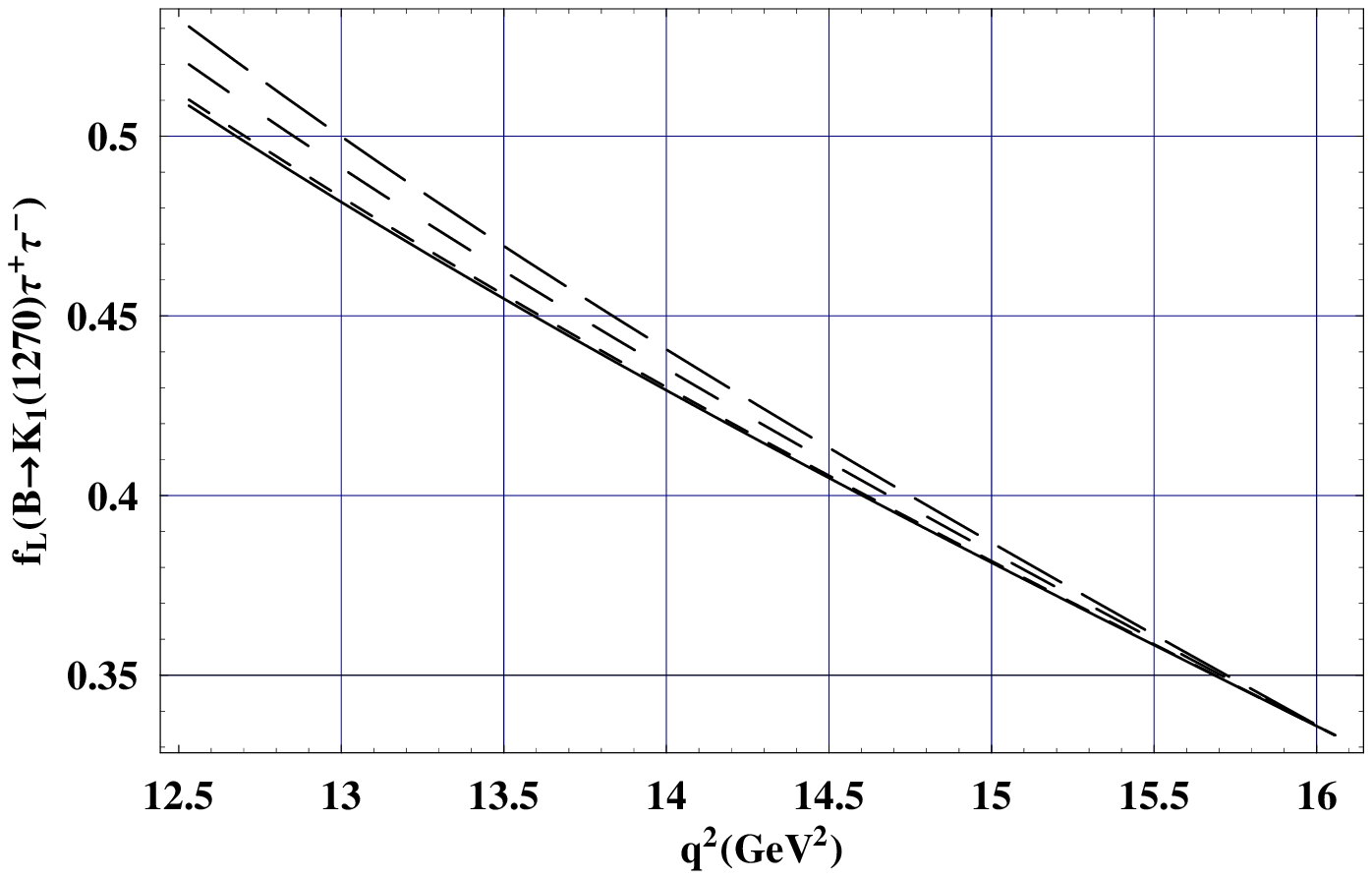} \ \ \
& \ \ \  \includegraphics[scale=0.4]{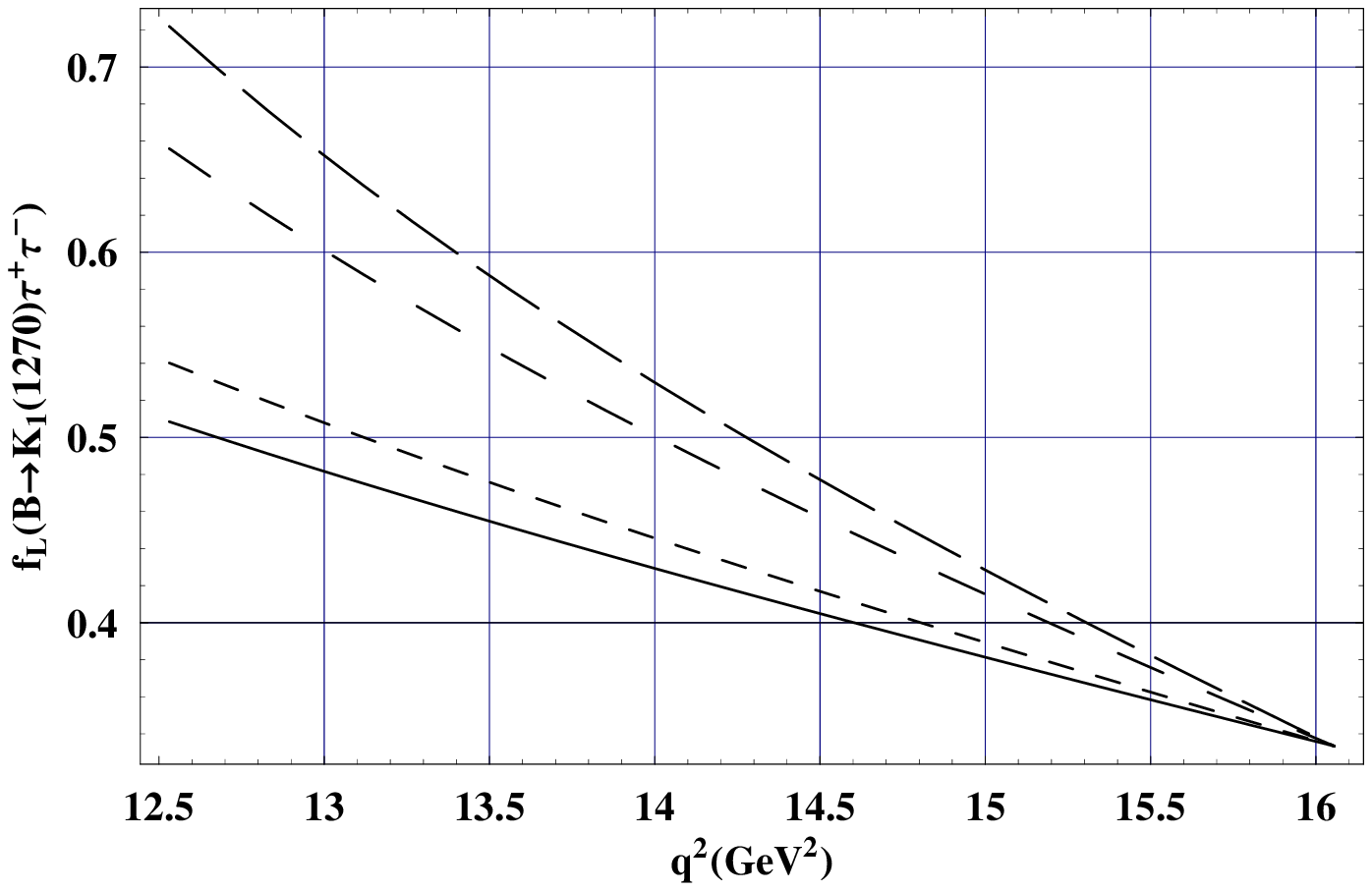}\\
\hspace{.6cm}($\mathbf{c}$)&\hspace{1.2cm}($\mathbf{d}$)\\
\includegraphics[scale=0.4]{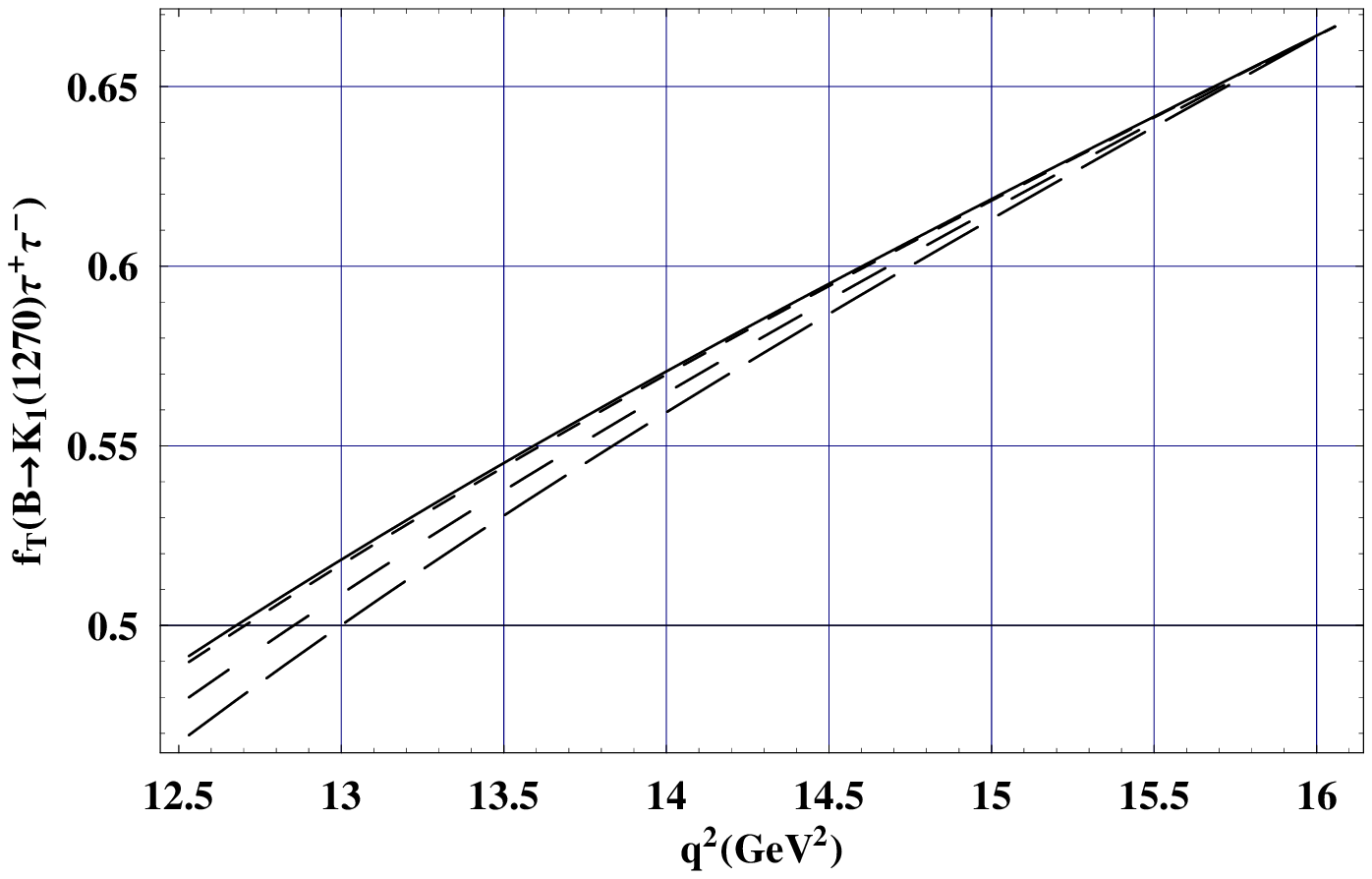} \ \ \
& \ \ \  \includegraphics[scale=0.4]{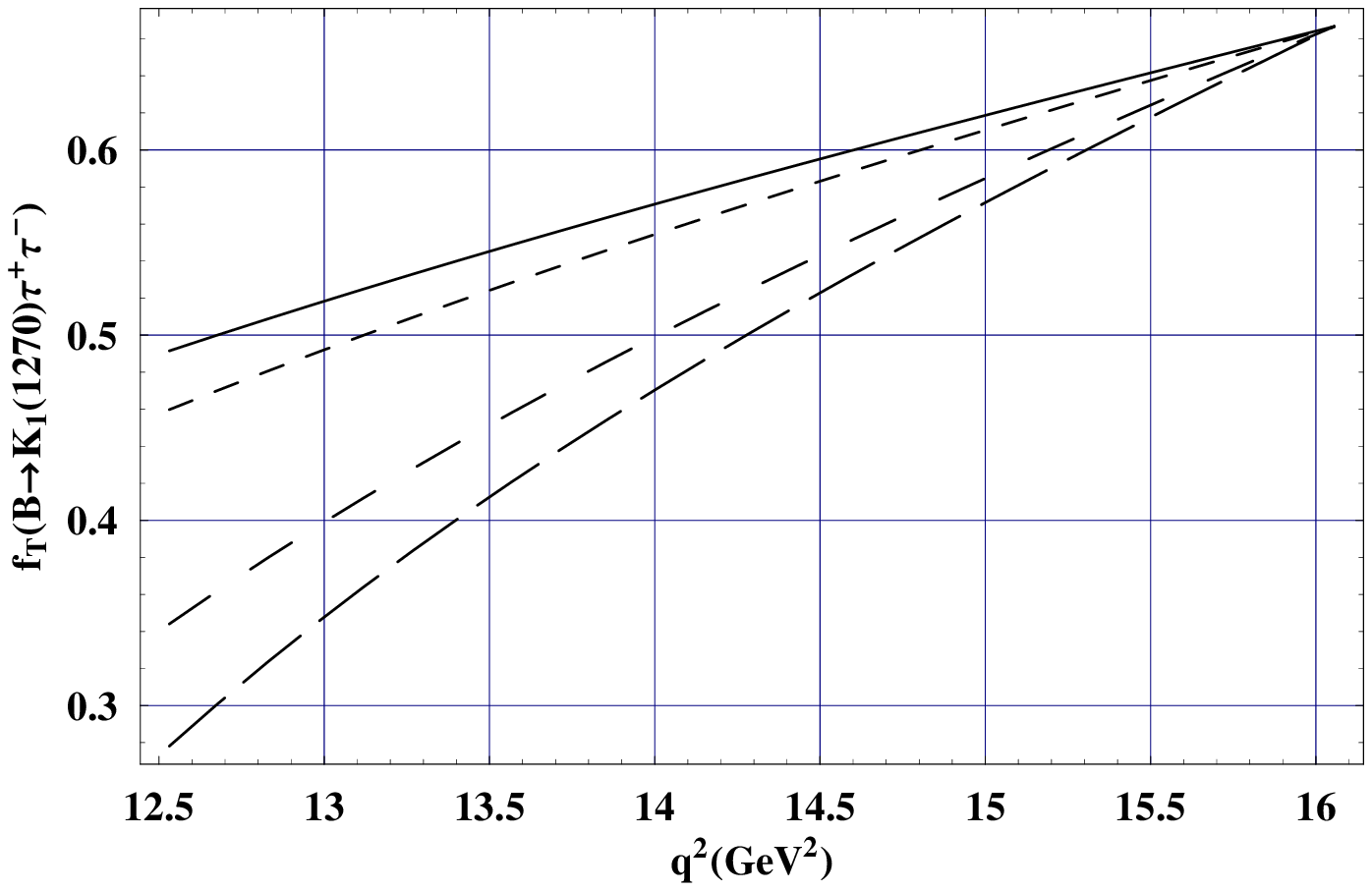}\end{tabular}
\caption{The dependence the probabilities of the longitudinal $(a,\ b)$ and transverse $(c,\ d)$ helicity fractions, $f_{L,T}$, of $K_{1}$ in $B\to K_{1}(1270)\tau^{+}\tau^{-}$ decays on $q^{2}$ for different values of $m_{t^{\prime}}$ and $\left\vert V^{\ast}_{t^{\prime}b}V_{t^{\prime}s}\right\vert$. The legends and the values
of fourth generation parameters are same as in Fig. \ref{HFL for muons 12}.}  \label{HFL for tauons 12}
\end{figure*}

\begin{figure*}[ht]
\begin{tabular}{cc}
\hspace{.6cm}($\mathbf{a}$)&\hspace{1.2cm}($\mathbf{b}$)\\
\includegraphics[scale=0.4]{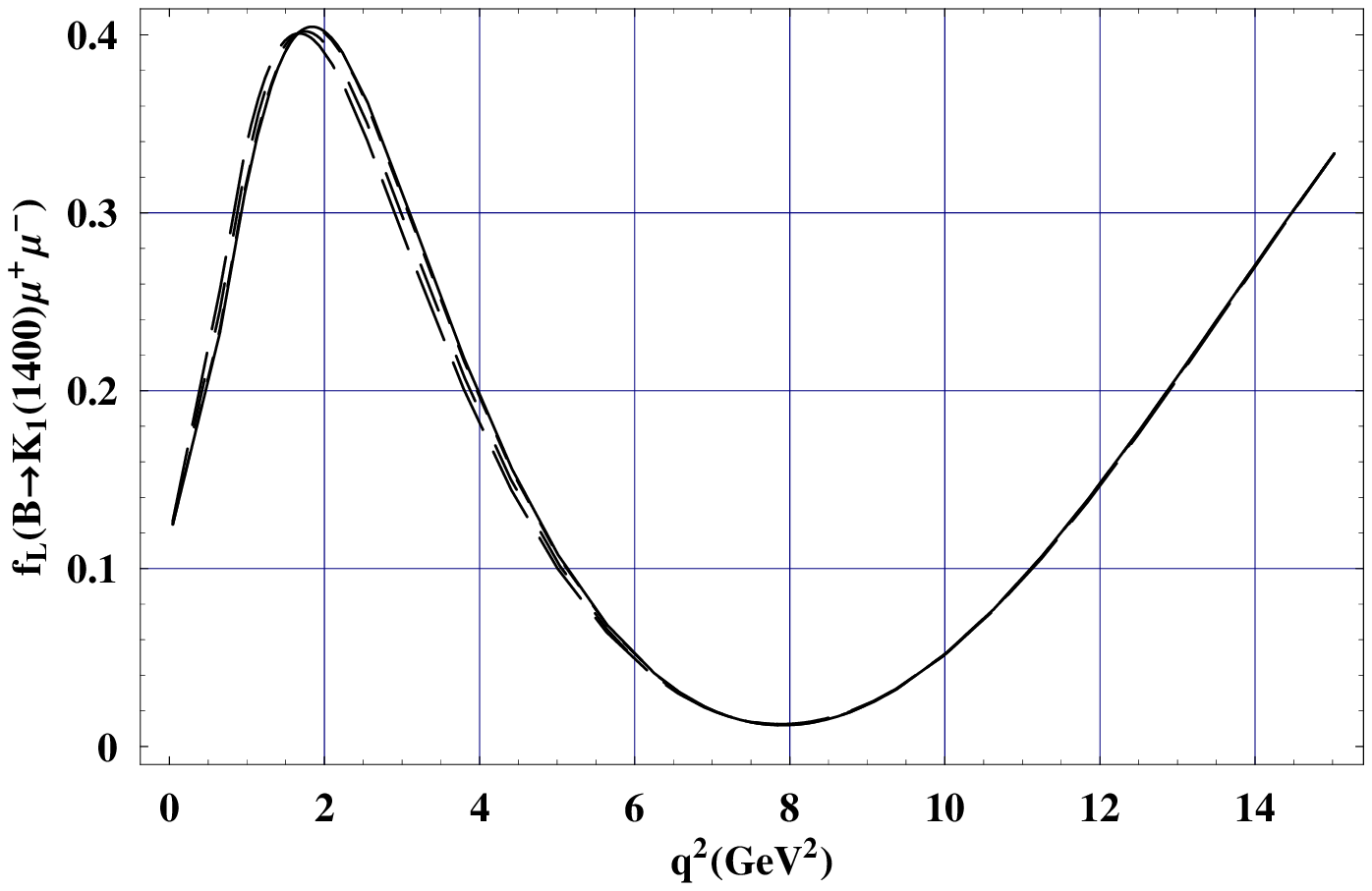} \ \ \
& \ \ \  \includegraphics[scale=0.4]{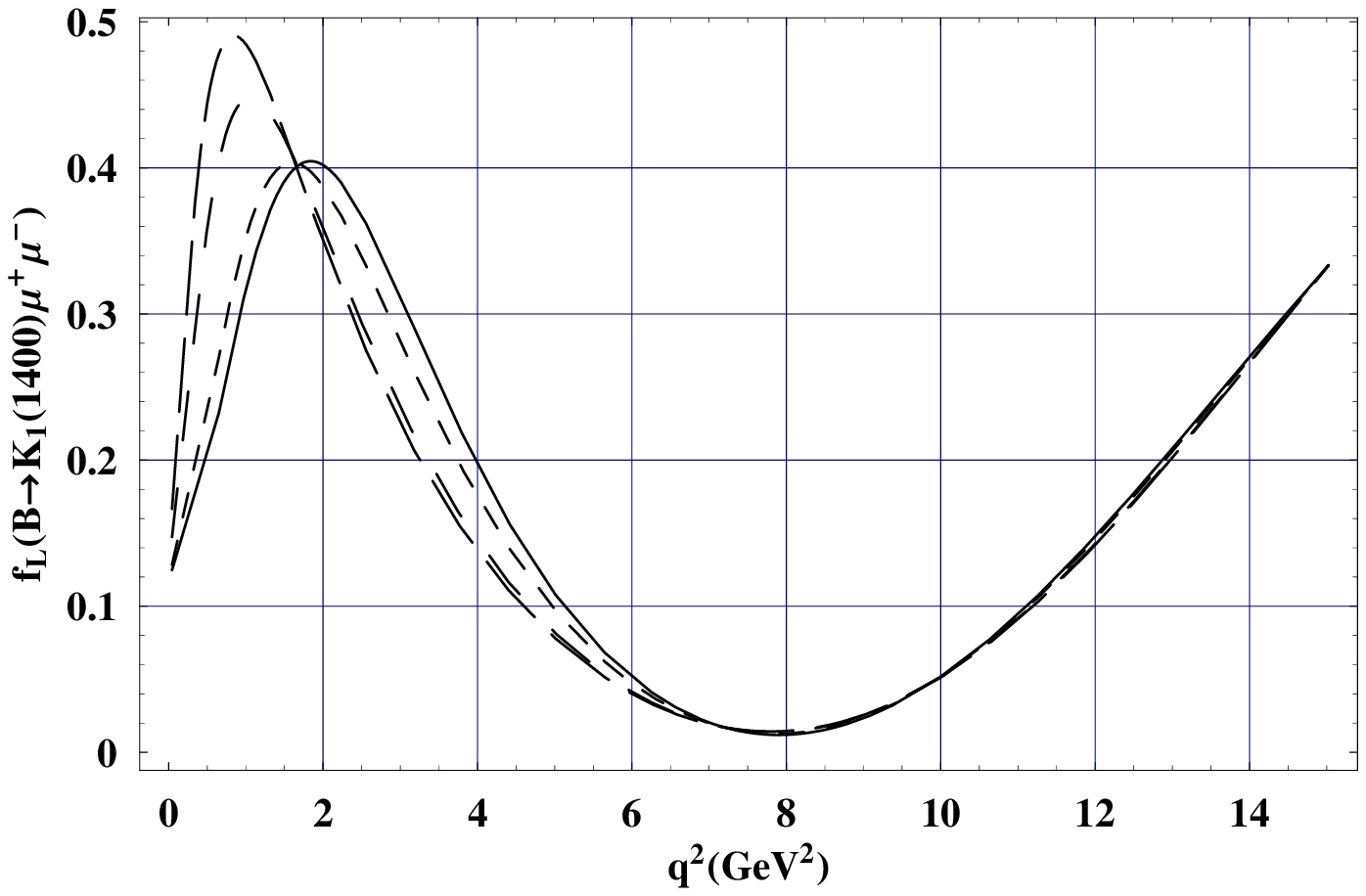}\\
\hspace{.6cm}($\mathbf{c}$)&\hspace{1.2cm}($\mathbf{d}$)\\
\includegraphics[scale=0.4]{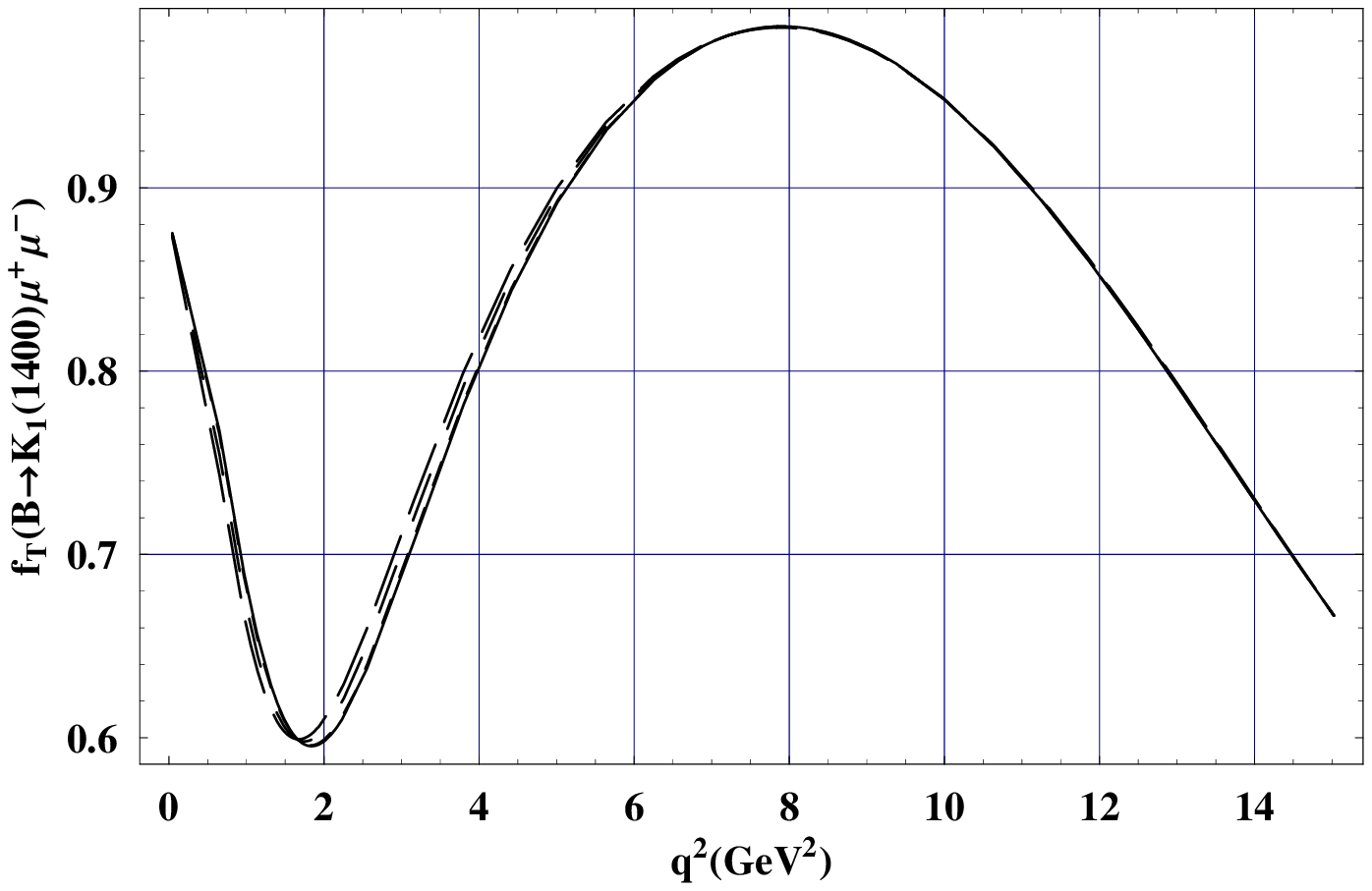} \ \ \
& \ \ \  \includegraphics[scale=0.4]{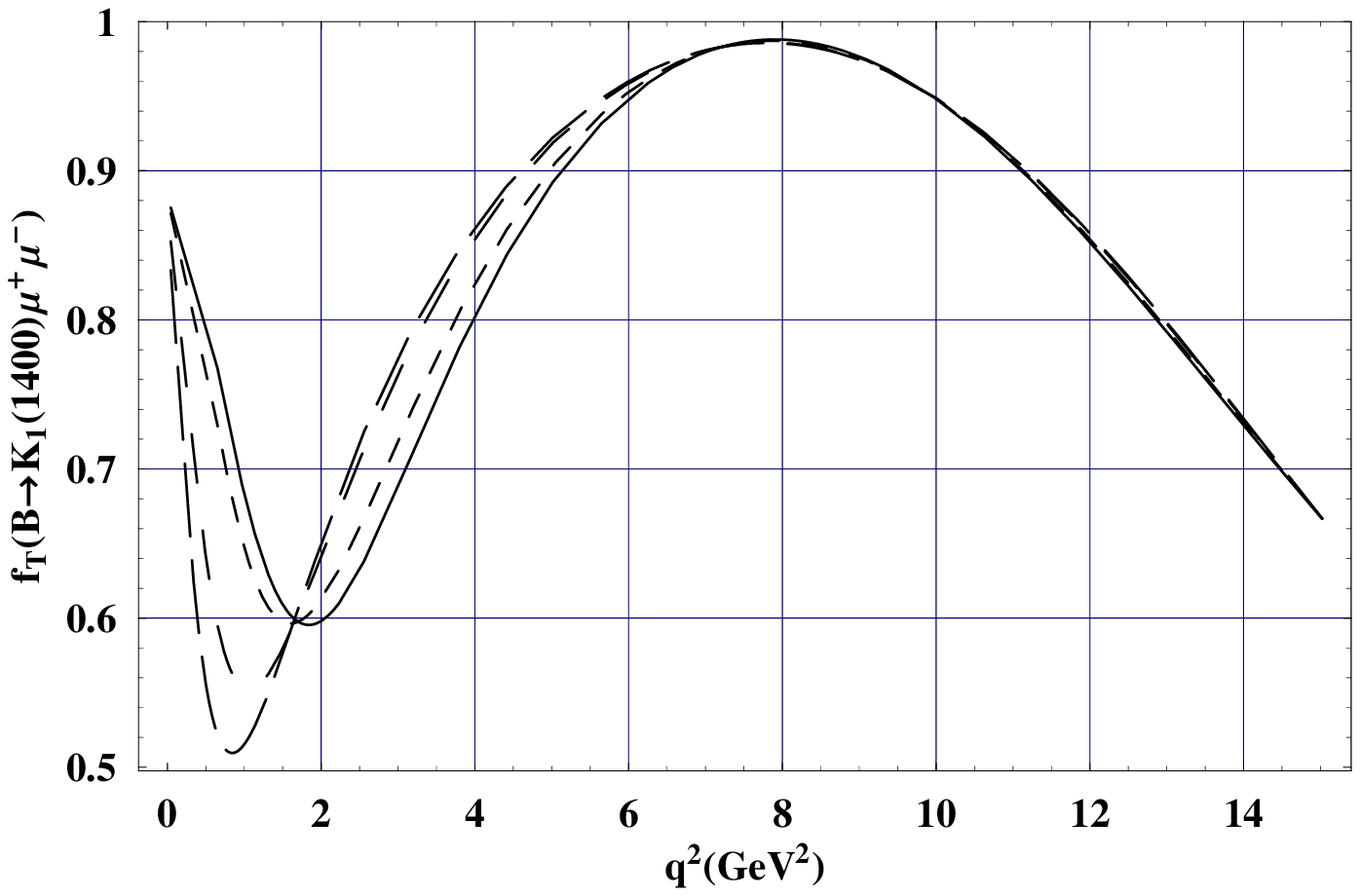}\end{tabular}
\caption{The dependence the probabilities of the longitudinal $(a,\ b)$ and transverse $(c,\ d)$ helicity fractions, $f_{L,T}$, of $K_{1}$ in $B\to K_{1}(1400)\mu^{+}\mu^{-}$ decays on $q^{2}$ for different values of $m_{t^{\prime}}$ and $\left\vert V^{\ast}_{t^{\prime}b}V_{t^{\prime}s}\right\vert$. The legends and the values
of fourth generation parameters are same as in Fig. \ref{HFL for muons 12}.}  \label{HFL for muons 14}
\end{figure*}

\begin{figure*}[ht]
\begin{tabular}{cc}
\hspace{.6cm}($\mathbf{a}$)&\hspace{1.2cm}($\mathbf{b}$)\\
\includegraphics[scale=0.4]{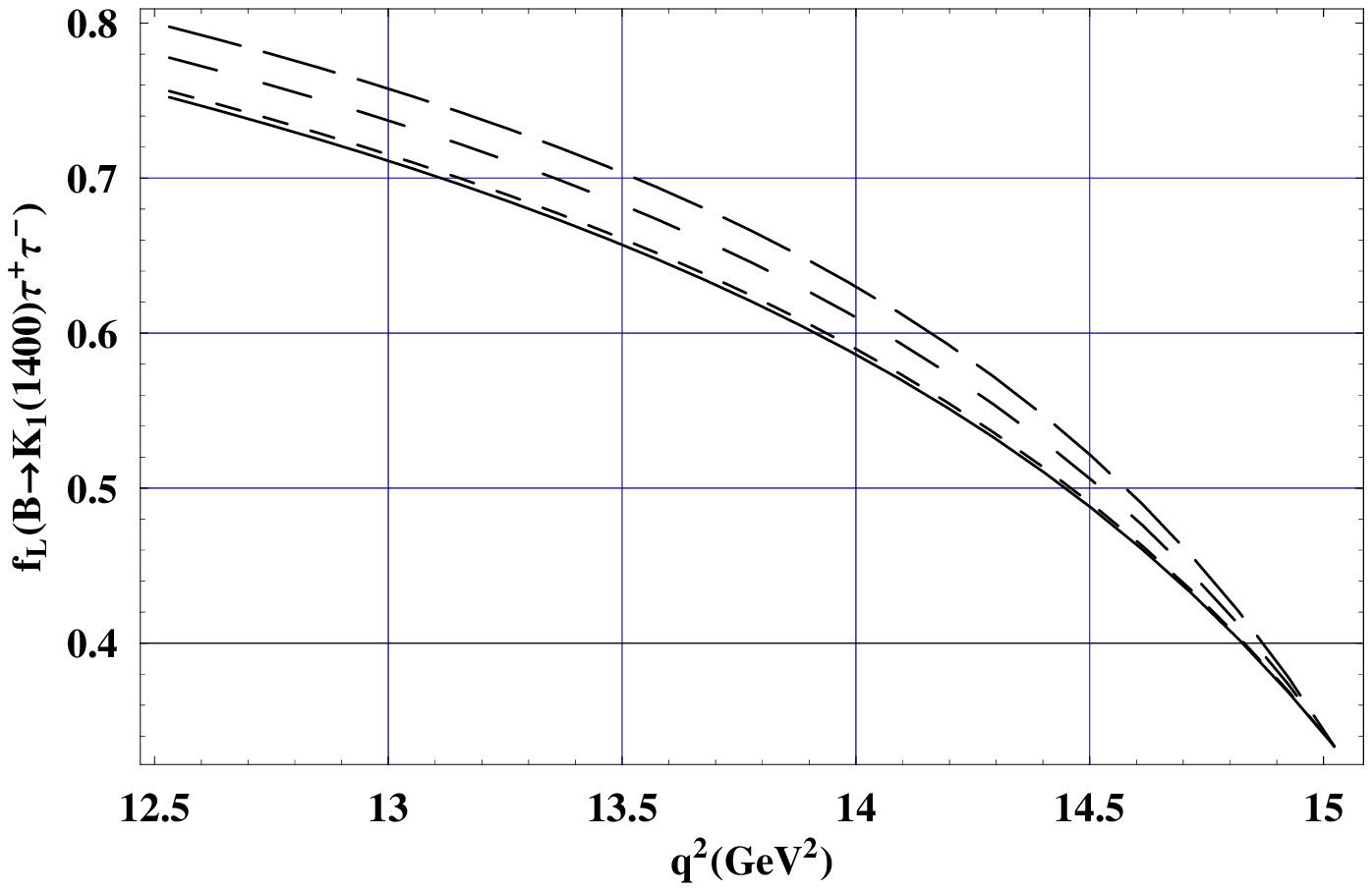} \ \ \
& \ \ \  \includegraphics[scale=0.4]{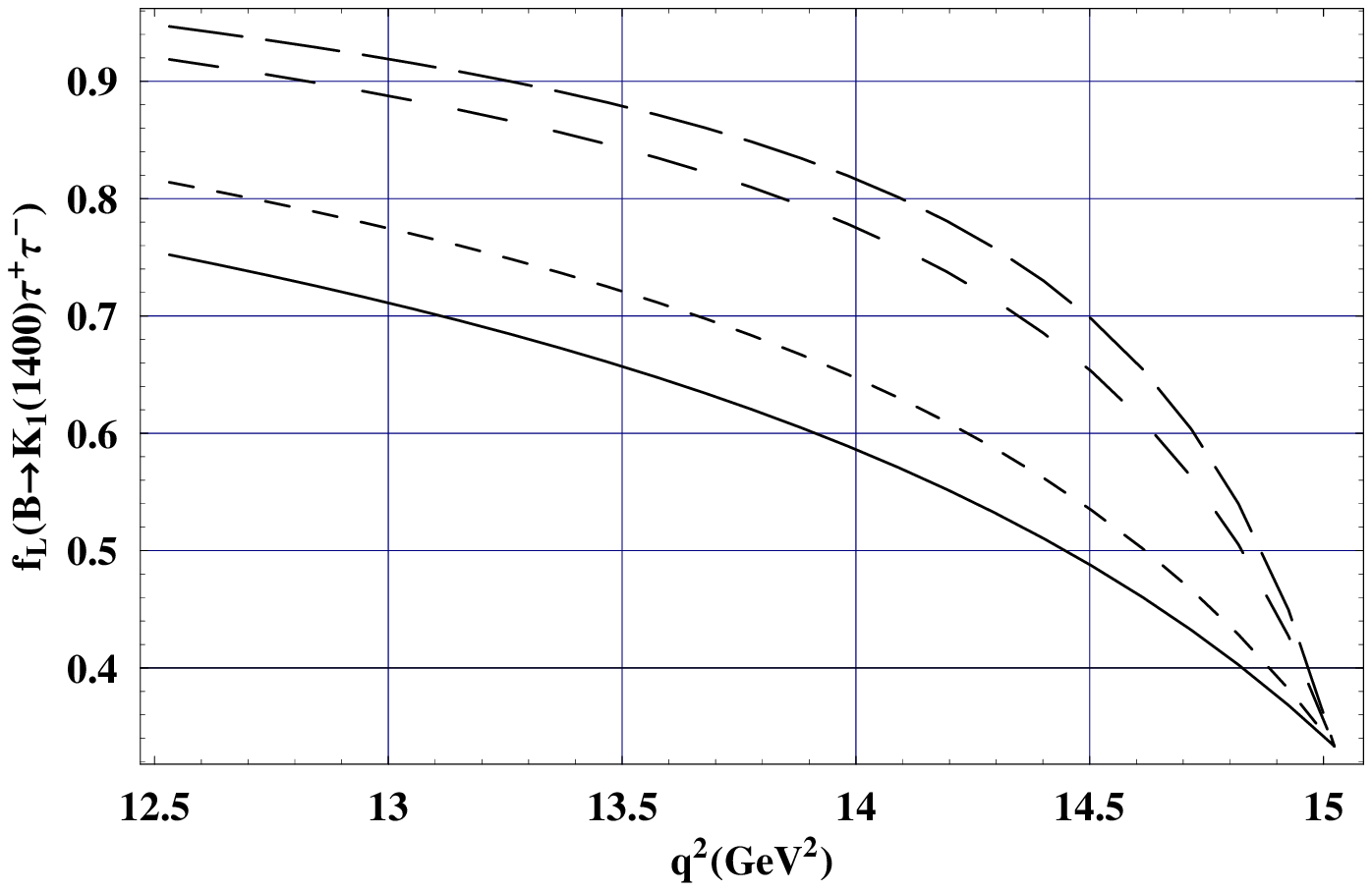}\\
\hspace{.6cm}($\mathbf{c}$)&\hspace{1.2cm}($\mathbf{d}$)\\
\includegraphics[scale=0.4]{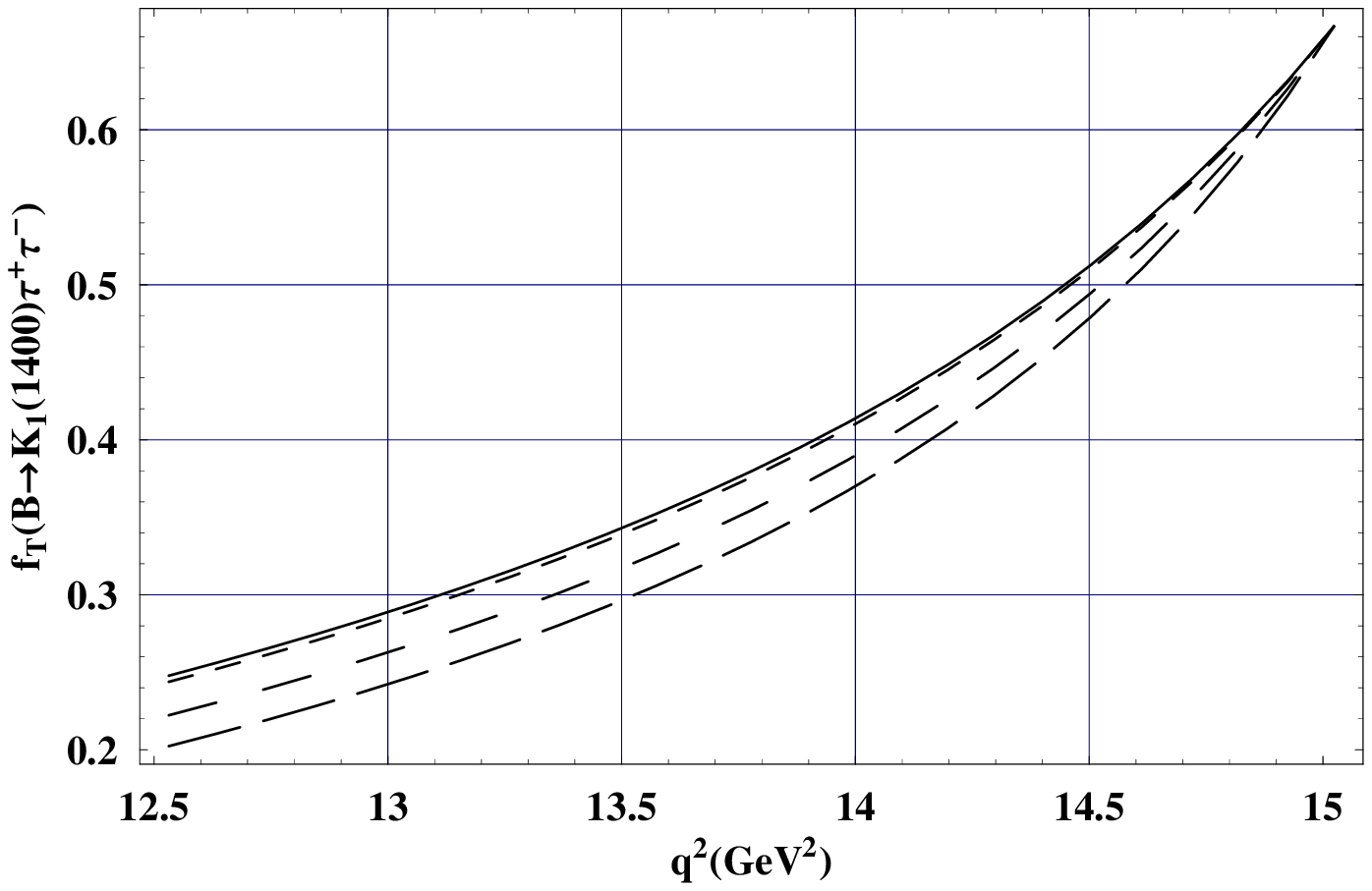} \ \ \
& \ \ \  \includegraphics[scale=0.4]{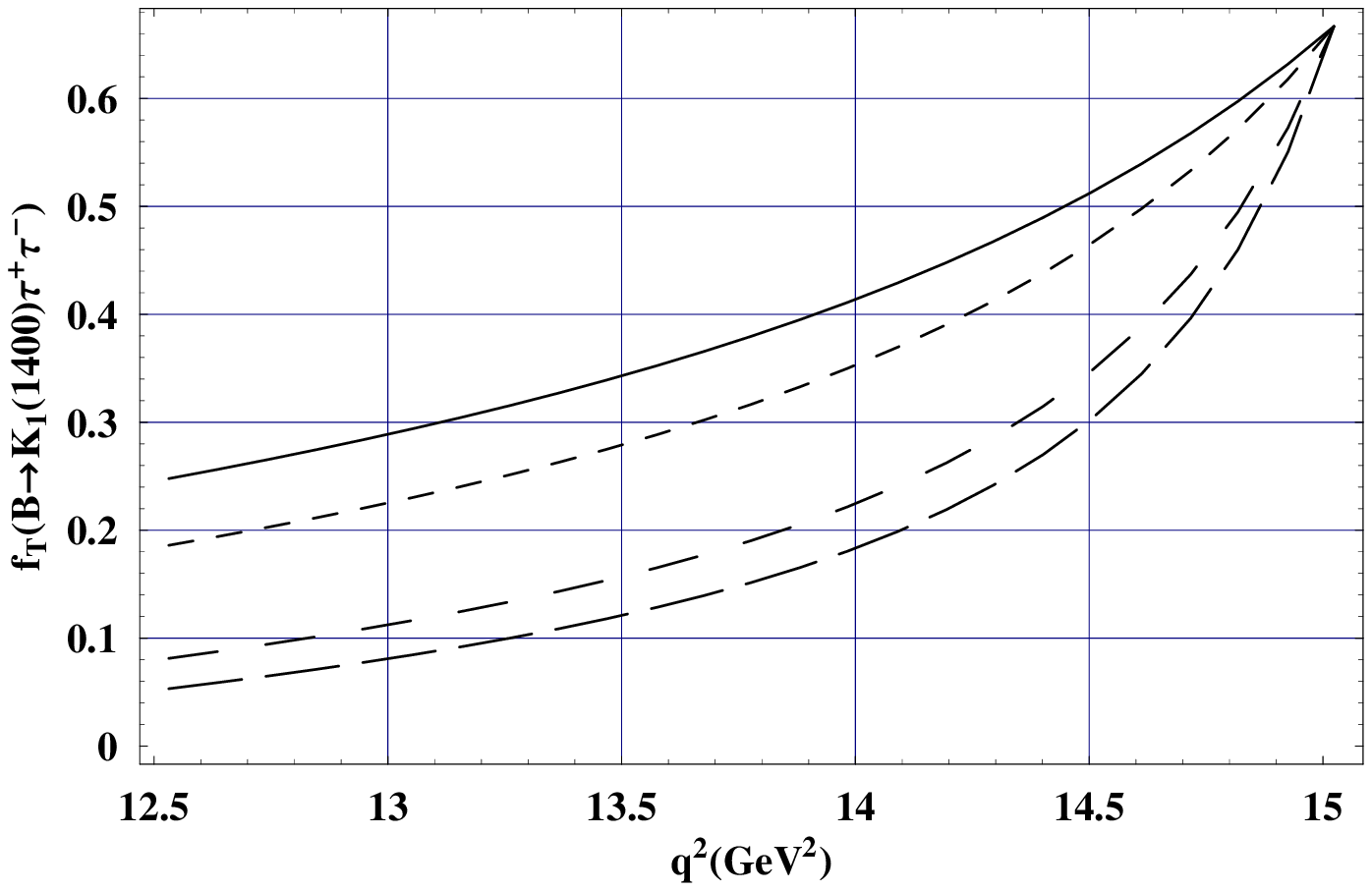}\end{tabular}
\caption{The dependence the probabilities of the longitudinal $(a,\ b)$ and transverse $(c,\ d)$ helicity fractions, $f_{L,T}$, of $K_{1}$ in $B\to K_{1}(1400)\tau^{+}\tau^{-}$ decays on $q^{2}$ for different values of $m_{t^{\prime}}$ and $\left\vert V^{\ast}_{t^{\prime}b}V_{t^{\prime}s}\right\vert$. The legends and the values
of fourth generation parameters are same as in Fig. \ref{HFL for muons 12}.}  \label{HFL for tauons 14}
\end{figure*}

Now we turn our attention to the case, where tauns are the final
state leptons and for this case the helicity fractions of
$K_{1}(1270, 1400)$ are plotted in Figs. \ref{HFL for tauons 12} and
\ref{HFL for tauons 14}. One can easily see that in contrast to the
case of muons, there is no shift in the position of the extremum
values of the helicity fractions, and are fixed at $q^{2}=12.5$
GeV$^{2}$. However, the change in the maximum (minimum) value of
longitudinal (transverse) is more prominent as compare to the
previous case where the muons are the final state leptons. These
figures have also enlightened the variation in the extremum values of
helicity fractions from the SM due to the change in the fourth
generation parameters. The change in extremum values are very well
marked up as compare to the uncertainties due to the mixing angle
$\theta_{K}$ and the hadronic matrix element. For $B\to
K_{1}(1270)\tau^{+}\tau^{-}$, the maximum setting of the fourth
generation parameters the maximum (minimum) value of longitudinal
(transverse) helicity fraction is changed from its SM value $0.51
(0.49)$ to $0.72 (0.28)$ and for $B\to K_{1}(1400)\tau^{+}\tau^{-}$
is changed from  $0.76 (0.24)$ to $0.92 (0.06)$ which is suitable
amount of change to measure.

The numerical analysis of helicity fractions shows that the
measurement of the maximum (minimum) values of $f_{L}$ and $f_{T}$
and its position in the case of $B\to K_{1}(1270,
1400)\tau^{+}\tau^{-}$ and $B\to K_{1}(1270, 1400)\mu^{+}\mu^{-}$
respectively can be used as a good tool in studying the NP beyond
the SM and the existence of the fourth generation quarks.

\section{Conclusion}\label{conc}

In our study on the rare $B\rightarrow
K_{1}(1270, 1400)\ell^{+}\ell^{-}$ decays with $\ell=\mu$, $\tau$, we have calculated branching ratio ($\mathcal{BR}$), the
forward backward asymmetry $\mathcal{A}_{FB}$ and helicity fractions
$f_{L,T}$ of the final state mesons and analyzed the implications of
the fourth generation effects on these observable for the said decays.

We have found a strong dependency of the $\mathcal{BR}$ on the fourth
generation parameters $V_{t^{\prime} b}V_{t^{\prime} s}$ and
$m_{t^{\prime}}$. The study has shown that the $\mathcal{BR}$ is an
increasing function of these parameters. At maximum values of these
parameters, i.e. $|V_{t^{\prime} b}V_{t^{\prime} s}|=0.015$ and
$m_{t^{\prime}}=600$ GeV, the values of $\mathcal{BR}$ increases
approximately 6 to 7 times larger than that of SM values when the
final leptons are muons and for the case of of tauns these values
are enhanced 3 to 4 times to the SM value. Hence the accurate
measurement of the $\mathcal{BR}s$ value for these decays is very
important tool to say something about the physics beyond the three
generation of SM.

Besides the $\mathcal{BR}$, our analysis shown that $\mathcal{A}_{FB}$
is also a very good observable to check the existence of the fourth
generation quarks, especially the zero position of the
$\mathcal{A}_{FB}$. We have found that the value of the
$\mathcal{A}_{FB}$ decreases with increases in the values of
$V_{t^{\prime} b}V_{t^{\prime} s}$ and $m_{t^{\prime}}$. Moreover,
the decrement in the values of the $\mathcal{A}_{FB}$ from the SM
values are important imprints of NP and also the shift in the zero
position of $\mathcal{A}_{FB}$ (which is towards low $q^{2}$ region)
provides a prominent signature of the NP fourth generation quarks.

To comprehend the fourth generation effects on these decays, we have
calculated the helicity fractions $f_{L,T}$ of final state mesons.
We have first calculated these helicity fractions of final state
mesons in the SM and then analyzed their extension to the fourth generation
scenario. The study has shown that the deviation from the SM values
of the helicity fractions are quite large when we set tauons as a
final state of leptons. It is also shown that there is a noticeable
change due to fourth generation in the position of the extremum values
of the longitudinal and transverse helicity fractions of $K_{1}$
meson for the case of muons as a final state leptons. Therefore, the
helicity fraction of $K_{1}$ meson can be a stringent test in
finding the status of the fourth generation quarks.

Another attraction to consider the decay channel $B\to K_{1}\ell^{+}\ell^{-}$ is to get the complimentary information about the parameters of fourth generation SM to that of the information obtained from other experiments such as the inclusive $B\to X_{s}\ell^{+}\ell^{-}$ and the exclusive $B\to M(K,K^{\ast})\ell^{+}\ell^{-}$ decays. It is also worth mentioning here that the information obtained about the fourth generation parameters from the other experiments can be used to fix the mixing angle $\theta_{K}$ between the $K_{1}$ states in our process. Therefore, the fourth generation SM information obtained from the other experiments will not only compliment our results but can be useful to understand the mixing nature of $K_{1}(1270)$ and $K_{1}(1400)$ mesons.

To sum up, the more data to be available from Tevatron and
LHCb will provide a powerful testing ground for the SM and the
possible existence of the fourth generation quarks and also put some constraints on the fourth generation parameters such as
$V_{t^{\prime} b}V_{t^{\prime} s}$ and $m_{t^{\prime}}$. Our
analysis of the fourth generation on the observables for $B\to K_{1}\ell^{+}\ell^{-}$ decays are
useful for probing or refuting the existence of fourth family of
quarks.

\section*{Acknowledgments}
The authors would like to thank Prof. Riazuddin and Prof. Fayyazuddin for their
valuable guidance and useful discussions.

\end{document}